\documentclass[5p,twocolumn]{elsarticle}

\usepackage[utf8]{inputenc}  
\usepackage[T1]{fontenc}     
\usepackage{lmodern}         
\usepackage{microtype}           
\usepackage[scaled]{helvet} 
\usepackage[english]{babel}  
\usepackage{graphicx}        
\usepackage[]{hyperref}            
\usepackage{comment}         
\biboptions{sort&compress}   

\journal{acmjacm}

\usepackage{amsmath,amssymb,latexsym}
\usepackage{stmaryrd}
\usepackage{hyperref}
\usepackage{paralist}
\usepackage[nounderscore]{syntax} 
\usepackage{subfig}
\usepackage{listings}
\usepackage{multirow}

\usepackage{etoolbox}
\AtBeginEnvironment{grammar}{\small}

\newcommand{\nospellcheck}[1]{#1}

\usepackage[x11names]{xcolor}
\usepackage{todonotes}

\newcommand{\sem}[2]{%
\ensuremath{\llbracket #1 : #2 \rrbracket}%
}
\newcommand{\tr}[1]{%
\ensuremath{\overline{#1}}%
}
\newcommand{\dom}[1]{%
\ensuremath{\mathbb{#1}}%
}

\newcommand{\decimate}[1]{%
\ensuremath{\pitchfork_{#1}}%
}

\newsavebox{\verbbox}

\newcommand{\esql}{eSQL}

\newcommand{\COMBINE}{\textsf{COMBINE}}

\newcommand{\FILTER}{\textsf{FILTER}}

\newcommand{\FROM}{\textsf{FROM}}

\newcommand{\ISA}{\textsf{IS A}}

\newcommand{\SELECT}{\textsf{SELECT}}

\newcommand{\WHEN}{\textsf{WHEN}}
\newcommand{\WHERE}{\textsf{WHERE}}
\newcommand{\WITH}{\textsf{WITH}}

\newcommand{\rpm}{\raisebox{.2ex}{$\scriptstyle\pm$}}

\usepackage{pifont}
\newcommand{\cmark}{\ding{51}}
\newcommand{\xmark}{\ding{55}}

\newcommand{\etal}{\textit{et al.}}

\newenvironment{codeblock}{}{}
\newenvironment{snippet}{\vskip 4pt\noindent}{\vskip 4pt}

\newcommand{\sz}{\phantom{s}}

\newcounter{querycounter}
\newenvironment{query}{%
\refstepcounter{querycounter}\begin{quote}\textbf{Query \thequerycounter.}%
}{\end{quote}}{}

\newcommand{\nt}[1]{\ensuremath{\langle\mbox{\textit{#1}}\rangle}}
\graphicspath{{fig/}}

\begin{document}

\begin{frontmatter}

\title{From Complex Event Processing to Simple Event Processing}

\author{Sylvain Hallé\fnref{label1}%
}
\fntext[label1]{Laboratoire d'informatique formelle, Université du Québec à Chicoutimi, Canada%
}
\begin{abstract}
Many problems in Computer Science can be framed as the computation of queries over sequences, or ``streams'' of data units called \emph{events}. The field of Complex Event Processing (CEP) relates to the techniques and tools developed to efficiently process these queries. However, most CEP systems developed so far have concentrated on relatively narrow types of queries, which consist of sliding windows, aggregation functions, and simple sequential patterns computed over events that have a fixed tuple structure. Many of them boast throughput, but in counterpart, they are difficult to setup and cumbersome to extend with user-defined elements.

This paper describes a variety of use cases taken from real-world scenarios that present features seldom considered in classical CEP problems. It also provides a broad review of current solutions, that includes tools and techniques going beyond typical surveys on CEP. From a critical analysis of these solutions, design principles for a new type of event stream processing system are exposed. The paper proposes a simple, generic and extensible framework for the processing of event streams of diverse types; it describes in detail a stream processing engine, called BeepBeep, that implements these principles. BeepBeep's modular architecture, which borrows concepts from many other systems, is complemented with an extensible query language, called \esql{}. The end result is an open, versatile, and reasonably efficient query engine that can be used in situations that go beyond the capabilities of existing systems.

\vskip 8pt
\noindent
\textbf{Keywords:} event processing, software testing, query languages, runtime verification

\end{abstract}
\end{frontmatter}



\newsavebox{\siddhicodebox}
\begin{lrbox}{\siddhicodebox}
\begin{lstlisting}[numbers=left]
public class SiddhiExample {
 public static void main(String[] args) {
  SiddhiManager man = new SiddhiManager();
  String query = "foo";
  ExecutionPlanRuntime epr = man.createExecutionPlanRuntime(query);
  epr.addCallback("query", new SiddhiCallback());
  InputHandler inputHandler = epr.getInputHandler("trace");
  epr.start();
  epr.shutdown();
  man.shutdown();
 }

 class MyCallback extends QueryCallback {
   public void receive(long ts, Event[] in_e, Event[] rm_e) {
   }
 }
\end{lstlisting}
\end{lrbox}

\newsavebox{\espercodebox}
\begin{lrbox}{\espercodebox}
\begin{lstlisting}[numbers=left]
public class EsperExample {

 public static void main(String[] args) {
  Configuration conf = new Configuration();
  conf.addEventType("TupleEvent", TupleEvent.class.getName());
  EPServiceProvider epService = 
    EPServiceProviderManager.getProvider("MyURI", conf);
  epService.initialize();
  EPStatement statement = 
    epService.getEPAdministrator().createEPL("foo");
  statement.addListener(new EsperListener());
  Scanner scanner = new Scanner(new File("trace.csv"));
  while (scanner.hasNextLine()) {
    String[] parts = line.split(",");
    TupleEvent e = new TupleEvent(scanner.nextLine());
    epService.getEPRuntime().sendEvent(e);
  }
  scanner.close();
 }
 
 class TupleEvent {
  int a;
  int b;

  public TupleEvent(String line) {
   String[] parts = line.trim().split(",");
   a = Integer.parseInt(parts[0]);
   b = Integer.parseInt(parts[1]);
  }
  
  public int getA() { return a; }
  public int getB() { return b; }
 }
 
 class MyListener extends UpdateListener {
   public void update(EventBean[] in_e, EventBean[] ol_e) {
     // Process output event here
   }
 }
\end{lstlisting}
\end{lrbox}


\newcommand{\machinestring}{\href{M1.0}{Intel Xeon E5-2630L v2 2.40GHz running Ubuntu 14.04}}

\newcommand{\jvmram}{\href{M1.1}{2499}}

\newcommand{\numexperiments}{\href{M1.2}{45}}

\newcommand{\numdatapoints}{\href{M1.3}{86}}

\newcommand{\factorEsper}{\href{M1.4}{1.0}}

\newcommand{\factorSase}{\href{M1.5}{1.0}}

\newcommand{\factorSiddhi}{\href{M1.6}{1.0}}

\newcommand{\timeout}{\href{M1.7}{600}}

\newcommand{\longestTrace}{\href{M1.8}{$10^\{6\}$}}

%

\newsavebox{\rawThroughput}
\begin{lrbox}{\rawThroughput}
\begin{tabular}{|c|c|c|}
\hline
\textbf{name} & \textbf{tool} & \textbf{throughput}\\
\hline\hline
 \multirow{3}{*}{\href{T1.0.0}{S1}} & {\href{T1.14.1}{BeepBeep}} & {\href{T1.14.2}{390396}}\\
\cline{2-3}
 & {\href{T1.0.1}{Esper}} & {\href{T1.0.2}{500847}}\\
\cline{2-3}
 & {\href{T1.1.1}{Siddhi}} & {\href{T1.1.2}{546887}}\\
\hline
 \multirow{3}{*}{\href{T1.3.0}{S2}} & {\href{T1.15.1}{BeepBeep}} & {\href{T1.15.2}{543113}}\\
\cline{2-3}
 & {\href{T1.2.1}{Esper}} & {\href{T1.2.2}{784530}}\\
\cline{2-3}
 & {\href{T1.3.1}{Siddhi}} & {\href{T1.3.2}{427206}}\\
\hline
 \multirow{3}{*}{\href{T1.16.0}{S3}} & {\href{T1.16.1}{BeepBeep}} & {\href{T1.16.2}{520514}}\\
\cline{2-3}
 & {\href{T1.4.1}{Esper}} & {\href{T1.4.2}{735082}}\\
\cline{2-3}
 & {\href{T1.5.1}{Siddhi}} & {\href{T1.5.2}{512954}}\\
\hline
 \multirow{3}{*}{\href{T1.6.0}{S4}} & {\href{T1.17.1}{BeepBeep}} & {\href{T1.17.2}{473715}}\\
\cline{2-3}
 & {\href{T1.6.1}{Esper}} & {\href{T1.6.2}{306207}}\\
\cline{2-3}
 & {\href{T1.7.1}{Siddhi}} & {\href{T1.7.2}{473939}}\\
\hline
 \multirow{3}{*}{\href{T1.19.0}{S5}} & {\href{T1.19.1}{BeepBeep}} & {\href{T1.19.2}{82295}}\\
\cline{2-3}
 & {\href{T1.8.1}{Esper}} & {\href{T1.8.2}{166974}}\\
\cline{2-3}
 & {\href{T1.9.1}{Siddhi}} & {\href{T1.9.2}{391988}}\\
\hline
 \multirow{3}{*}{\href{T1.10.0}{S6}} & {\href{T1.18.1}{BeepBeep}} & {\href{T1.18.2}{319479}}\\
\cline{2-3}
 & {\href{T1.10.1}{Esper}} & {\href{T1.10.2}{657733}}\\
\cline{2-3}
 & {\href{T1.11.1}{Siddhi}} & {\href{T1.11.2}{501652}}\\
\hline
 \multirow{3}{*}{\href{T1.12.0}{S7}} & {\href{T1.20.1}{BeepBeep}} & {\href{T1.20.2}{58105}}\\
\cline{2-3}
 & {\href{T1.12.1}{Esper}} & {\href{T1.12.2}{138060}}\\
\cline{2-3}
 & {\href{T1.13.1}{Siddhi}} & {\href{T1.13.2}{167100}}\\
\hline
 \multirow{4}{*}{\href{T1.22.0}{Temporal query}} & {\href{T1.21.1}{BeepBeep}} & {}\\
\cline{2-3}
 & {\href{T1.23.1}{Esper}} & {}\\
\cline{2-3}
 & {\href{T1.24.1}{MySQL}} & {}\\
\cline{2-3}
 & {\href{T1.22.1}{Siddhi}} & {}\\

\hline
\end{tabular}
\end{lrbox}

\newsavebox{\throughputByTool}
\begin{lrbox}{\throughputByTool}
\begin{tabular}{|c|c|c|c|c|}
\hline
\textbf{name} & \textbf{MySQL} & \textbf{BeepBeep} & \textbf{Esper} & \textbf{Siddhi}\\
\hline\hline
{\href{T2.0.0}{S1}} & {\href{T2.0.1}{}} & {\href{T2.0.2}{390396}} & {\href{T2.0.3}{500847}} & {\href{T2.0.4}{546887}}\\
\hline
{\href{T2.1.0}{S2}} & {\href{T2.1.1}{}} & {\href{T2.1.2}{543113}} & {\href{T2.1.3}{784530}} & {\href{T2.1.4}{427206}}\\
\hline
{\href{T2.2.0}{S3}} & {\href{T2.2.1}{}} & {\href{T2.2.2}{520514}} & {\href{T2.2.3}{735082}} & {\href{T2.2.4}{512954}}\\
\hline
{\href{T2.3.0}{S4}} & {\href{T2.3.1}{}} & {\href{T2.3.2}{473715}} & {\href{T2.3.3}{306207}} & {\href{T2.3.4}{473939}}\\
\hline
{\href{T2.4.0}{S5}} & {\href{T2.4.1}{}} & {\href{T2.4.2}{82295}} & {\href{T2.4.3}{166974}} & {\href{T2.4.4}{391988}}\\
\hline
{\href{T2.5.0}{S6}} & {\href{T2.5.1}{}} & {\href{T2.5.2}{319479}} & {\href{T2.5.3}{657733}} & {\href{T2.5.4}{501652}}\\
\hline
{\href{T2.6.0}{S7}} & {\href{T2.6.1}{}} & {\href{T2.6.2}{58105}} & {\href{T2.6.3}{138060}} & {\href{T2.6.4}{167100}}\\
\hline
{\href{T2.7.0}{Temporal query}} & {} & {} & {} & {}\\

\hline
\end{tabular}
\end{lrbox}

\newsavebox{\relativeThroughput}
\begin{lrbox}{\relativeThroughput}
\begin{tabular}{|c|c|c|c|c|}
\hline
\textbf{name} & \textbf{MySQL} & \textbf{BeepBeep} & \textbf{Esper} & \textbf{Siddhi}\\
\hline\hline
{\href{T3.0.0}{S1}} & {\href{T3.0.1}{}} & {\href{T3.0.2}{1.0}} & {\href{T3.0.3}{1.2829204}} & {\href{T3.0.4}{1.400852}}\\
\hline
{\href{T3.1.0}{S2}} & {\href{T3.1.1}{}} & {\href{T3.1.2}{1.271314}} & {\href{T3.1.3}{1.8364209}} & {\href{T3.1.4}{1.0}}\\
\hline
{\href{T3.2.0}{S3}} & {\href{T3.2.1}{}} & {\href{T3.2.2}{1.0147382}} & {\href{T3.2.3}{1.4330369}} & {\href{T3.2.4}{1.0}}\\
\hline
{\href{T3.3.0}{S4}} & {\href{T3.3.1}{}} & {\href{T3.3.2}{1.5470417}} & {\href{T3.3.3}{1.0}} & {\href{T3.3.4}{1.5477732}}\\
\hline
{\href{T3.4.0}{S5}} & {\href{T3.4.1}{}} & {\href{T3.4.2}{1.0}} & {\href{T3.4.3}{2.028969}} & {\href{T3.4.4}{4.7632055}}\\
\hline
{\href{T3.5.0}{S6}} & {\href{T3.5.1}{}} & {\href{T3.5.2}{1.0}} & {\href{T3.5.3}{2.0587676}} & {\href{T3.5.4}{1.570219}}\\
\hline
{\href{T3.6.0}{S7}} & {\href{T3.6.1}{}} & {\href{T3.6.2}{1.0}} & {\href{T3.6.3}{2.3760433}} & {\href{T3.6.4}{2.8758283}}\\
\hline
{\href{T3.7.0}{Temporal query}} & {} & {} & {} & {}\\

\hline
\end{tabular}
\end{lrbox}

\newsavebox{\codebaseSize}
\begin{lrbox}{\codebaseSize}
\begin{tabular}{|c|c|}
\hline
\textbf{tool} & \textbf{size}\\
\hline\hline
{\href{T4.0.0}{BeepBeep}} & {\href{T4.0.1}{317}}\\
\hline
{\href{T4.2.0}{Esper}} & {\href{T4.2.1}{5870}}\\
\hline
{\href{T4.3.0}{SASE}} & {\href{T4.3.1}{183}}\\
\hline
{\href{T4.1.0}{Siddhi}} & {\href{T4.1.1}{7140}}\\

\hline
\end{tabular}
\end{lrbox}

\newsavebox{\plumbing}
\begin{lrbox}{\plumbing}
\begin{tabular}{|c|c|}
\hline
\textbf{Number of pipes} & \textbf{Throughput}\\
\hline\hline
{\href{T5.0.0}{0}} & {\href{T5.0.1}{1878992}}\\
\hline
{\href{T5.1.0}{4}} & {\href{T5.1.1}{649350}}\\
\hline
{\href{T5.2.0}{8}} & {\href{T5.2.1}{380430}}\\
\hline
{\href{T5.3.0}{12}} & {\href{T5.3.1}{289318}}\\
\hline
{\href{T5.4.0}{16}} & {\href{T5.4.1}{205829}}\\
\hline
{\href{T5.5.0}{20}} & {\href{T5.5.1}{159708}}\\
\hline
{\href{T5.6.0}{24}} & {\href{T5.6.1}{133944}}\\
\hline
{\href{T5.7.0}{28}} & {\href{T5.7.1}{123186}}\\
\hline
{\href{T5.8.0}{32}} & {\href{T5.8.1}{103111}}\\
\hline
{\href{T5.9.0}{36}} & {\href{T5.9.1}{93477}}\\
\hline
{\href{T5.10.0}{40}} & {\href{T5.10.1}{83450}}\\
\hline
{\href{T5.11.0}{44}} & {\href{T5.11.1}{72442}}\\
\hline
{\href{T5.12.0}{48}} & {\href{T5.12.1}{71552}}\\
\hline
{\href{T5.13.0}{52}} & {\href{T5.13.1}{66963}}\\
\hline
{\href{T5.14.0}{56}} & {\href{T5.14.1}{60245}}\\
\hline
{\href{T5.15.0}{60}} & {\href{T5.15.1}{54637}}\\

\hline
\end{tabular}
\end{lrbox}

%

\newsavebox{\plotthroughputByTool}
\begin{lrbox}{\plotthroughputByTool}
\href{P1.0}{\includegraphics[page=1,width=\linewidth]{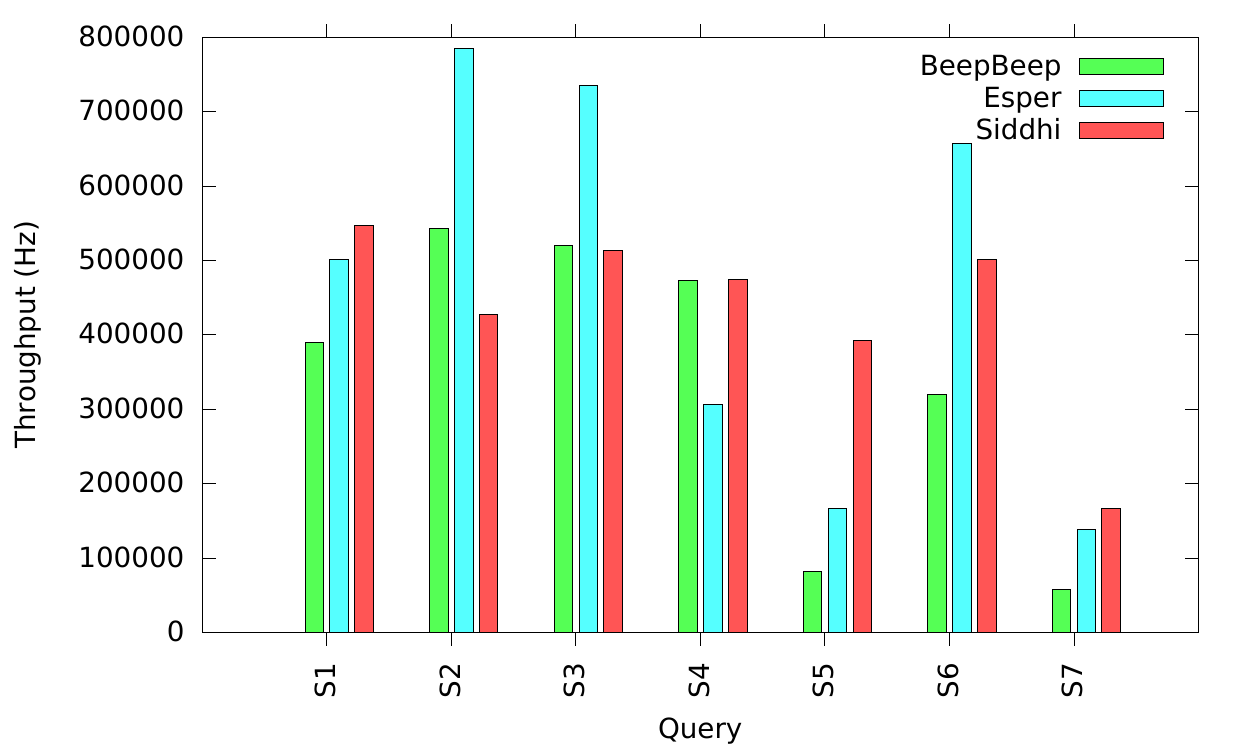}}
\end{lrbox}

\newsavebox{\plotcodebaseSize}
\begin{lrbox}{\plotcodebaseSize}
\href{P2.0}{\includegraphics[page=2,width=\linewidth]{labpal-plots.pdf}}
\end{lrbox}

\newsavebox{\plotplumbing}
\begin{lrbox}{\plotplumbing}
\href{P3.0}{\includegraphics[page=3,width=\linewidth]{labpal-plots.pdf}}
\end{lrbox}

\section{Introduction}\label{sec:intro} 

Event streams have become an important part of the mass of data produced by computing systems. They can be generated by a myriad of sources such as sensors \cite{Wang2013,Jia2009,DBLP:conf/aaai/HalleGB16}, business process logs \cite{process-mining}, instrumented software \cite{DBLP:conf/icst/CalvarTH12,DBLP:journals/jlp/LeuckerS09,DBLP:conf/icse/JinMLR12}, financial transactions \cite{Adi2006}, healthcare systems \cite{DBLP:conf/edoc/BerryM13}, and network packet captures \cite{DBLP:conf/aina/LaFC16}. The ability to collect and process these event streams can be put to good use in fields as diverse as software testing, data mining, and compliance auditing.

Event stream processing typically involves computations that go beyond the evaluation of simple functions on individual events. Of prime importance is the possibility to perform correlations between events, either at multiple moments in time within a single stream, or even between events taken from different event streams. The term \emph{Complex Event Processing} (CEP) has been coined to refer to computations of this nature. One of the goals of CEP is to create aggregated (i.e.\ ``complex'') events using data fetched from one or more lower-level events \cite{DBLP:books/daglib/0017658}. This computation can be executed in cascade, with the output streams of one process becoming the input streams of the next, leading to events of increasingly higher levels of abstraction. Section \ref{sec:scenarios} starts this paper by presenting a wide range of examples taken from domains as varied as bug detection in video games and network intrusion detection.

This potent concept has spawned an impressive amount of work in the past twenty years. As we will see in Section \ref{sec:related}, there exist literally dozens of competing systems claiming the CEP label, ranging from academic proofs-of-concept to commercial cloud frameworks such as Apache Spark or Microsoft Azure. These systems are based on a commensurate number of research papers, technical reports and buzzword-laden whitepapers introducing a plethora of incompatible formalizations of the problem. 
This reveals that CEP has never been a single problem, but rather a family of related problems sharing a relatively blurry common ground.
Many of these systems and frameworks, however, have in common the fact that they deserve the epithet ``complex''. They often rely on an intricate definition of seemingly simple concepts; some of them don't even state them formally, making their available implementation a \textit{de facto} specification. Many of their built-in query languages are quirky outgrowths of SQL, whose syntax seldom preserves backward-compatibility for operations identical to those found in relational databases. Almost none of them provides comprehensive means for extending their language syntax with user-defined constructs. Finally, some suffer from high setup costs, requiring hours if not days of arcane configuration editing and boilerplate code to run even the smallest example.

While it is obvious that convincing use cases motivate the existence of these systems, the current state of things leaves a potential user between two uncomfortable extremes: embrace a complex Event Processing system, with all its aforementioned shortcomings, or do without and fall back to low-level scripting languages, such as Perl or Python, to write menial trace-crunching tasks. What seems to be missing is a ``Simple Event Processing'' engine, in the same way that a spreadsheet application like Microsoft Excel is often a satisfactory middle ground between a pocket calculator and a full-blown accounting system. Such a system should provide higher abstraction than hand-written scripts, an easy to understand computational model, zero-configuration operation and reasonable performance for light- to medium-duty tasks.

This paper presents a detailed description of such a Simple Event Processing engine, called BeepBeep. In Section \ref{sec:principles}, we first describe the fundamental design principles behind the development of this system. Some of these principles purposefully distance themselves from trends followed by current CEP solutions, and are aimed at making the intended system both simpler and more versatile. Section \ref{sec:theory} then formally describes BeepBeep's computational model. This simple formalization completely describes the system's semantics, making it possible for alternate implementations to be independently developed. 

One of the key features of BeepBeep is its associated query language, called \esql{}, which is described in Section \ref{sec:language}. Substantial effort has been put in making \esql{} simple and coherent; in accordance to BeepBeep's design principles, it strives towards \emph{relational transparency}, meaning that queries that perform computations similar to relational database operations are written in a syntax that is backwards-compatible with SQL. Section \ref{sec:palettes} then describes the various means of extending BeepBeep's built-in functionalities. A user can easily develop new event processing units in a handful of lines of code, and most importantly, define arbitrary grammatical extensions to \esql{} to use these custom elements inside queries. Extensions can be bundled in dynamically-loaded packages called \emph{palettes}; we describe a few of the available palettes, allowing BeepBeep to manipulate network captures, tuples, plots, and temporal logic operators, among others.

Equipped with these constructs, Section \ref{sec:examples} then proceeds to showcase BeepBeep's functionalities. An experimental comparison of BeepBeep's performance with respect to a selection of other CEP engines is detailed in Section \ref{sec:experiments}. To the best of our knowledge, this is the first published account of such an empirical benchmark of CEP engines on the same input data. These experiments reveal that, on a large number of examples, BeepBeep's versatility makes it able to tackle problems difficult to express with existing solutions; moreover, its simple formal foundations result in queries that are both easy to read, and are computed with reasonable throughput.


\section{Use Cases for Event Stream Processing}\label{sec:scenarios} 

Complex Event Processing (CEP) can loosely be defined as the task of analyzing and aggregating data produced by event-driven information systems \cite{DBLP:books/daglib/0017658}. A key feature of CEP is the possibility to correlate events from multiple sources, occurring at multiple moments in time. Information extracted from these events can be processed, and lead to the creation of new, ``complex'' events made of that computed data. This stream of complex events can itself be used as the source of another process, and be aggregated and correlated with other events.



Event processing distinguishes between two modes of operation. In \emph{online} (or ``streaming'') mode, input events are consumed by the system as they are produced, and output events are progressively computed and made available. It is generally assumed that the output stream is monotonic: once an output event is produced, it cannot be ``taken back'' at a later time. In contrast, in \emph{offline} (or ``batch'') mode, the contents of the input streams are completely known in advance (for example, by being stored on disk or in a database). Whether a system operates online or offline sometimes matters: for example, offline computation may take advantage of the fact that events from the input streams may be indexed, rewinded or fast-forwarded on demand. Recently, the hybrid concept of ``micro-batching'' has been introduced in systems like Apache Spark Streaming (cf.\ Section \ref{subsubsec:apache}). It is a special case of batch processing with very small batch sizes.

Guarantees on the delivery of events in a CEP system can also vary. ``At most once'' delivery entails that every event may be sent to its intended recipient, but may also be lost. ``At least once'' delivery ensures reception of the event, but at the potential cost of duplication, which must then be handled by the receiver. In between is perfect event delivery, where reception of each event is guaranteed without duplication. These concepts generally matter only for distributed event processing systems, where communication links between nodes may involve loss and latency.

In the following, we proceed to describe a few scenarios where event streams are produced and processed. 

\subsection{Stock Ticker}\label{subsec:stocks}

A recurring scenario used in CEP to illustrate the performance of various tools is taken from the stock market \cite{DBLP:conf/cidr/ChandrasekaranDFHHKMRRS03}. One considers a stream of stock quotes, where each event contains attributes such as a stock symbol, the price of the stock at various moments (such as its minimum price and closing price), as well as a timestamp. A typical stream of events of this nature is shown in Figure \ref{fig:stock}. This figure shows that events are structured as tuples, with a fixed set of attributes, each of which taking a scalar value. We shall see that many use cases have events structured as tuples, and that many event stream engines and query languages take for granted that events have a tuple structure.

\begin{figure}
\centering
\begin{tabular}{c|c|c|c}
\textbf{Timestamp} & \textbf{Symbol} & \textbf{Min.\ price} & \textbf{Closing price}\\
\hline
1 & \textsc{msft} & 1024 & 1111 \\
1 & \textsc{appl} &  562 &  666 \\
2 & \textsc{gogl} & 1234 & 1244  
\end{tabular}
\caption{The continuous output of a stock ticker can be assimilated to an event stream; each line of the table corresponds to an event with a number of attributes and values.}
\label{fig:stock}
\end{figure}

This simple example can be used to illustrate various queries that typically arise in an event stream processing scenario. A first, simple type of query one can compute over such a trace is called a \emph{snapshot} query, such as the following:

\begin{query}\label{q:snapshot}
Get the closing price of \textsc{msft} for the first five trading days.
\end{query}

The result of that query is itself a trace of tuples, much in the same way the relational \textsf{SELECT} statement on a table returns another table. 
A refinement of the snapshot query is the \emph{landmark} query, which 
returns only events that satisfy some criterion, such as:

\begin{query}\label{q:landmark}
Select all the days after the hundredth trading day, on which the closing price of \textsc{msft} has been greater than \$50. 
\end{query}


This simple query highlights the fact that, in online mode, outputting a tuple may require waiting until more of the input trace is made available ---and that waiting time is not necessarily bounded. In the worst case, \textsc{msft} may be the last stock symbol for which the price is known on a given day, and all events of that day must somehow be retained before knowing if they must be output in the result or discarded.

In \emph{window queries}, a computation is repeatedly made on a set of successive events. The size of that set is called the \emph{width} of the window; the width is specified as a number of events or as a time interval. A \emph{sliding query} is a particular case of window query where, after each computation, the window moves forward into the trace and a new set of successive events is considered. Often, as is the case in this example, the computation applied to the contents of the window is an aggregate function, such as a sum or an average. Systems such as LinQ \cite{linq} propose other types of window queries, such as the \emph{hopping query} (also called a \emph{tumble} window by \cite{DBLP:conf/vldb/CarneyCCCLSSTZ02}), where the window moves forward by exactly its width, such that no two windows ever overlap. For example:

\begin{query}\label{q:sliding}
On every fifth trading day starting today, calculate the average closing price of \textsc{msft} for the five most recent trading days. 
\end{query}

Other windows include the \emph{latch}, which maintains an internal state between window calculations. This is useful for calculations that are cumulative from the beginning of the stream.

A \emph{join} query involves the comparison of multiple events together. In the stock ticker example, a possible join query could be:

\begin{query}\label{q:band-join}
For the five most recent trading days starting today, select all stocks that closed higher than \textsc{msft} on a given day. 
\end{query}

When computing the result of such a query, a tuple is added to the output result depending on its relationship with respect to the price of \textsc{msft} for the same day. In most CEP systems, this is done by an operation similar to the \textsf{JOIN} operator in relational databases: the input stream is joined with itself, producing pairs of tuples $(t_1,t_2)$ where $t_1$ belongs to the first ``copy'' of the stream, and $t_2$ belongs to the second. The join condition, in our example, is that the timestamps of $t_1$ and $t_2$ must be equal. Since traces are potentially infinite, join operations require bounds of some kind to be usable in practice; for example, the join operation may only be done on events of the last minute, or on a window of $n$ successive events.

\subsection{Medical Records Management}\label{subsec:hl7}

We now move to the field of medical record management, where events are messages expressed in a structured format called HL7 \cite{hl7}. An HL7 message is a text string composed of one or more segments, each containing a number of fields separated by the pipe character (\texttt{|}). The possible contents and meaning of each field and each segment is defined in the HL7 specification. Figure \ref{fig:hl7} shows an example of an HL7 message; despite its cryptic syntax, this messages has a well-defined, machine-readable structure. However, it slightly deviates from the fixed tuple structure of our first example: although all messages of the same type have the same fixed structure, a single HL7 stream contains events of multiple types.

\begin{figure}
\centering
\begin{minipage}{3in}
\begin{verbatim}
1234567890^DOCLAST^DOCFIRST^M^
^^^^NPI|OBR|1|||80061^LIPID
PROFILE^CPT-4||20070911||||||||||OBX|1|
NM|13457-7^LDL (CALCULATED)^LOINC|
49.000|MG/DL| 0.000 - 100.000|N|||F|
OBX|2|NM|
2093-3^CHOLESTEROL^LOINC|138.000|
MG/DL|100.000 - 200.000|N|||F|OBX|3|
NM|2086-7^HDL^LOINC|24.000|MG/DL|
45.000 - 150.000|L|||F|OBX|4|NM|
2571-8^TRIGLYCERIDES^LOINC|324.000|
\end{verbatim}
\end{minipage}
\caption{An excerpt from a message in HL7 format.}
\label{fig:hl7}
\end{figure}

HL7 messages can be produced from various sources: medical equipment producing test results, patient management software where individual medical acts and procedures are recorded, drug databases, etc. For a given patient, the merging of all these various sources produces a long sequence of HL7 messages that can be likened to an event stream. The analysis of HL7 event traces produced by health information systems can be used, among other things, to detect significant unexpected changes in data values that could compromise patient safety \cite{DBLP:conf/edoc/BerryM13}. 

In this context, a general rule, which can apply to any numerical field, identifies whenever a data value starts to deviate from its current trend:

\begin{query}
Notify the user when an observed data field is three standard deviations above or below its mean.
\end{query}
We call such computations \emph{trend} queries, as they relate a field in the current event to an aggregation function applied on the past values of that field. Trend queries can be made more complex, and correlate values found in multiple events, such as the following:

\begin{query}\label{q:trend}
Notify the user when two out of three successive data points lie more than two standard deviations from the mean on the same side of the mean line.
\end{query}

Although our example query does not specify it, this aggregation can be computed over a window as defined in our previous use case, such as the past 100 events, or events of the past hour.

A \emph{slice} query is the application of the same computation over multiple subsets (slices) of the input stream. In the present use case, assuming that the HL7 stream contains interleaved messages about multiple patients, a possible slice query could be to perform the outlier analysis mentioned above for each patient.

Other applications of CEP in healthcare have been studied by Wang \etal{} \cite{DBLP:journals/pvldb/WangREW10}.

\subsection{Online Auction}\label{subsec:auction}

Our next use case moves away from traditional CEP scenarios, and considers a log of events generated by an online auction system \cite{DBLP:conf/fm/BarringerFHRR12}. In such a system, when an item is being sold, an auction is created and logged using the $\mbox{\textsl{start}}(i,m,p)$ event, where $m$ is the minimum price the item named $i$ can be sold for and $p$ is the number of days the auction will last. The passing of days is recorded by a propositional \nospellcheck{\textsl{endOfDay}} event; the period of an auction is considered over when there have been $p$ number of \nospellcheck{\textsl{endOfDay}} events.

The auction system generates a log of events similar to Figure \ref{fig:auction-events}. Although the syntax differs, events of this scenario are similar to the HL7 format: multiple event types (defined by their name) each define a fixed set of attributes.

\begin{figure}
\centering
\begin{minipage}{2in}
\textsl{start}(vase,3,15).\\
\textsl{bid}(vase,15).\\
\textsl{start}(ring,5,30).\\
\textsl{endOfDay}.\\
\textsl{bid}(ring,32).\\
\textsl{bid}(ring,33).\\
\textsl{bid}(vase,18).\\
\textsl{sell}(vase).
\end{minipage}
\caption{A log of events recorded by an online auction system.}
\label{fig:auction-events}
\end{figure}

One could imagine various queries involving the windows and aggregation functions mentioned earlier. However, this scenario introduces special types of queries of its own. For example:

\begin{query}\label{q:price-increase}
Check that every bid of an item is higher than the previous one, and report to the user otherwise.
\end{query}

This query expresses a pattern that correlates values in pairs of successive bid events: namely, the price value in any two bid events for the same item $i$ must increase monotonically. Some form of slicing, as shown earlier, is obviously involved, as the constraint applies separately for each item; however, the condition to evaluate does not correspond to any of the query types seen so far. A possible workaround would be to add artificial timestamps to each event, and then to perform a join of the stream with itself on $i$: for any pair of bid events, one must then check that an increasing timestamp entails an increasing price.

Unfortunately, in addition to being costly to evaluate in practice, stream joins are flatly impossible if the interval between two bid events is unbounded. A much simpler ---and more practical--- solution would be to simply ``freeze'' the last \textsl{Price} value of each item, and to compare it to the next value. For this reason, queries of that type are called \emph{freeze} queries.

The previous query involved a simple sequential pattern of two successive bid events. However, the auction scenario warrants the expression of more intricate patterns involving multiple events and multiple possible orderings:

\begin{query}\label{q:forbidden-bid}
List the items that receive bids outside of the period of their auction.
\end{query}

As one can see, this query refers to the detection of a pattern that takes into account the relative positioning of multiple events in the stream: an alarm should be raised if, for example, a bid for some item $i$ is seen before the start event for that same item $i$. Simiarly, an occurrence of a bid event for $i$ is also invalid if it takes place $n$ \nospellcheck{endOfDay} events after its opening, with $n$ being the \textsl{Duration} attribute of the corresponding start event. We call such query a \emph{lifecycle} query, as the pattern it describes corresponds to a set of event sequences, akin to what a finite-state machine or a regular expression can express.

\subsection{Electric Load Monitoring}\label{subsec:electric}

\begin{figure}
\centering
\includegraphics[scale=0.6]{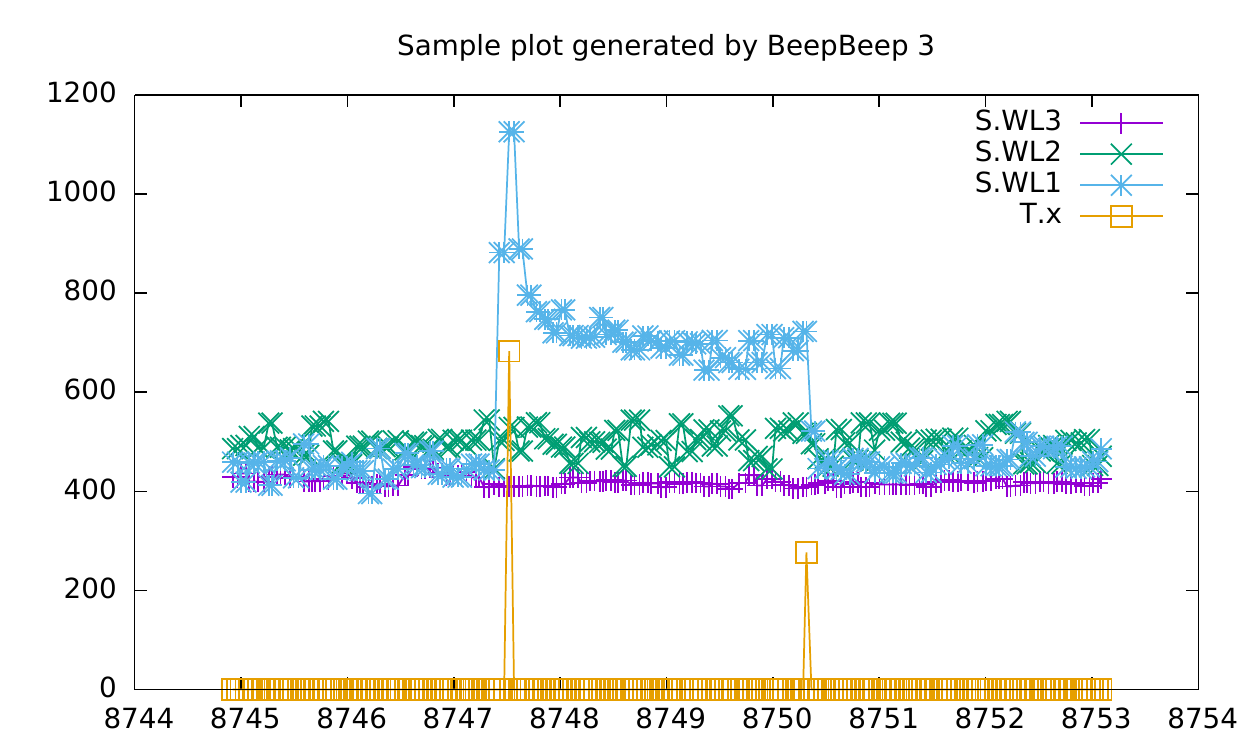}
\caption{The top three lines represent three components of the electrical signal when an electrical appliance is used. In orange, the output of a peak detector taking the electrical signal as its input.}
\label{fig:blender}
\end{figure}

The next scenario touches on the concept of ambient intelligence, which is a multidisciplinary approach that consists of enhancing an environment (room, building, car, etc.) with technology (e.g.\ infrared sensors, pressure mats, 
etc.), in order to build a system that makes decisions based on real-time information and historical data to benefit the users within this environment. A main challenge of ambient intelligence is activity recognition, which consists in raw data from sensors, filter it, and then transform that into relevant information that can be associated with a patient's activities of daily living using Non-Intrusive Appliance Load Monitoring (NIALM) \cite{ZE11}. 
Typically, the parameters considered are the voltage, the electric current and the power (active and reactive). This produces a stream similar to Figure \ref{fig:blender}. An event consists of a timestamp, and numerical readings of each of the aforementioned electrical components.



The NIALM approach attempts to associate a device with a load signature extracted from a single power meter installed at the main electrical panel. This signature is made of abrupt variations in one or more components of the electrical signal, whose amplitude can be used to determine which appliance is being turned on or off \cite{DBLP:conf/aaai/HalleGB16}. An example of query in this context could be:

\begin{query}\label{q:toaster}
Produce a ``Toaster On'' event whenever a spike of 1,000$\pm$200~W is observed on Phase 1 and the toaster is currently off.
\end{query}

Again, this scenario brings its own peculiarities. Here, events are simple tuples of numerical values, and slicing is applied in order to evaluate each signal component separately; however, the complex, higher-level events to produce depend on the application of a peak detection algorithm over a window of successive time points. Moreover, elements of a lifecycle query can also be found: the current state of each appliance has to be maintained, as the same peak or drop may be interpreted differently depending on whether a device is currently operating or not.

While this scenario certainly is a case of event stream processing in the strictest sense of the term, it hardly qualifies as a typical CEP scenario, as per the available tools and their associated literature. As a matter of fact, we shall see later that no CEP engine directly provide the appropriate machinery to tackle a problem such as this one.

\subsection{Runtime Verification}\label{subsec:pingus}

Our last use case considers event streams produced by the execution of a piece of software. \emph{Runtime verification} is the process of observing a sequence of events generated by a running system and comparing it to some formal specification for potential violations \cite{DBLP:journals/jlp/LeuckerS09}. It was shown how the use of a runtime monitor can speed up the testing phase of a system, such as a video game under development, by automating the detection of bugs when the game is being played \cite{nous-acm-cie}.

We take as an example the case of a game called \textit{Pingus}, a clone of Psygnosis' \textit{Lemmings} game series. The game is divided into levels populated with various kinds of obstacles, walls, and gaps. Between 10 and 100 autonomous, penguin-like characters (the Pingus) progressively enter the level from a trapdoor and start walking across the area. 
The player can give special abilities to certain Pingus, allowing them to modify the landscape to create a walkable path to the goal. For example, some Pingus can become Bashers and dig into the ground; others can become Builders and construct a staircase to reach over a gap. Figure \ref{subfig:pingus-screenshot} shows a screenshot of the game.

\newsavebox{\pingusxml}
\savebox{\pingusxml}{%
\begin{minipage}{2in}
$<$characters$>$\\
\phantom{i}$<$character$>$\\
\phantom{i}\phantom{i}$<$id$>$0$<$/id$>$\\
\phantom{i}\phantom{i}$<$action$>$faller$<$/action$>$\\
\phantom{i}\phantom{i}$<$isalive$>$true$<$/isalive$>$\\
\phantom{i}\phantom{i}$<$position$>$\\
\phantom{i}\phantom{i}\phantom{i}$<$x$>$1121$<$/x$>$$<$y$>$393$<$/y$>$\\
\phantom{i}\phantom{i}$<$/position$>$\\
\phantom{i}\phantom{i}$<$velocity$>$\\
\phantom{i}\phantom{i}\phantom{i}$<$x$>$0$<$/x$>$$<$y$>$3.6$<$/y$>$\\
\phantom{i}\phantom{i}$<$/velocity$>$\\
\phantom{i}\phantom{i}$<$groundtype$>$earth$<$/groundtype$>$\\
\phantom{i}$<$/character$>$\\
\phantom{i}...\\
$<$/characters$>$
\end{minipage}
}
\begin{figure}
\centering
\subfloat[]{%
\begin{minipage}{2in}
\includegraphics[width=2in]{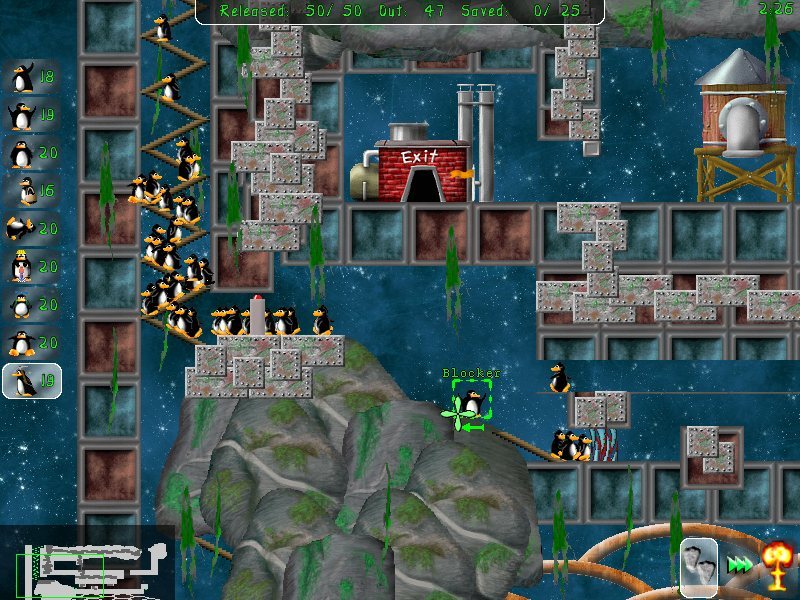}
\end{minipage}\label{subfig:pingus-screenshot}}~~~
\subfloat[]{\scalebox{0.75}{\usebox{\pingusxml}}\label{subfig:pingus-xml}}
\caption{The Pingus video game. (a) A screenshot of the game; (b) A small portion of XML event produced by an instrumented version of the code.}
\label{fig:pingus}
\end{figure}

When running, the game updates the playing field about 150 times per second; each cycle of the game's main loop produces an XML snapshot of its state similar to the one shown in Figure \ref{subfig:pingus-xml}. Hence, analyzing the execution of the game can be assimilated to processing the stream of individual XML events it generates. The abnormal execution of the game can be expressed as event stream query, looking for a pattern corresponding to bugs in the game. An example of an incorrect execution pattern could be:

\begin{query}\label{q:turn-around}
Make sure that a walking Pingu that encounters a Blocker turns around and starts walking in the other direction.
\end{query}

This query is special in at least two respects. First, the Pingus use case introduces a new type of event unseen in previous examples. Indeed, the XML events produced by the game are not fixed tuples of name-value pairs, but rather contain nested substructures. Hence, in each event, the \verb+<character>+ element is repeated for as many Pingus as there are on the playing field; each such element contains the data (position, velocity, skills) specific to one character. It does not make sense, in this context, to talk about ``the'' ID inside an event, as it contains multiple such IDs. The contents of XML documents must therefore be accessed using a more sophisticated querying mechanism, such as XPath expressions. Moreover, events are unusually large: a single event can contain as much as ten kilobytes of XML data. 

Second, in order to detect this pattern of events, one must correlate the x-y position of two distinct Pingus (a Walker and a Blocker), and then make sure that the distance between the two increases over the next couple of events (indicating a turnaround).\footnote{One cannot simply look for a change of sign in velocity, as the turnaround may lag the ``collision'' by a few cycles of the game loop.} These computations go beyond the basic slicing and lifecycle queries studied in the previous examples.

Furthermore, various kinds of analyses can also be conducted on the execution of the game. For example, one may be interested in watching the realtime number of Pingus possessing a particular skill, leading to a query such as:
\begin{query}\label{q:ratio}
Determine the realtime proportion of all active Pingus that are Blockers.
\end{query}

Such a query involves, for each event, the counting of all Pingus with a given skill with respect to the total number of Pingus contained in the event. Going even further, one may also divide the playing field into square cells of a given number of pixels, and count the Pingus that lie in each cell at any given moment, producing a form of ``heat map'':

\begin{query}\label{q:heatmap}
Produce a heat map of the location of Pingus across the game field; update this map every three seconds.
\end{query}

This last query outputs a stream of events of an unusual type, namely two-dimensional arrays of numerical values. Such arrays could then be passed to a plotting program that could display a graph in real time.

\subsection{Other Use Cases}\label{subsec:other-use-cases}

It is probably clear at this point that a large number of diverse problems can be re-framed as a form of computation over event streams of various kinds. Moreover, the last few examples have shown queries and event types that stretch what is generally meant by CEP in both research and practice.

There exist many other use cases of event stream processing, which we mention only in passing. Microsoft's StreamInsight tutorial \cite{linq} considers toll booths along a road sending out \texttt{TollReading} events whenever a car passes through the booth. 
Research on the Twitter platform has led to the development of TweeQL, a streaming SQL-like interface to the Twitter API, making common tweet processing tasks simpler \cite{DBLP:journals/sigmod/MarcusBBKMM11}. Event streams have also been used to detect intrusions in a network infrastructure \cite{DBLP:conf/debs/Kumaran13}, identify non-compliant behaviour of aircraft within a regulated airspace \cite{DBLP:conf/nfm/PielBLBK16}, monitor an electrical grid \cite{Balis:2011:RGM:2009738.2009875}.

The Runtime Verification community has defined a number of use cases with intricate sequential patterns over events produced by a running system. In addition to the online auction described above, past works have considered: the correct interleaving of method calls on a Java object according to its API \cite{DBLP:journals/fmsd/KimVKLS04,DBLP:conf/icse/JinMLR12}; the analysis of commands and responses sent by a spacecraft under test for the detection of bugs \cite{DBLP:conf/ftscs/HavelundJ14}; the analysis of real-world web service XML payloads \cite{DBLP:journals/tsc/HalleV12}; the detection of fraudulent activity in an event log \cite{DBLP:journals/fmsd/BasinKMZ15}; the analysis of system calls on traces of assembly instructions \cite{DBLP:conf/isola/KhouryHW16}. 


\section{State of the Art in Event Stream Processing}\label{sec:related} 

We shall now provide an overview of the available solutions for event stream processing. Recent and extended surveys of CEP engines already exist \cite{DBLP:journals/csur/CugolaM12}; the goal of this paper is not to replicate such efforts. The main distinguishing point of this review is that it divides these solutions in two families: first, tools and research projects that have been developed as Complex Event Processing systems, and recognized as such; second, tools that have been developed by the Runtime Verification (RV) community, which present a significant overlap with event stream processing, but have been consistently overlooked by traditional reviews on CEP.

\subsection{Tools for Event Stream Processing}

A large number of CEP and related engines have been developed over the past decade. We describe some of them by emphasizing their distinguishing features. 

\subsubsection{Aurora and Borealis}

One of the earliest systems is \emph{Aurora} \cite{DBLP:conf/vldb/CarneyCCCLSSTZ02}. It defines eight primitive operations on streams of tuples. The \emph{window} operates over sets of consecutive tuples, and applies a user-defined function to each window; four types of windows are supported (sliding, latch, tumble and \emph{resample}, which interpolates new tuples between the original tuples of an input stream). Four other operators act on a single tuple at a time:  the \emph{filter} operator screens tuples in a stream for those that satisfy some predicate; \emph{map} applies an input function to every tuple in a stream; \emph{group by} partitions tuples across multiple streams into new streams whose tuples contain the same values over some
input set of attributes; \emph{join} pairs tuples from input streams whose difference in timestamps falls within some given interval. These primitive functions can be composed through a graphical user interface, where a user can create and connect \emph{boxes}.

Aurora can perform a run-time optimization of a graph of boxes. This is done by identifying a subset of the graph, hold all input messages at upstream connection points and drain the subgraph of events through all downstream
connection points. Then, a few optimization strategies can be attempted. One of them is to analyze the attributes of the tuples considered in that part of the graph, and to insert a projection operation to remove from the input tuples all attributes that are not necessary. In addition, pairs of boxes can sometimes be merged into a single box for more efficiency, such as when two filtering operations are applied in sequence. In the same way as filters and projections are preferably applied first in a relational database query, the same shuffling of operations can also be attempted in an Aurora query in order to reduce the number or the size of tuples that need to be processed downstream. 

A global \emph{scheduler} decides on what box is allowed to perform an execution step at each point in time. Boxes may be allocated to multiple threads. The decisions of the scheduler are informed by a Quality of Service (QoS) monitor, which can trigger load shedding strategies (such as dropping tuples) when QoS for a specific query degrades past a user-defined level.

Aurora was followed by a multi-processor version called \emph{Borealis} \cite{DBLP:conf/cidr/AbadiABCCHLMRRTXZ05}. In Borealis, boxes are provided with special control lines in addition to their standard data input lines; these lines can carry information (such as new sets of parameters) that may change the box's behaviour during the execution of the query. Moreover, in Borealis an event stream can contain deletion messages, indicating that a tuple previously inserted in the stream is to be removed, and replacement messages that revise values for a tuples already inserted in the stream. Hence, streams are no longer monotonic, and query results can be updated following a corresponding update of the input stream. This is done by sending out one or more revision messages, computed from the input revision messages received. This feature is unique among all the systems considered in this review.

References on Aurora and Borealis do not explicitly describe a query language, other than the boxes-and-arrows model described above. It is reported, however, that these projects led to the creation of SQLstream, an extension of the SQL language for the manipulation of tuple event streams. SQLstream can query relational databases using regular SQL syntax;  to specify a query on a stream, one must use a keyword called \textsf{STREAM} immediately after \textsf{SELECT}. The \textsf{OVER} construct can be used to define windows and apply the standard SQL aggregation functions over that window. For example:

\begin{snippet}
\textsf{SELECT STREAM} o.orderid, \textsf{SUM}(t.amount)\\
\textsf{FROM} OrderStream \textsf{OVER} (\textsf{RANGE CURRENT ROW}) \textsf{AS} o
\textsf{LEFT JOIN} TradeStream\\
 \textsf{OVER} (\textsf{RANGE INTERVAL} '1' \textsf{HOUR FOLLOWING}) \textsf{AS} t
 \textsf{ON} o.orderid = t.tradeid\\
\textsf{GROUP BY} \textsf{FLOOR}(OrderStream.\textsf{ROWTIME TO HOUR}), o.orderid\\
\textsf{HAVING} o.amount <> \textsf{SUM}(t.amount);
\end{snippet}

SQLstream also provides a special object called a \emph{pump}. A pump provides a continuously running \textsf{INSERT INTO} \textit{stream} \textsf{SELECT} \textit{query} functionality, thereby enabling the results of a query to be continuously entered into another stream. In other words, a pump pulls data from a stream, and pushes a transformed version into another stream.

SQLstream is supported by a commercial product called SQLstream Blaze, which is part of the Amazon Kinesis platform.

\subsubsection{TelegraphCQ}

Another early system is TelegraphCQ \cite{DBLP:conf/cidr/ChandrasekaranDFHHKMRRS03}. 
It originates from the Telegraph project, which began almost twenty years ago with the goal of developing an Adaptive Dataflow Architecture for supporting a wide variety of data-intensive, networked applications. It consists of an extensible set of composable dataflow modules or operators that produce and consume records in a manner analogous to the operators used in traditional database query engines, or the modules used in composable network routers. Query processing is performed by routing tuples through query modules. These modules are pipelined versions of standard relational database operators such as joins, selections, projections, grouping and aggregation, and duplicate elimination. 

\emph{Eddies} are modules that adaptively decide how to route data to other query operators on a tuple-by-tuple basis. Each Eddy is responsible for the processing of tuples by a set of commutative query modules. Based on the current state of the system, an Eddy may dynamically decide on the order in which tuples are handled by each of the query modules. When one of the modules processes a tuple, it can generate other tuples and send them back to the Eddy for further routing. A tuple is sent to the Eddy's output if all the modules connected to the Eddy have successfully handled it.

The glue that binds the various modules together to form a query plan is an inter-module communications API that is called \textit{Fjords}. It allow pairs of modules to be connected by various types of queues. For example, a pull-queue is implemented using a blocking dequeue on the consumer side and a blocking enqueue on the producer side. A push-queue is implemented using non-blocking enqueue and dequeue; control is returned to the consumer when the queue is empty. This allows the system to efficiently deal with slow or unresponsive data sources, which would otherwise suspend the execution of the system. 

Dataflows are initiated by clients either via an ad hoc query language (a basic version of SQL) or by an equivalent scripting language for creating dataflow graphs. 
It supports much more general windows than the landmark and sliding windows described above. This is done using a for-loop construct to declare the sequence of windows over which the user desires the answers to the query: a variable t moves over the timeline as the for-loop iterates, and the left and right ends (inclusive) of each window in the sequence, and the stopping condition for the query can be defined with respect to this variable $t$. The syntax of the for-loop is as follows:

\begin{verbatim}
for (t = initial_value; 
  continue_condition(t); change(t)){
  WindowIs(Stream A, left_end(t), right_end(t));
  WindowIs(Stream B, left_end(t), right_end(t));
  ...
}
\end{verbatim}
For example, here is an example of the landmark query mentioned in the previous section, expressed in TelegraphCQ's query language:

\begin{verbatim}
SELECT closingPrice, timestamp
FROM ClosingStockPrices
WHERE stockSymbol = ‘\textsc{msft}’
for (; t==0; t = -1 ){
  WindowIs(ClosingStockPrices, 1, 5);
}
\end{verbatim}
The authors of TelegraphCQ explicitly state that such a for-loop ``is intended as a powerful, \emph{low-level} mechanism rather than a user-level query language construct'' \cite[p.\ 32, emphasis added]{DBLP:conf/cidr/ChandrasekaranDFHHKMRRS03}. Unfortunately the description of a user-level equivalent of this loop is not discussed.

TelegraphCQ was implemented in C/C++, by reusing a good amount of code from the existing PostgreSQL relational database engine and adapting it for continuous queries.

\subsubsection{SASE}\label{subsubsec:sase} This system \cite{DBLP:conf/sigmod/WuDR06} was brought as a solution to meet the needs of a range of RFID-enabled monitoring applications. In contrast with the window and join queries that were the focus of Aurora, Borealis and TelegraphCQ, SASE rather deals with pattern queries, which describe a sequence of events that occur in temporal order and are further correlated based on their attribute
values. A pattern query looks like this:

\begin{verbatim}
PATTERN SEQ(TaskStart a,CPU b,TaskFinish c,CPU d)
WHERE a.taskId = c.taskId AND
      b.nodeId = a.nodeId AND
      d.nodeId = a.nodeId AND
      b.value > 95% AND
      d.value <= 70% AND
     skip_till_any_match(a, b, c, d)
WITHIN 15 seconds
RETURN a, b, c
\end{verbatim}

The \textsf{PATTERN} clause describes the pattern of events to be observed; the \textsf{WHERE} clause further expresses conditions on the events' attributes for the pattern to be considered. Since the events relevant to the pattern are not necessarily in contiguous positions in the input stream, this clause can also specify an event selection strategy. For example, the ``skip till next match'' strategy specifies that in the pattern matching process, irrelevant events are skipped until an event matching the next pattern component is encountered. If multiple events in the stream can match the next pattern component, only the first of them is considered. Finally, the \textsf{WITHIN} clause restricts the pattern to a time period, while the \textsf{RETURN} clause selects the events to be included in the pattern match. 

SASE deals with the particular problem of uncertain timestamps affixed to incoming events. An event in SASE's model has the following format: (type, id, [lower, upper], attributes), where \textit{type} specifies the attributes allowed in the events of this type and \textit{id} is the unique event identifier. For example, $a_1=(A, 1, [5, 9], (v_1, v_2, v_3))$ represents an event of type A, with an id 1, an uncertainty interval from time 5 to time 9, and three required attributes named $v_1$, $v_2$ and $v_3$. The fact that timestamps are only known within some precision bounds obviously complicates the process of pattern matching. At every point $t$, the system collects each event $e$ from the input whose uncertainty interval spans $t$, and injects to a new stream a point event that replaces $e$'s uncertainty interval with a fixed timestamp $t$. This is possible under the hypothesis that if $e_1$ arrives before $e_2$, then with respect to the occurrence time, $e_1$ either completely precedes $e_2$ or overlaps with $e_2$. Unfortunately, the version of the SASE system available at the time of this writing does not handle these imprecise timestamps. It does support the processing of pattern queries.

\subsubsection{Cayuga}\label{subsubsec:cayuga} Cayuga is a complex event monitoring system for high speed data streams \cite{DBLP:conf/debs/BrennaGHJ09}. It provides a simple query language for composing stateful queries with a scalable query processing engine based on non-deterministic finite state automata with buffers. Each event stream has a fixed relational schema, and events in the stream are treated as relational tuples. Each event has two timestamps, a start time and a detection time, modeling the fact that events can have a non-zero but finite duration.

A Cayuga query has three parts; the \textsf{SELECT} clause chooses the attributes to include in the output events, the \textsf{FROM} clause describes a pattern of events that have to be matched, and the \textsf{PUBLISH} gives a name to the resulting output stream. For example, the following expression is a Cayuga query that creates an output event whenever  there are at least ten input events whose \textsl{summary} attribute contains the word ``iPod'' within the same 24-hour interval:

\begin{verbatim}
SELECT * FROM
(SELECT *, cnt AS 1 FROM
FILTER {contains(summary,”iPod”)=1}(webfeeds))
FOLD {TRUE, cnt>10 AND dur<1 day, cnt AS cnt+1}
(SELECT * FROM
FILTER {contains(summary,”iPod”)=1}(webfeeds))
PUBLISH ipod popularity
\end{verbatim}

Note how this query itself involves two sub-queries. The FILTER\{$\theta$\} operator selects events from the input stream that satisfy the predicate $\theta$. The FOLD operator looks for patterns comprising two or more events; it defines the condition for the iteration, a  stopping condition for iteration, and a mapping between iteration steps. In this case, every event matching the FILTER condition will increment a variable called \verb+cnt+, until this value reaches 10.

Each query is internally converted into a non-deterministic finite state automaton with buffers. Each vertex of the automaton is associated with a specific schema; a transition between two states $P$ and $Q$ is labelled by a triple $\langle S, \theta, f\rangle$, where $S$ identifies an input stream, $\theta$ is a predicate over the joint schemas of $P$ and $S$, and $f$ is a function mapping these schemas to the schema of $Q$. The details of the transformation are given in \cite{DBLP:conf/cidr/DemersGPRSW07}.

\subsubsection{Siddhi}\label{subsubsec:siddhi}

Siddhi is the query engine used in the WSO2 Complex Event Processor \cite{DBLP:conf/sc/SuhothayanGNCPN11}, an open source engine for real-time analytics of events. It supports classical event detection patterns like filters, windows, joins and event patterns and more advanced features like event partitions, mapping database values to events, etc.

Siddhi represents events using a tuple data structure. Its architecture consists of \emph{processors} connected through event queues. Incoming events are placed in event queues, and processors listening to those event queues process those events and place any matching events into output queues of that processor, which will then be processed by other processors or send to end users as event notifications. As we shall see later, our proposed system follows a similar high-level design.

It differs, however, in how processors execute their computations. Each processor is composed of several \emph{executors} that express the query conditions; each executor processes the incoming events and produces a Boolean output indicating whether the event has matched. Non-matching events are discarded, and matching events are processed by logical executors downstream. Communication between processors is done through a ``publish-subscribe'' mechanism, with downstream processors registering to receive events produced from upstream processors.

When a processor is connected to multiple input streams, Siddhi employs an original model to handle the incoming events. It uses a single input event queue and multiplexes all the events together. This is done in order to reduce the complexity of checking all input queues and keeping track of which events are yet to be processed. Each event is affixed with the ID of the stream it belongs to, making it possible for the processor to make sense of all mixed events and process them correctly.

In terms of query capabilities, Siddhi supports the computation of typical aggregation functions (e.g.\ sum, average) over windows. Moreover, it can also express sequential patterns of events, similar to SASE's, but using a different syntax. The following query shows an example of such a pattern:

\begin{verbatim}
select f.symbol, p.accountNumber, f.accountNumber
from pattern [every f=FraudWarningEvent2 ->
p=PINChangeEvent2(accountNumber = f.accountNumber)]
\end{verbatim}

It relates two events \verb+f+ and \verb+p+, such that \verb+p+ must follow \verb+f+ and the \textsl{accountNumber} attribute of both must be identical. When such a pattern occurs, the query produces an output event containing the symbol and account number identifying this pattern.

Contrarily to many CEP engines, Siddhi tries to bring in stream processing aspects like multi-threading and pipelining, although these aspects do not seem document in research papers.

\subsubsection{Esper}\label{subsubsec:esper}

Esper is probably the most complete and versatile of the CEP engines included in this review. First, Esper's events may contain rich and nested domain-specific information. In particular, an event's property may itself be composed of other events; Esper uses the term fragment for such event pieces. Each portion of a query is also associated with a \emph{context}; a context takes a cloud of events and classifies them into one or more sets, called \emph{context partitions}. An event processing operation that is associated with a context operates on
each of these context partitions independently.

Esper's query language (EQL) is an extension of SQL that supports windows and patterns over streams. 
%
%
A pattern may appear anywhere in the from clause of an EPL statement including joins and subqueries. There are four types of pattern operators:
\begin{inparaenum}
\item Operators that control pattern sub-expression repetition: \texttt{every}, \texttt{every-distinct}, \texttt{[num]} and \texttt{until}
\item Boolean connectives
\item A single ``followed-by'' temporal operator that operates on event order
\item Where-conditions that control the lifecycle of sub-expressions (such as \texttt{timer:within}).
\end{inparaenum}

For example, the following query, taken from Esper's documentation, selects a total price per customer over pairs of events (a \textsl{ServiceOrder} followed by a \textsl{ProductOrder} event for the same customer id within one minute), occurring in the last two hours, in which the sum of price is greater than 100, and using a \textsl{where} clause to filter on the customer's name:

\begin{verbatim}
select a.custId, sum(a.price + b.price)
from pattern [every a=ServiceOrder ->
b=ProductOrder(custId = a.custId)
where timer:within(1 min)].win:time(2 hour)
where a.name in ('Repair', b.name)
group by a.custId
having sum(a.price + b.price) > 100
\end{verbatim}

The commercial product Oracle CEP uses Esper as its internal query engine.

\subsubsection{The Apache Ecosystem}\label{subsubsec:apache}

We now move our focus distributed event processing frameworks. A first observation that can be made from these systems is that they generally focus on the routing and load balancing of event streams across a multi-machine infrastructure.  In counterpart, we shall see that they offer much fewer functionalities for the actual processing of these streams, which is often left to the user as procedural (i.e.\ Java or Python) code. Due to their distributed nature, they also involve a much more complex setup than the solutions detailed so far.

The Apache Foundation hosts several (sometimes competing) projects related to the processing of events. Apache \emph{Samza} is a distributed stream processing framework \cite{samza}. It provides a very simple callback-based ``process message'' API comparable to MapReduce. As such, it is more an environment in which jobs can be deployed, coordinated and dispatched across multiple machines, than a system providing facilities to actually perform these computations. Each separate ``job'' in Samza still has to be written in low-level code, such as Java or Scala. 
It is reasonable to think, however, that many other CEP engines mentioned above could operate within a Samza infrastructure at the job level, making these two kinds of systems complementary.

Closer to our topic is Apache \emph{S4}, a platform for massive scale processing of data streams using the actor model \cite{DBLP:conf/icdm/NeumeyerRNK10}. It is, however, unable to express queries that span multiple events, which hardly qualifies it as a CEP engine. Apache \emph{Spark} \cite{DBLP:journals/cacm/ZahariaXWDADMRV16} is yet another distributed batch processing platform, similar to Hadoop: its core provides memory management and fault recovery, scheduling, distributing and monitoring jobs on a cluster, and functionalities for interacting with storage systems. The main data structure in Spark is a distributed set of objects called the Resilient Distributed Dataset (RDD). No specific type is imposed on the contents of an RDD. Operations on RDDs include \emph{transformations}, which take an RDD as their input and produce another RDD as their output; examples of transformations include map, join, filter, etc. The second type of operation is \emph{actions} that run a computation on an RDD and return a value; examples of actions include counting and aggregation functions. Transformations in Spark are said to be ``lazy'', in the sense that the input data for a transformation or an action is not computed until the output of that transformation is requested.

Spark provides a few relatively low-level constructs for processing RDDs; similarly to S4 and Samza, it focuses on the distributed dispatching of jobs. It can be completed with extensions that provide more elaborate facilities for expressing computations on events. One of them is SparkSQL, which allows querying RDDs as if they are relational tables, using SQL. Of interest to this paper is Spark Streaming, an API that allows Spark to handle streams of data. 
For example, here is a Scala code example that computes a sliding window average over a stream:

\begin{lstlisting}[language=scala]
val inputsStream = ssc.socketStream(...)
val windowStream1 = inputStream.window(Seconds(4))
val w = Window.partitionBy("id").orderBy("cykle").rowsBetween(-2, 2)
val x = windowStream1.select(\$"id",\$"cykle",avg(\$"value").over(w))
\end{lstlisting}

While Spark provides out-of-the-box functionalities for computing windows, aggregate functions, filters and map-reduce jobs, it seems to lack similar constructs for handling sequential patterns, such as those considered by SASE, Siddhi and Esper.

%
%

\emph{Storm} \cite{storm} is another distributed processing platform supported by the Apache Foundation. In 2014, it earned the title of the fastest open source engine for sorting a petabyte of data in the 100 TB Daytona GraySort contest. In Storm, events are immutable sets of key-value pairs called \emph{tuples}, while a stream is a potentially infinite sequence of tuples. Each stream is given an ID (manually set by the user), and is associated to a fixed schema for the tuples it contains. ``Bolts'' are units that take input streams and produce output streams, with initial tuple sources being called ``spouts''. Distributed computation is achieved by configuring Storm so that multiple instances of the same bolt can be run, each on a different fragment of an input stream. 

This splitting and merging of streams is configured manually by the user, although libraries like Trident can simplify their management \cite{trident-api}. Trident also provides higher-level objects, such as \emph{functions}. Functions have a special semantics, similar to a form of tuple join: the fields of the output tuple they produce are appended to the original input tuple in the stream. If the function emits no tuples, the original input tuple is filtered out. Otherwise, the input tuple is duplicated for each output tuple.

\begin{lstlisting}
class HashTagNormalizer extends BaseFunction {
  public void execute(TridentTuple tuple, TridentCollector col) {
    String s = tuple.getStringByField("foo");
    s = s.trim();
    col.emit(new Values(s));
  }
}
\end{lstlisting}

Additional Trident constructs include \emph{filters}, which take in a tuple as input and decide whether or not to keep that tuple or not; \emph{map}, which returns a stream consisting of the result of applying the given mapping function to the tuples of the stream; min/max which returns the minimum (resp.\ maximum) value on each partition of a batch of tuples; and a number of classical windowing and aggregation functions.

All these infrastructures provide a relatively low-level API for manipulating events; besides, apart from SparkSQL (which only works for relational queries on static RDDs), none of these systems provides an actual query language that would abstract implementation concerns.

\subsubsection{Other Systems}

Due to space considerations, several other systems have to be left out of this presentation, such as Cordies \cite{DBLP:conf/debs/KochKR10}, Flink \cite{DBLP:journals/vldb/AlexandrovBEFHHKLLMNPRSSHTW14}, LogCEP \cite{logcep}, and SPA \cite{DBLP:conf/edoc/DijkmanPH16}. They all provide functionalities similar in nature to that of one of the tools described above. Other early works on stream databases include \cite{DBLP:conf/cidr/MotwaniWABBDMORV03,DBLP:conf/sigmod/SeshadriLR94,Arasu:ilprints641}. Also outside of this review are systems peripheral to the actual processing of event streams, such as ``event brokers'' like Apache Kafka, Flume, Twitter, ZeroMQ, etc. There also exist dozens of commercial tools claiming CEP features with widely varying levels of detail, and for which it is hard to provide information without detailed documentation and a freely available implementation. We only mention in passing Amazon Kinesis, StreamBase SQL \cite{streambase}, StreamInsight \cite{linq}, and SAS Event Stream Processing Studio. For the sake of completion, we finally mention log analysis systems that provide very simple, Grep-like filtering capabilities, such as EventLog Analyzer\footnote{\url{www.manageengine.com/EventLogAnalyzer}} and Lumberjack\footnote{\url{https://fedorahosted.org/lumberjack/}}.

\subsection{Tools for Runtime Verification}

Perhaps lesser known to mainstream CEP users is the existence of another field of research, called Runtime Verification (RV). In RV, a \emph{monitor} is given a formal \emph{specification} of some desirable property that a stream of events should fulfill. The monitor is then fed events, either directly from the execution of some instrumented system or by reading a pre-recorded file, and is responsible for providing a verdict, as to whether the trace satisfies or violates the property.

Classical RV problems are centered around properties that deal with the way events are ordered. For example, the canonical ``HasNext'' property stipulates that, whenever an Iterator object is used in a program, any call to its \verb+next()+ method must be preceded by a call to \verb+hasNext()+ that returns the value \verb+true+. Consequently, the languages used by monitors to specify properties all have a strong temporal or sequential component: this includes finite-state automata, temporal logic, $\mu$-calculus, and multiple variations thereof

There are clear ties between CEP and RV, which have been revealed in a recent paper \cite{DBLP:conf/rv/Halle16}. Both fields consider sequences of events, which must be processed to provide some output. In both cases, the analysis is generally done in a streaming fashion. Despite these similarities, contributions from one community have been largely overlooked by the other, and \textit{vice versa}. In the following, we describe a few popular RV systems developed over the years, and present them in the light of their event processing capabilities.

\subsubsection{LOLA}

LOLA is a specification language and a set of algorithms for the online and offline monitoring of synchronous systems including circuits and embedded systems \cite{DBLP:conf/time/DAngeloSSRFSMM05}. It resembles synchronous programming languages such as LUSTRE \cite{DBLP:journals/tse/HalbwachsLR92}, 
but provides additional functionalities for expressing properties about the future.

\begin{figure}
\begin{eqnarray*}
s_1 & = & \mbox{true}\\
s_2 & = & t_3\\
s_3 & = & t_1 \vee (t_3 \leq 1)\\
s_4 & = & ((t_3)^2 + 7) \mod 15\\
s_5 & = & \mbox{ite}(s_3; s_4; s_4 + 1)\\
s_6 & = & \mbox{ite}(t_1; t_3 \leq s_4; \neg s_3)\\
s_7 & = & t_1[+1; \mbox{false}]\\
s_8 & = & t_1[-1; \mbox{true}]\\
s_9 & = & s_9[-1;0] + (t_3 \mod 2)\\
s_{10} & = & t_2 \vee (t_1 \wedge s_{10}[1; \mbox{true}])
\end{eqnarray*}
\caption{An example of a LOLA specification showing various features of the language}
\label{fig:lola}
\end{figure}

A LOLA specification is a set of equations over typed stream variables. Figure \ref{fig:lola} shows an example of a LOLA specification, summarizing most of the language's features. It defines ten streams, based on three independent variables $t_1$, $t_2$ and $t_3$. A stream expression may involve the value of a previously defined stream. The values of the streams corresponding to $s_3$ to $s_6$ are obtained by evaluating their defining expressions place-wise at each position. The expression ``ite'' represents an if-then-else construct: the value returned depends on whether the first operand evaluates to true. The stream corresponding to $s_7$ is obtained by taking at each position $i$ the value of the stream corresponding to $t_1$ at position $i+1$, except at the last position, which assumes the default value false.

The specification can also declare certain output Boolean variables as \emph{triggers}. Triggers generate notifications at instants when their corresponding values become true. Hence, the property ``the number of events where $b$ holds never exceeds the number of events where $a$ holds'' can be written in LOLA as:

\begin{align*}
& s = s[-1; 0] + \mbox{ite}((a \wedge \neg b); 1; 0)
+ \mbox{ite}((b \wedge \neg a); -1; 0)\\
& \mbox{trigger}(s \leq 0)
\end{align*}

A LOLA specification is said to be \emph{efficiently monitorable} its worst case memory requirement is constant in the size of the trace. The introductory paper on LOLA shows how these bounds can be computed. Many features of CEP can be accommodated by this notation, such as windows and simple aggregations. However, basic LOLA has only primitive support for event data manipulation.

\subsubsection{MOP} The Monitoring Oriented Programming (MOP) project \cite{chen-jin-meredith-rosu-2009-icicis} is a programming infrastructure where monitors are automatically synthesized from properties and integrated into the original system to check its behaviour at runtime.

JavaMOP is an instance of MOP targeted at Java programs \cite{DBLP:conf/icse/JinMLR12}; it relies on concepts of Aspect-Oriented Programming (AOP) \cite{DBLP:journals/csur/Kiczales96} to fetch relevant events from the execution of a system. In JavaMOP, an event corresponds to a \emph{pointcut}, such as a function call, a function return, the creation of an object, the assignment of a field, etc. The behaviour of the system is expressed as properties over these events. JavaMOP supports several formalisms for expressing these properties: ERE (extended regular expressions), FSM (finite state machines), CFG (context free grammars), PTLTL (past time linear temporal logic), FTLTL (future time linear temporal logic), and ptCaRet (past time linear temporal logic with calls and returns).

\newsavebox{\javamop}
\begin{lrbox}{\javamop}
\begin{lstlisting}
HasNext(Iterator i) {

  event hasnexttrue after(Iterator i) returning(boolean b) : 
              call(* Iterator.hasNext()) 
              && target(i) && condition(b) { } 
  event hasnextfalse after(Iterator i) returning(boolean b) : 
              call(* Iterator.hasNext()) 
              && target(i) && condition(!b) { } 
  event next before(Iterator i) : 
              call(* Iterator.next()) 
              && target(i) { } 

  ltl: [](next => (*) hasnexttrue)

  @violation { System.out.println("ltl violated!");}
}
\end{lstlisting}
\end{lrbox}
\begin{figure}
\centering
\usebox{\javamop}
\caption{A JavaMOP specification.}
\label{fig:mop}
\end{figure}

Figure \ref{fig:mop} shows an example of a JavaMOP specification using Linear Temporal Logic. Three atomic events (\verb+hasnexttrue+, \verb+hasnextfalse+ and \verb+before+) are created from pointcuts corresponding to method calls on objects of type \verb+Iterator+. An LTL specification then stipulates that every \verb+next+ event must be preceded by \verb+hasnexttrue+. The \verb+@violation+ section can contain arbitrary Java code that is to be executed when the specification becomes violated. The whole specification is enclosed in a declaration that is parameterized by $i$; this is an example of a technique called \emph{parametric slicing}. Concretely, one monitor instance will be created for each iterator manipulated by the program; JavaMOP takes care of dispatching the relevant events to the appropriate monitor instances.

Given a specification in any of the supported languages, JavaMOP transforms it into an optimized AspectJ code for a monitor, which is program-independent. This AspectJ code can then be \emph{weaved} with any Java program; the end result is that the monitor will check that the program conforms with specification at runtime.

\subsubsection{LARVA}

LARVA is another runtime monitoring architecture specifically aimed at the verification of Java programs \cite{CPS09larva}. LARVA uses as its main specification language a dynamic form of communicating automata with timers and events, called DATEs. In this context, events can be visible system actions (such as method calls or exception handling), timer events, channel synchronization (through which different automata may synchronize) or a combination of these elements.

\begin{figure}
\centering
\includegraphics[width=2in]{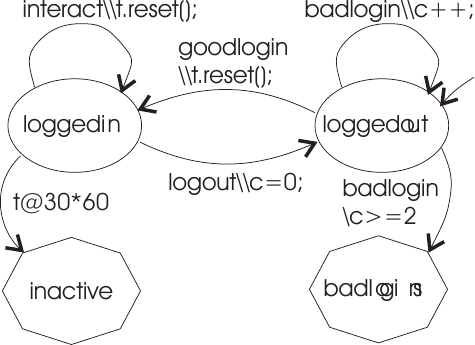}
\caption{An example of automaton used by LARVA.}
\label{fig:larva}
\end{figure}

Figure \ref{fig:larva} shows an example of DATE, for a property that monitors bad logins occurring in a system. Each transition is guarded by conditions on the input event (such as the event's name); optionally, a transition may also update internal variables specific to each automaton instance, such as counters.  Of interest in DATEs is the possibility to define timeout events; the ``\verb+t@30*60+'' guard on the leftmost transition indicates that this transition is to be taken automatically, if no other transition has fired in the past 30 minutes. Although timeouts and clocks have been used in model checking software such as UPPAAL \cite{DBLP:conf/hybrid/BengtssonLLPY95}, LARVA is one of the only runtime (i.e.\ streaming) tools supporting them.

DATEs are actually input into the LARVA system using a textual representation of the automaton. Such a representation allows a DATE to be nested within a \verb+foreach+ construct, which allows multiple instances of an automaton to be created for each value of the specified parameter encountered during the execution of the program. Recently, a tool called LarvaStat has also extended LARVA with the possibility of computing statistics on the execution of the program \cite{DBLP:conf/rv/ColomboGP10}. These statistics are exposed as ``statistical events'', and properties can be expressed in terms of these statistics.

\subsubsection{MarQ}

MarQ \cite{DBLP:conf/fm/BarringerFHRR12} is a runtime monitoring tool that deals with parametric properties, in which events may carry data. A parametric event is a pair of an event name and a list of data values (such as shown in Figure \ref{fig:auction-events}), and a parametric trace is a finite sequence of parametric events. A parametric \emph{property} denotes a set of parametric traces, in the same way that a regular expression describes a set of symbol sequences.

\begin{figure}
\centering
\includegraphics[width=3in]{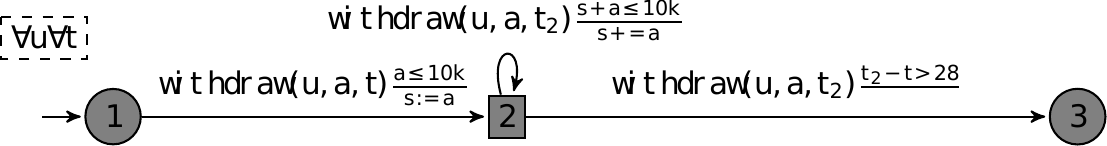}
\caption{An example of Quantified Event Automaton, used by MarQ.}
\label{fig:qea}
\end{figure}

Quantified event automata (QEA) is a notation for describing parametric properties. Figure \ref{fig:qea} shows an example of such an automaton, corresponding to the property that a user must withdraw less than \$10,000 in a 28-day period \cite{DBLP:conf/tacas/RegerCR15}. It highlights the various features of QEAs. First, an automaton can be parameterized by universal and existential quantifiers. These quantifiers will create as many \emph{slices} from the original trace, corresponding to the possible variable bindings encountered along the stream. Each QEA instance also has internal variables; guards on transitions of the automaton can refer to the values of the current event, and also of the current values of these internal variables. Moreover, these variables can be updated when a transition is taken.

One advantage of QEAs is the fact that they admit an efficient monitoring algorithm via an approach called \emph{parametric trace slicing} \cite{DBLP:conf/tacas/ChenR09}. In the Runtime Verification community, MarQ has consistently fared among the fastest runtime monitors available, such as at the latest Competition on Runtime Verification \cite{DBLP:conf/rv/RegerHF16}.

\subsubsection{LogFire}

The LogFire system was developed for verifying execution traces against formal specifications. Instead of automata, like in LARVA and MarQ, it uses a different formalism for specifying the expected behaviour of a system, based on the concept of rules \cite{DBLP:conf/rv/Havelund13}. In this respect, it shares similarities with a popular rule engine called Drools\footnote{\url{http://www.jboss.org/drools}} and Jess\footnote{\url{http://herzberg.ca.sandia.gov}}. Basic events follow the same structure as in MarQ; these events, along with additional \emph{facts}, can be written to a dynamic structure called a fact memory. LogFire implements a domain-specific language (DSL) based on the Scala programming language, to allow the expression of rules that correlate these events and facts. Each rule has the form:
\[
\mbox{name} -- \mbox{condition}_1 \wedge \dots \wedge \mbox{condition}_n |-> \mbox{action}
\]

A rule is defined by a name, a left hand side consisting of a conjunction of conditions, and a right hand side consisting of an action to be executed if all the conditions match the fact memory. An action can be adding facts, deleting facts, or generally be any Scala code to be executed when a match for the left-hand side is found. For example, Figure \ref{fig:logfire} shows a simple Scala block of code for a \verb+Monitor+ object that checks the property: a resource can only be granted to one task (once) at any point in time, and must eventually be released by that task. Rule \verb+r1+, for example, is fired when the current event's name is ``grant'' with parameters $t$ and $r$, and there is no fact in the memory called \verb+Granted+ with the same parameters $t$ and $r$. If such is the case, the rule fires, and its action consists of adding a new fact $\mbox{Granted}(t,r)$ in the fact memory.

\newsavebox{\logfire}
\begin{lrbox}{\logfire}
\begin{lstlisting}[language=scala]
class ResourceProperties extends Monitor {
  val grant, release , end = event
  val Granted = fact
  "r1" -- grant('t, 'r) & not(Granted('t, 'r)) |-> Granted('t, 'r)
  "r2" -- Granted('t, 'r) & release('t, 'r) |-> remove(Granted)
  "r3" -- Granted('t, 'r) & grant(' , 'r) |-> fail("double grant")
  "r4" -- Granted('t, 'r) & end() |-> fail("missing release")
  "r5" -- release('t,'r) & not(Granted('t,'r)) 
    |-> fail("bad release")
}
\end{lstlisting}
\end{lrbox}
\begin{figure}
\centering
\usebox{\logfire}
\caption{A set of rules used by LogFire, written in Scala.}
\label{fig:logfire}
\end{figure}

The readability of the rules is enhanced by a ``trick'': each rule definition in the monitor is actually an implicit chain of method calls which gets executed when the class is first instantiated. To this end, the \verb+Monitor+ class declares implicit functions; these functions are applied by the Scala compiler in cases where type checking of an instruction fails but where it succeeds if one such (unique) implicit function can be applied. One implicit function is called \verb+R+; it takes a string as an argument and returns an object of a class. This object, in turn, defines a function called \verb+--+, which takes a condition, and returns another object, this time defining a method called \verb+&+, and so on. Hence, once implicit functions are inserted by the compiler, each rule actually becomes a plain Scala statement that instantiates objects and calls their methods.

To determine what rules may fire upon an incoming event, LogFire implements a pattern matching algorithm called Rete \cite{DBLP:journals/ai/Forgy82}. The DSL allows domain specific constructs to be mixed with Scala code, making the notation very expressive and convenient for practical purposes. When an error is detected, the system produces an error trace illustrating what events caused what rules to fire, allowing the user to understand the cause of the violation.

Another system, T-REX, uses a rule-based language called \textsc{Tesla} \cite{DBLP:journals/jss/CugolaM12}. Instead of a Rete-based algorithm, \textsc{Tesla} rules are evaluated through a conversion into finite-state automata.

\subsubsection{MonPoly}

MonPoly is another tool for the evaluation of logical properties over event logs \cite{DBLP:journals/fmsd/BasinKMZ15}. Its specification language is called Metric First-Order Temporal Logic (MFOTL), and is an extension of Linear Temporal Logic with predicates and first-order quantifiers.

In MonPoly, each event is viewed as a mini ``database'' that can be queried by means of predicates. For example, an expression like $\mbox{withdraw}(u;a)$ is true if the current event represents a withdrawal made by user $u$ for amount $a$. In addition to Boolean connectives, basic assertions can also include temporal operators. The ``globally'' modality, noted $\Box \varphi$, signifies that $\varphi$ must hold for every suffix of the trace starting at the current point. The ``eventually'' modality, noted $\lozenge \varphi$, stipulates that $\varphi$ must hold for some suffix of the trace starting at the current point. These two modalities also have their past equivalents, represented by black symbols. Temporal operators can also be parameterized by an interval; for example, $\lozenge_{[a,b]} \varphi$ says that $\varphi$ must hold at some point between $a$ and $b$ time units from the current moment.

Special care has been taken in MFOTL to handle aggregate functions over multiple events. An expression of the form $[\omega_t \overline{z}.\psi](y;\overline{g})$ is called an aggregation formula. Here, $\overline{g}$ is a list of attributes on which grouping is performed, $t$ is the term on which the aggregation operator $\omega$ is applied, and $y$ is the attribute that stores the result. 
Supported aggregation operators include sum, average, minimum and maximum. Finally, MFOTL also supports first-order quantifiers $\forall$ and $\exists$, which are interpreted in the standard way.

This makes it possible to express rich properties involving both sequential patterns and aggregation functions. For example, the following MFOTL property checks that for each user, the number of withdrawal peaks in the last 31 days does not exceed a threshold of five, where a withdrawal peak is a value at least twice the average over the last 31 days:
\begin{multline*}
\Box \forall u: \forall c: [\mbox{\textsf{CNT}}_j\, v; p; \kappa: [\mbox{\textsf{AVG}}_a\, a; \tau.\blacklozenge_{[0;31)}  \\
\mbox{withdraw}(u; a) \wedge \mbox{ts}(\tau)](v; u) \wedge\\
\blacklozenge_{[0; 31)} \mbox{withdraw}(u; p) \wedge \mbox{ts}(\kappa) \wedge 2 \cdot \vee \prec p](c; u) \rightarrow c \preceq 5
\end{multline*}

Experimental evaluation of an implementation of MonPoly revealed that MFOTL queries are easier to maintain than their equivalent (and significantly longer) MySQL queries, and that the runtime performance is in the same order of magnitude than the STREAM system.

\subsubsection{Other systems}

Other runtime monitors and log analysis tools developed in the past include J-Lo \cite{DBLP:journals/entcs/StolzB06}, Mufin \cite{DBLP:conf/tacas/DeckerHS0T16}, PoET \cite{DBLP:conf/sp/ErlingssonS00}, PQL \cite{DBLP:conf/oopsla/MartinLL05}, PTQL \cite{DBLP:conf/oopsla/GoldsmithOA05}, RuleR \cite{DBLP:journals/logcom/BarringerRH10}, SEQ.OPEN \cite{DBLP:conf/spin/GaravelM04}, SpoX \cite{DBLP:conf/pldi/HamlenJ08}, and Tracematches \cite{DBLP:journals/logcom/BoddenHLLN10}.
Their specification languages can be related to one of the aforementioned systems.


\section{Desiderata for a Stream Query Engine}\label{sec:principles} 

The previous section has given a broad picture of the event processing landscape. We now make a few observations on the relative strengths and weaknesses of these solutions. Many of them will becom design goals warranting the development of a new and (hopefully) complementary event processing system.

In the realm of relational databases, desirable properties of a potential query language have been collated into a document called the Third Manifesto (3M) \cite{DBLP:journals/sigmod/DarwenD95}. 
In the following, we list a number of observations and choices that should be taken into account when designing a query language for ESP. These design choices will be reflected in the implementation of our event query engine, BeepBeep, and its companion query language, \esql{}.

\subsection{No Unique Event Type}

All CEP tools, with exception of Esper, 
assume events to be \emph{tuples}. In relational databases, the 3M (prescriptions 6--10) also enforces this rigid data model. In such a case every tuple of a trace must have the same fixed set of attributes, and events must differ only in the values they define for each attribute. Moreover, these values must be scalar. A query can transform an input trace into a different output, but the resulting trace will still be made of tuples, with possibly different attributes. RV tools have slightly more diverse events. At one extreme, JavaMOP events are atomic symbols, but at the other, MonPoly events are mini-databases that can be queried with predicates. Most other tools lie in between, and assume an event structure that can be mapped to lines of a CSV file (i.e.\ a form of tuple).

Yet, we have seen in Section \ref{sec:scenarios} how the tuple datatype is not appropriate for all possible queries. This is especially true of the use case of Section \ref{subsec:pingus}, where events produced by the running systems have a nested data structure where the same element names can occur multiple times. This issue has been raised in the Competition on Runtime Verification \cite{DBLP:conf/rv/RegerHF16}: to translate these events into flat tuples, the organizers had to introduce an event per character object, with the other metadata being copied between these new events. They report that flattening the events in such a way led to more complex specifications that needed to deal with the arbitrary ordering of events that should be observed at the same point. Query \ref{q:heatmap} is even further away from the tuple model. 

A truly generic event processing system should not presuppose that any single type of events is appropriate for all problems. Rather, each type of event should come with its own set of \emph{event manipulation functions} (EMF) to extract data, manipulate and create new events of that type. These functions should be distinct from \emph{stream manipulation functions} (SMF), which, in contrast, should make very limited assumptions on the traces they manipulate. This clear separation of EMF and SMF should make it possible to easily mix events of different types into queries. 
It should also help avoid the ``square peg in a round hole'' problem, where one must write an overly complicated expression simply to work around the limitations of the single available event type.

\subsection{Modularity and Composition}

A similar problem also arises with respect to the specification (or query) language of each tool. First, some tools (such as Apache Storm) have no query language: computations can only be achieved through code. The database foundations of ESP have led many solutions to compute everything through a tentacular \textsf{SELECT} statement, with optional constructs attempting to account for every possible use case.
A modular event processing framework should alleviate this problem by proposing a set of basic processing units that can be freely composed. Therefore, rather than proposing a single, all-encompassing query language, it should accommodate multiple query languages, along with lightweight syntactical ``glue'' to allow for their composition. Hence every step of the computation to be expressed in the notation most appropriate for it. 

Moreover, such a framework should provide, at the implementation level, easy means for extending it. This means both allowing the user to define new processing units, and also new ways for the language to accommodate them. In this respect, existing systems do not fare very well. With the exception of MOP, which lets users define new plugins, RV tools have a completely fixed specification language that cannot be extended. CEP languages sometimes allow the user to define new function symbols, but these new symbols can only be invoked in the traditional notation ``\textsl{function}(\textit{arguments})''.

\subsection{Relational Transparency}

A recurring problem with RV systems is that their specification language is seen as repulsive by potential end users. In contrast, virtually every CEP system touts an ``SQL-like'' query language, which has the reassuring effect of tapping into concepts that practitioners already know. Unfortunately, while they indeed borrow keywords from the SQL language, their syntax is almost invariably incompatible with SQL. For example, in the Cayuga language \cite{DBLP:conf/sigmod/BrennaDGHOPRTW07}, selecting all events where attribute \texttt{cnt} is greater than 10 is written as:

\begin{snippet}
\SELECT{} * \FROM{} \FILTER{} \{cnt $>$ 10\}(webfeeds)
\end{snippet}
\noindent
and in Siddhi as
\begin{snippet}
\SELECT{} * \FROM{} webfeeds(cnt $>$ 10)
\end{snippet}

\noindent
while extracting the same data from a database would be written as the following SQL query:

\begin{snippet}
  \SELECT{} * \FROM{} webfeeds \WHERE{} cnt $>$ 10
\end{snippet}

\noindent
Even SQLstream's syntax is not compatible, as it distinguishes between querying a stream and querying a table; the former requires the addition of the \textsf{STREAM} keyword. The only exception is Esper, whose basic \textsf{SELECT} statement is identical to SQL's.

When the context allows an event trace to be interpreted as an ordered relation whose events are tuples, then the SQL query computing some result over that relation should be a valid event stream query as well; we call this concept \emph{relational transparency}. Conversely, this means that standard relational tables should be able to be used as drop-in replacements for event traces anywhere in an expression where tuples are expected. This statement, in line with 3M's ``Very strong suggestion'' \#9, is in itself a distinguishing point to virtually every other ESP system around.

\subsection{Circumscribed Procedural Escapes}\label{subsec:escapes}

All event processing should be made through the combination of relatively high-level, declarative language constructs, without resorting to procedural statements or traditional code. A counter-example would a TelegraphCQ expression like this one:

\begin{codeblock}
\begin{verbatim}
Select AVG(closingPrice)
From ClosingStockPrices
Where stockSymbol = ‘\textsc{msft}’
for (t = ST; t < ST+50, t+= 5) {
  WindowIs(ClosingStockPrices, t - 4, t);
}
\end{verbatim}
\end{codeblock}

One can see that part of its processing is done through the use of a C-style \texttt{for} loop. There are many reasons why such an approach is undesirable. Besides being inelegant, it pollutes the declarative style of SQL with procedural statements which arguably should not occur in a query language. This, in turn, makes the semantics of the language very hard to define, and the actual meaning of a query difficult to grasp. There is also an implicit coupling between the value 5 that increments the loop counter, and the value 4 subtracted in the \textsf{\nospellcheck{WindowIs}} segment. 

In contrast, we expect an event stream query language to be fully declarative. Syntactically, this entails that no procedural constructs (if-then blocks, loops, variables) should be offered to the user. This point of view is actually stricter than SQL, as most SQL engines extend the language with such constructs. This also contradicts 3M's prescription 5, (which requires the presence of if-then blocks. 

This does not mean that the resulting system should not support user extensions. However, it should support them in a way that whatever procedural code that needs to be written can then be accessed through \emph{extensions} to the query language's syntax, thereby creating a Domain-Specific Language (DSL). While new processing units are made of (potentially Turing-complete) code, users should not have the possibility of writing procedural code inside their queries, thus preserving their declarative nature. 

%

\subsection{Increased Expressiveness}

In terms of the expressiveness of their respective input languages, RV and CEP systems have complementary strengths. Compared to RV, CEP tools are far less advanced in terms of evaluating sequential patterns of events. In many of their input languages, the only way of correlating an event with past or future events is through a \textsf{JOIN} of the trace with itself ---an expensive operation, which can only be done in restricted ways (such as by bounding the window of events that are joined). Intricate sequential relationships, such as those easily expressible with a finite-state machine notation common to many monitoring systems, are very hard to state in existing CEP tools. In a few cases, a language offers the possibility to describe primitive sequencing patterns (using a form of regular expression, or simple ``A follows B'' instructions). These patterns are very restricted in their use (for example, they do not allow negation) and, as empirical testing will reveal, costly to evaluate. Some systems like Cayuga transform their queries internally into finite-state machines, but their query language does not allow a user to directly specify FSMs. It shall also be noted that most CEP tools disallow queries that necessitate an unbounded number of future events to compute their result.

This is in sharp contrast with RV systems, where the sequential aspect of event traces is central. Since the specification language of monitors is based on logic, it is also natural to find a form of first-order quantification in many of them. This quantification occurs in problems where some specific pattern must hold ``for all elements''. A few CEP systems allow a trace to be split into various slices, but as a rule, no true equivalent of universal and existential quantification is supported.

In counterpart, CEP calculate the result of a \emph{query} on a trace of events, and the output of that query can itself be a sequence of events with data-rich contents, which can be reused as the input of another query. In contrast, a monitor evaluates a \emph{property} over a trace. Intermediate results of its computation are seldom exposed or expected to be consumable, and its output (most often a single truth value) is not reusable as the input of another monitor for further processing. There do exist monitors whose specification language involves more advanced data computing capabilities (numerical aggregation functions, mostly), but they still compute the answer to what is fundamentally a yes/no question.

As a consequence of the previous observation, it can be noted that CEP problems feature data-rich events, over which complex transformations and computations can made. Such functionalities are considered standard for a CEP language. Indeed, the \textsf{SELECT} construct provided by most CEP engines makes it possible to produce output tuples made of attributes from multiple input tuples, coming from potentially different input traces, combine them and apply various built-in functions (mostly numerical). In contrast, most monitors do support events with data fields, but only allow basic (again, Boolean) comparisons ($=$, $\leq$, etc.) between values of these fields. The handling of aggregation functions and other forms of computation over event data is not a common feature in RV, and only a handful of monitors so far support them \cite{DBLP:journals/fmsd/BasinKMZ15,DBLP:conf/rv/ColomboGP10,DBLP:conf/time/DAngeloSSRFSMM05,DBLP:journals/fmsd/FinkbeinerSS05,DBLP:journals/logcom/BarringerRH10}. 

Obviously, one should aim for the best of both worlds, with a system allowing the expression of rich data manipulation operations, rich pattern specifications, and more.

\subsection{Limiting Boilerplate Code and Configuration}

Many of the systems mentioned earlier, and in particular distributed CEP systems, require high amounts of setup and preparation before running even the smallest example. The ``Hello World'' example for Apache S4 requires setting up a cluster, editing half-a-dozen source and configuration files, and typing about as many arcane commands at the command line; the whole example requires six pages of documentation that one can hardly describe as user-friendly \cite{hw-s4}. Even when a CEP system is a reasonably stand-alone application, running a query on a simple input stream may still require non-trivial amounts of boilerplate code. Figure \ref{fig:mwe} shows an example of this situation. It displays the minimal Java code for reading tuples from a CSV file, running an Esper query that computes the sum of attributes \textsl{a} and \textsl{b} and printing the output events from that query one by one. Several observations can be made from this excerpt. First, about a dozen statements are required to instantiate all the required objects: a \verb+Configuration+, a \verb+ServiceProvider+, instances of \verb+EPAdministrator+, \verb+EPStatement+, \verb+EPRuntime+, and finally a user-defined \verb+UpdateListener+ to catch the output of the query. As is the case in other tools such as Siddhi, some of these objects must be passed to others, initialized, started, shutdown, reset, etc.\footnote{Figure \ref{fig:jdbc} shows the same code using BeepBeep's JDBC interface.}

Second, Esper does not provide a generic ``tuple'' event; an event type (here class \verb+TupleEvent+) must be explicitly created for each tuple type ---a tuple with different attributes would require yet another class declaration. Moreover, each field of the tuple must have a public getter method, and Esper even imposes the name it should have: the value of a field called \verb+foo+ must be accessed through a method called \verb+getFoo+. 

\begin{figure}
\centering
\usebox{\espercodebox}
\caption{Minimum working example for Esper}
\label{fig:mwe}
\end{figure}

Besides being cumbersome, this also goes against our first design requirement, as events cannot be arbitrary objects. For example, a trace of numbers cannot use Java's \verb+Number+ class for its type; because of the above conventions, the number would have to be encapsulated in a user-defined \verb+EventNumber+ class to be processed by Esper. Otherwise, events can be queried as JavaBeans, but this again imposes restrictions on their structure; for example, a primitive type is still not a JavaBean object. 

As a side note, the system also does not provide means to read events from a source; lines 10--12 and 16 must take care of this manually by reading the lines of the file, and lines 24--26 further break each text line to extract the attributes they contain. 

\subsection{Balancing Throughput and Other Considerations}

Most event stream processing systems emphasize their performance first and foremost. Virtually all commercial-grade event stream engines available contain the words ``fast'' or ``high throughput'' on their web sites and in their documentation. 
Recently, the term ``Fast Data'' has even emerged as the hyped successor of the ``Big Data'' trend \cite{zuho}.

There is without question a market for high-load, high-throughput solutions such as those described earlier. However, one of they key motivations of the present work is to put primary focus on the definition of a simple and versatile formal semantics for an event processing system, followed by the design of a readable and fully declarative query language. Performance considerations are relegated to third place; in particular, our system should not gain performance at the price of readability or simplicity, or succumb to premature optimization: ``first do it right, then do it fast''.

Case in point, in some of the use cases described in Section \ref{sec:scenarios}, the challenge is not high event load. A NIALM system generates readings at the same frequency as the power grid, i.e.\ 60 Hz; the \textit{Pingus} video game produces one event at each cycle of its game loop, which is approximately 150 Hz. Such a throughput can easily be processed with custom shell scripts. What one gains from using an event stream engine, rather than these scripts, is ease of use, flexibility, and maintainability of the queries ---not computation speed. In the same way, an Excel spreadsheet is not preferred by users because it is faster than a pocket calculator on raw arithmetical calculations, but because it eases the expression, maintenance and presentation of these calculations.

This standpoint is a minority voice in the current heavy trend focusing on performance. 
This, however, is not to be taken as an excuse for the bad performance of our engine. As a matter of fact, our empirical analysis at the end of this paper shows that for some classes of queries, our proposed tool has a performance commensurate with other CEP systems.


\section{Computational Framework}\label{sec:theory} 

The observations made in the previous section motivated the design of BeepBeep~3, a new event stream processing engine that aims to reconcile RV and CEP by supporting functionalities of both. As its name implies, it is the third incarnation of the BeepBeep line of monitoring software. Earlier versions of BeepBeep used a first-order extension of Linear Temporal Logic as their specification language. 

BeepBeep was designed with the goal of borrowing and improving on concepts found in a variety of other software. It fragments the processing of traces into pipelined computational units that generalize Trident's \emph{functions}, Aurora's \emph{boxes} and Siddhi's \emph{processors}. It supports both a ``push'' and ``pull'' semantics that resembles SQLstream's \emph{pumps}. Similar to Esper, its events can be objects of arbitrary types. Extensions to BeepBeep's core can handle finite-state machines like MarQ's, and a form of first-order temporal logic akin to MonPoly's. It provides yet another SQL-like query language, but which maintains backwards compatibility with SQL and can easily be extended by user-defined grammatical constructs.

BeepBeep can be used either as a Java library embedded in another application's source code, or as a stand-alone query interpreter running from the command-line. Versions of BeepBeep~3 are publicly available for download, and its code is released under an open source license.\footnote{\url{https://liflab.github.io/beepbeep-3}} Thanks to the simplicity of its formal foundations, the core of BeepBeep~3 is implemented using slightly less than 10,000 lines of Java code.

In this section, we describe the formal foundations of BeepBeep's computational model. In this model, the evaluation of a query is performed synchronously in discrete steps by computation units called \emph{processors}.

\subsection{Events, Functions and Processors}

Let $\dom{T}$ be an arbitrary set of elements. An \emph{event trace of type $\dom{T}$} is a sequence $\tr{e} = e_0 e_1 \dots$ where $e_i \in \dom{T}$ for all $i$. The set of all traces of type $\dom{T}$ is denoted $\dom{T}^*$. In the following, event types are written in double strike (e.g.\ $\dom{T}, \dom{U}, \dots$) and can refer to any set of elements. 
In line with the observations made previously, BeepBeep makes no assumption whatsoever as to what an event can be. Event types can be as simple as single characters or numbers, or as complex as matrices, XML documents, plots, logical predicates, polynomials or any other user-defined data structure. In terms of implementation, an event can potentially be any descendent of Java's \texttt{Object} class.

A \emph{function} is an object that takes zero or more events as its input, and produces zero or more events as its output. The \emph{arity} of a function is the number of input arguments and output values they have. Borrowing terminology from the theory of relations \cite{Quine45}, a function accepting one event as input will be called \emph{monadic} (or unary), while one accepting two events will be called \emph{dyadic} (or binary), and so on. Functions accepting no input are called \emph{medadic}, or more simply \emph{constants}. Since functions both have input and output, they must be qualified according to both ---one may hence talk about a dyadic-input, monadic-output function, or more succinctly a 2:1 function.
For example, the addition function $+ : \mathbb{R}^2 \rightarrow \mathbb{R}$ is the 2:1 function that receives two real numbers as its input, and returns their sum as its output. While functions with an output arity of 2 or more are rare, they do occur in some situations; for example, one can imagine the function $f : \mathbb{C} \rightarrow \mathbb{R}^2$ which, given a complex number, returns both its real and imaginary parts as two distinct outputs. In BeepBeep, functions are first-class objects; they all descend from an abstract ancestor named \texttt{Function}, which declares a method called \texttt{evaluate()} so that outputs can be produced from  a given array of inputs.

A \emph{processor} is an object that takes zero or more event \emph{traces}, and produces zero or more event \emph{traces} as its output. The difference between a function and a processor is important. While a function is stateless, and operates on individual events, 
a processor is a stateful device: for a given input, its output may depend on events received in the past. Processors in BeepBeep all descend from the abstract class \texttt{Processor}, which provides a few common functionalities, such as obtaining a reference to the \nospellcheck{$n$-th} input or output, getting the type of the \nospellcheck{$n$-th} input or output, etc. Processors are similar in their nature to other concepts in CEP systems, such as ``bolts'' in Apache Storm, or to the similarly-named objects in Siddhi.

We shall use a formal notation that defines the output trace(s) of a processor in terms of its input trace(s). Let $\tr{e}_1, \dots, \tr{e}_n$ be $n$ input traces, and $\varphi$ be a processor. The expression $\sem{\tr{e}_1, \dots, \tr{e}_n}{\varphi}$ will denote the output trace produced by $\varphi$, given these input traces. As a simple example, let us consider a processor, noted \decimate{n}, that outputs every \nospellcheck{$n$-th} event of its input and discards the others (this process is called \emph{decimation}). This can be defined as:
\[
\sem{\tr{e}}{\,\decimate{n}}_i \equiv e_{ni}
\]
The expression states that the $i$-th event of the output stream is the $(n \times i)$-th event of the input stream.

Each processor instance is also associated with a \emph{context}. A context is a persistent and modifiable map that associates names to arbitrary objects. When a processor is duplicated, its context is duplicated as well. If a processor requires the evaluation of a function, the current context of the processor is passed to the function. Hence the function's arguments may contain references to names of context elements, which are replaced with their concrete values before evaluation. Basic processors, such as those described in Section \ref{subsec:builtin-proc}, do not use context. However, some special processors defined in extensions to BeepBeep's core (the Moore machine and the first-order quantifiers, among others) manipulate their \texttt{Context} object.

For a given input event, a processor can produce any number of output events. For example, one can imagine a stuttering processor $\psi_n$ that repeats each input event $n$ times, and defined as follows:
\[
\sem{\tr{e}}{\,\psi_n}_i \equiv e_{\lfloor\frac{i}{n}\rfloor }
\]

\subsection{Streaming, Piping and Buffering}

A processor produces its output in a \emph{streaming} fashion. However, a processor can require more than one input event to create an output event, and hence may not always output something. This can be seen in the case of the decimation processor described above. Given a trace $e_0 e_1, \dots$, the processor outputs $e_0$ immediately after reading it. However, it does not produce any output after consuming $e_1$; it will only produce another output after having consumed $n$ inputs.

\begin{figure}
\centering
\includegraphics[scale=0.4]{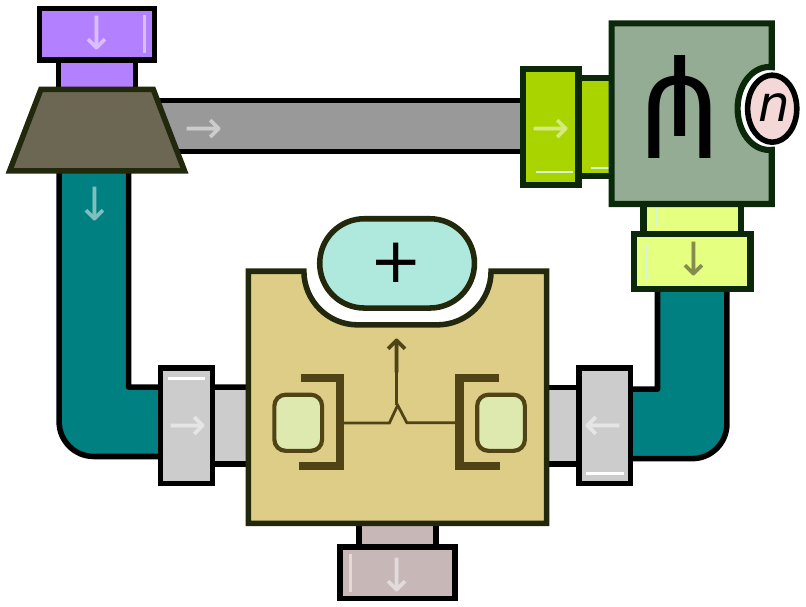}
\caption{A simple composition of processors, represented graphically}
\label{fig:simple-pipe}
\end{figure}

Processors can be composed (or ``piped'') together, by letting the output of one processor be the input of another. Another important characteristic of BeepBeep is that this piping is possible as long as the type of the first processor's output matches the second processor's input type.
%
The piping of processors can be represented graphically, as Figure \ref{fig:simple-pipe} illustrates. In this case, an input trace (of numbers) is duplicated into two copies; the first is sent as the first input of a 2:1 processor labelled ``+''; the second is first sent to the decimation processor, whose output is connected to the second input of ``+''. The end result is that output event $i$ will contain the value $e_i + e_{ni}$.

When a processor has an arity of 2 or more, the processing of its input is done synchronously. A \emph{front} is a tuple of input events with matching positions in each input stream. A computation step will be performed if and only if a complete front is available, i.e.\ an event can be consumed from each input trace. This is a strong assumption; many other CEP engines allow events to be processed asynchronously, meaning that the output of a query may depend on what input trace produced an event first. One can easily imagine situations where synchronous processing is not appropriate. However, in use cases where it is suitable, assuming synchronous processing greatly simplifies the definition and implementation of processors. The output result is no longer sensitive to the order in which events arrive at each input, or to the time it takes for an upstream processor to compute an output.\footnote{The order of arrival of events from the same input trace, obviously, is preserved.} As a result, given the formal definition of each processor in a query, a ``pen and paper'' calculation will always yield the same result as the implementation.

This hypothesis entails that processors must implicitly manage \emph{buffers} to store input events until a complete front can be consumed. Consider the case of the processor chain illustrated in Figure \ref{fig:simple-pipe}. When $e_0$ is made available in the input trace, both the top and bottom branches output it immediately, and processor ``+'' can compute their sum right away. When $e_1$ is made available, the first input of ``+'' receives it immediately. However, the decimation processor produces no output for this event. Hence ``+'' cannot produce an output, and must keep $e_1$ in a queue associated to its first input. Events $e_2$, $e_3$, \dots will be accumulated into that queue, until event $e_n$ is made available. This time, the decimation processor produces an output, and $e_n$ arrives at the second input of ``+''. Now that one event can be consumed from each input trace, the processor can produce the result (in this case, $e_0 + e_n$) and remove an event from both its input queues. 

Note that while the queue for the second input becomes empty again, the queue for the first input still contains $e_2, \dots e_{n}$. The process continues for the subsequent events, until $e_{2n}$, at which point ``+'' computes $e_2 + e_{2n}$, and so on. In this chain of processors, the size of the queue for the first input of ``+'' grows by one event except when $i$ is a multiple of $n$.

This buffering is implicit in the formal definition of processors, and is absent from the graphical representation of their piping. Nevertheless, the concrete implementation of a processor must take care of these buffers in order to produce the correct output. In BeepBeep, this is done with the abstract class \texttt{SingleProcessor}; descendents of this class simply need to implement a method named \texttt{compute()}, which is called only when an event is ready to be consumed at each input. Examples will be given in Section \ref{subsec:custom-proc}.

The reader can observe that many advanced features present in other event stream engines (such as handling out-of-order events, fault tolerance, order of arrival, clock synchronization, or validity intervals for events) are \emph{deliberately} left out of this model. One may argue that this makes for a poor and unappealing system, in terms of the number of bleeding-edge research concepts it implements. This is counter-balanced by three factors. First, some of these features can be handled by the environment in which the system is running; this is particularly the case of fault tolerance (virtual machine infrastructures readily provide crash recovery) and synchronization (the Precision Time Protocol can timestamp with sub-microsecond accuracy across machines). Similarly, BeepBeep can easily be run within another CEP architecture, such as Apache Spark, and benefit from its reliability properties. These solutions are certainly far superior than any potential built-in replication of their functionalities within the engine. Second, there exist numerous use cases (such as the ones we presented in Section \ref{sec:scenarios}) where these features are simply not needed. For those use cases, a user actually benefits from a simpler computational model. Finally, we shall see that in counterpart, thanks to this simple model, BeepBeep implements many features that other CEP engines do not.

\subsection{``Pull'' vs.\ ``Push'' Mode}

A first such feature allows events to be generated in two modes. In \emph{pull} mode, the handling of events in the processor pipe is triggered by requesting for a new output event. In order to produce this output event, the processor may require itself to fetch new events from its input(s), which in turn may ultimately lead to fetching events from the original input streams. On the contrary, in \emph{push} mode, output events are produced by the arrival of new events at the input side of the processor pipe. Both modes of operation require processors to handle input and output buffers ---pull mode asks processors to pile up events into their output buffer, while push mode makes them stack events into their input buffer. The presence of both input and output queues is necessary only to accommodate both modes. A pull-only system could be designed with only output queues, while a push-only system would only require input queues.

The interaction with a \texttt{Processor} object is done through two interfaces: \texttt{Pullable} and \texttt{Pushable}.
A \texttt{Pullable} object queries events on one of a processor's outputs. For a processor with an output arity of $n$, there exist $n$ distinct pullables, namely one for each output stream. Every pullable works roughly like classical \texttt{Iterator}: it is possible to check whether new
output events are available (\texttt{hasNext()}), and get one new output event (\texttt{next()}). However, contrarily to iterators, a \texttt{Pullable} has two versions of each method: a ``soft'' and a ``hard'' version.

``Soft'' methods make a single attempt at producing an output event. Since processors are connected in a chain, this generally means pulling events from the input in order to produce the output. However, if pulling the input produces no event, no output event can be produced. In such a case, \texttt{hasNextSoft()} will return a special value (\texttt{MAYBE}), and \texttt{pullSoft()} will return \texttt{null}. Soft methods can be seen as doing ``one turn of the crank'' on the whole chain of processors ---whether or not this outputs something.

``Hard'' methods are actually calls to soft methods until an output event is produced: the ``crank'' is turned as long as necessary to produce something. This means that one call to, e.g.\ \texttt{pull()} may consume more than one event from a processor's input. Therefore, calls to \texttt{hasNext()} never return \texttt{MAYBE} (only \texttt{YES} or \texttt{NO}), and \texttt{pull()} returns \texttt{null} only if no event will ever be output in the future (this occurs, for example, when pulling events from a file, and the end of the file has been reached). For the same processor, mixing calls to soft and hard methods is discouraged. As a matter of fact, the \texttt{Pullable}'s behaviour in such a situation is left undefined.

Interface \texttt{Pushable} is the opposite of \texttt{Pullable}: rather than querying events form a processor's output (i.e. ``pulling''), it gives events to a processor's input. This has for effect of triggering the processor's computation and ``pushing'' results (if any) to the processor's output. If a processor is of input arity $n$, there exist $n$ distinct \texttt{Pullable}s: one for each input trace.

It shall be noted that in BeepBeep, any processor can be used in both push and pull modes. In contrast, CEP systems (with the exception of TelegraphCQ) and runtime monitors generally support a single of these modes. 
The ``lazy'' evaluation of Apache Spark is an example of pull mode: upstream data is only generated upon request from downstream consumers. In contrast, the ``publish-subscribe'' model adopted by some event brokers (like Apache Kafka), corresponds to BeepBeep's push mode: an application subscribes to an event source, and is then notified of incoming events by the platform. 

This is also the case of Esper and Siddhi, where a user must define a callback function that the system calls whenever new output events are ready to be processed. The reader is referred to Figure \ref{fig:mwe} for an example. Surprisingly, this mode of operation, favoured by most engines, is the opposite of what is typically done in classical relational databases; the following shows a Java code sample querying a database using SQL:
\begin{lstlisting}
ResultSet res = st.executeQuery("SELECT * FROM  mytable");
while (res.next()) {
  int i = res.getInt("a");
}
\end{lstlisting}
Once the query is executed, the \verb+while+ loop fetches the tuples one by one, which clearly is an example of pull mode. Conversely, the use of push mode in an RDBMS has seldom (if ever) been seen.

The notion of push and pull is also present in the field of event-based parsing of XML documents, where so-called ``SAX'' (push) parsers \cite{sax} are opposed to ``StAX'' (pull) parsers \cite{stax}. XQuery engines such as XQPull \cite{DBLP:conf/ximep/FegarasDW06} implement these models to evaluate XQuery statements over XML documents. The use of such streaming XQuery engines to evaluate temporal logic properties on event traces had already been explored in an early form in \cite{DBLP:conf/sac/HalleV09}. 

\subsection{Built-in Processors}\label{subsec:builtin-proc}

BeepBeep is organized along a modular architecture. The main part of BeepBeep is called the \emph{engine}, which provides the basic classes for creating processors and functions, and contains a handful of general-purpose processors for manipulating traces. The rest of BeepBeep's functionalities is dispersed across a number of \emph{palettes}. In the following, we describe the basic processors provided by BeepBeep's engine.

\subsubsection{Function Processors}

A first way to create a processor is by lifting any $m:n$ function $f$ into a $m:n$ processor. This is done by applying $f$ successively to each front of input events, producing the output events. The processor responsible for this is called a \texttt{FunctionProcessor}. A first example of a function processor was shown in Figure \ref{fig:simple-pipe}. A function processor is created by applying the ``+'' (addition) function, represented by an oval, to the left and right inputs, producing the output. Recall that in BeepBeep, functions are first-class objects. Hence the \texttt{Addition} function can be passed as an argument when instantiating the \texttt{FunctionProcessor}. Since this function is 2:1, the resulting processor is also 2:1. Formally, the function processor can be noted as:
\[
\sem{\tr{e}_1, \dots, \tr{e}_m}{f}_i \equiv f(\tr{e}_1[i], \dots, \tr{e}_m[i])
\]

Two special cases of function processors are worth mentioning. The first is the \texttt{Passthrough}, which is the function processor where $m=n$ and $f$ is the identity function. The passthrough merely relays to its output what it receives at its input.
The \texttt{Mutator} is a $m:n$ processor where $f$ returns the same output, no matter its input. Hence, this processor ``mutates'' whatever its input is into the same output. The \texttt{Fork} is a $1:n$ processor that simply copies its input to its $n$ outputs. 

A variant of the function processor is the \texttt{CumulativeProcessor}, noted $\Sigma_f^t$. Contrarily to the processors above, which are stateless, a cumulative processor is stateful. Given a binary function $f : \mathbb{T} \times \mathbb{U} \rightarrow \mathbb{T}$, a cumulative processor is defined as:
\[
\sem{\tr{e}_1, \tr{e}_2}{\Sigma_f^t}_i \equiv f(\sem{\tr{e}_1, \tr{e}_2}{\Sigma_f^t}_{i-1}, \tr{e}_2[i])
\]
Intuitively, if $x$ is the previous value returned by the processor, its output on the next event $y$ will be $f(x,y)$. The processor requires an initial value $t \in \mathbb{T}$ to compute its first output.

Depending on the function $f$, cumulative processors can represent many things. If $f : \mathbb{R}^2 \rightarrow \mathbb{R}$ is the addition and $0 \in \mathbb{R}$ is the start value, the processor outputs the cumulative sum of all values received so far. If $f : \{\top,\bot,?\}^2 \rightarrow \{\top,\bot,?\}$ is the three-valued logical conjunction and $?$ is the start value, then the processor computes the three-valued conjunction of events received so far, and has the same semantics as the LTL$_3$ ``Globally'' operator. 

These simple processors can already be mixed. For example, an ``average'' processor can be built by dividing the output of two streams: one produced by the cumulative sum processor, the other produced by a mutator into the constant 1 piped into another cumulative sum. The result is indeed the sum of events divided by their number.

\subsubsection{Trace Manipulating Processors}

A few processors can be used to alter the sequence of events received. We already mentioned the \emph{decimator}, formally named \texttt{CountDecimate}, which returns every \nospellcheck{$n$-th} input event and discards the others. The \texttt{Freeze} processor, noted $\downarrow$, repeats the first event received; it is formally defined as
\[
\sem{\tr{e}}{\downarrow} \equiv (\tr{e}_0)^*
\]

A processor generates new output events only when being fed an input front. Hence, the previous processor does not output an infinite stream of $\tr{e}_0$ right after receiving it; rather, it will output one event for each input event consumed.

Another operation that can be applied to a trace is trimming its output. Given a trace $\tr{e}$, the \texttt{Trim} processor, denoted as $\rhd_n$, returns the trace starting at its \nospellcheck{$n$-th} input event. This is formalized as follows:
\begin{eqnarray*}
\sem{\tr{e}}{\rhd_n} & \equiv & \tr{e}^n
\end{eqnarray*}

Events can also be discarded from a trace based on a condition. The \texttt{Filter} processor \textsc{f} is a $n:n-1$ processor defined as follows:
\[
\sem{\tr{e}_1, \dots, \tr{e}_{n-1}, \tr{e}_n}{\mbox{\textsc{f}}%
}_i \equiv \begin{cases}
\tr{e}_1[i], \dots, \tr{e}_{n-1}[i] & \mbox{if $\tr{e}_n[i] = \top$}\\
\epsilon & \mbox{otherwise}
\end{cases}
\]
The filter behaves like a passthrough on its first $n-1$ inputs, and uses its last input trace as a guard; the events are let through on its $n-1$ outputs, if the corresponding event of input trace $n$ is $\top$; otherwise, no output is produced. A special case is a binary filter, where its first input trace contains the events to filter, and the second trace decides which ones to keep.

This filtering mechanism, although simple to define, turns out to be very generic. The processor does not impose any particular way to determine if the events should be kept or discarded. As long as it is connected to something that produces Boolean values, any input can be filtered, and according to any condition ---including conditions that require knowledge of future events to be evaluated. Note also that the sequence of Booleans can come from a different trace than the events to filter. This should be contrasted with CEP systems, that allow filtering events only through the use of a \textsf{WHERE} clause inside a \textsf{SELECT} statement, and whose syntax is limited to a few simple functions.

%
%


\subsubsection{Window Processor}\label{subsubsec:window}

Many times, one wants to perform computations over a ``sliding window'' of all events received, such as the sum of each set of $n$ successive events. This would produce an output sequence where the first number is the sum of events $1, 2, 3, \dots, n$ in the input sequence, the second number is the sum of events $2, 3, 4, \dots, n+1$, and so on.

Let $\varphi : \dom{T}^* \rightarrow \dom{U}^*$ be a 1:1 processor. The \emph{window processor of $\varphi$ of width $n$}, noted as $\Upsilon_n(\varphi)$, is defined as follows:
\[
\sem{\tr{e}}{\Upsilon_n(\varphi)}_i \equiv \sem{\tr{e}^i}{\varphi}_{n}
\]
One can see how this processor sends the first $n$ events (i.e.\ events numbered 0 to $n-1$) to an instance of $\varphi$, which is then queried for its \nospellcheck{$n$-th} output event. The processor also sends events 1 to $n$ to a second instance of $\varphi$, which is then also queried for its \nospellcheck{$n$-th} output event, and so on. The resulting trace is indeed the evaluation of $\varphi$ on a sliding window of $n$ successive events. 

In existing CEP engines, window processors can be used in a restricted way, generally within a \textsf{SELECT} statement, and only a few simple functions (such as sum or average) can be applied to the window. In contrast, in BeepBeep, \emph{any} processor can be encased in a sliding window, provided it outputs at least $n$ events when given $n$ fronts. This includes stateful processors: for example, a window of width $n$ can contain a processor that increments a count whenever an event $a$ is followed by a $b$. The output trace hence produces the number of times $a$ is followed by $b$ in a window of width $n$.

\subsubsection{Slicer}

The \texttt{Slicer} is a 1:1 processor that separates an input trace into different ``slices''. It takes as input a processor $\varphi$ and a function $f : \mathbb{T} \rightarrow \mathbb{U}$, called the \emph{slicing function}. There exists potentially one instance of $\varphi$ for each value in the image of $f$. If $\mathbb{T}$ is the domain of the slicing function, and $\mathbb{V}$ is the output type of $\varphi$, the slicer is a processor whose input trace is of type $\mathbb{T}$ and whose output trace is of type $2^\mathbb{V}$.

When an event $e$ is to be consumed, the slicer evaluates $c = f(e)$. This value determines to what instance of $\varphi$ the event will be dispatched. If no instance of $\varphi$ is associated to $c$, a new copy of $\varphi$ is initialized. Event $e$ is then given to the appropriate  instance of $\varphi$. Finally, the last event output by every instance of $\varphi$ is collected into a set, and that set is the output event corresponding to input event $e$. The function $f$ may return a special value $\#$, indicating that no new slice must be created, but that the incoming event must be dispatched to \emph{all} slices.

As a simple example, one may be interested in computing the sum of all odd and even numbers in a trace separately. This can be done by defining the slicing function as $f: x \mapsto x~\mbox{mod}~2$, and $\varphi$ as the  processor $\Sigma_+^0$, which computes the cumulative sum. Let us consider the trace $2,3,5$. Upon receiving the first event, the slicer evaluates $f(2) = 0$; a new instance of $\varphi$ is created, and is fed the value 2. Then the last value of all instances of $\varphi$ is collected, which leads to the set $\{2\}$. The process is repeated for the next event, 3. This time, $f(3) = 1$; a new instance of $\varphi$ is created, and the output this time becomes $\{2,3\}$. When 5 is consumed, it is dispatched to the existing instance of $\varphi$ associated to $f(5) = 1$, and the output is now $\{2,8\}$.

A particular case of slicer is when $\varphi$ is a processor returning Boolean values; the output of the slicer becomes a set of Boolean values. Applying the logical conjunction of all elements of the set results in checking that $\varphi$ applies ``for all slices'', while applying the logical disjunction amounts to existential quantification over slices.

The Slicer is reminiscent of Esper's \emph{context partition} (cf.\ Section \ref{subsubsec:esper}). As a matter of fact, one can use for $f$ a function that depends on a processor's context, which may be modified from outside the processor. In such a case, events are dispatched to a slice depending on an external context.


\section{The Event Stream Query Language}\label{sec:language} 

BeepBeep provides multiple ways to create processor pipes and to fetch their results. The first way is programmatically, using BeepBeep as a library and Java as the glue code for creating the processors and connecting them. For example, the code snippet in Figure \ref{fig:java-glue} creates the processor chain corresponding to Figure \ref{fig:simple-pipe}.  A \texttt{Fork} is instructed to create two copies of its input. The first (or ``left'') output of the fork is connected to the ``left'' input of a processor performing an addition. The second (or ``right'') output of the fork is connected to the input of a decimation processor, which itself is connected to the ``right'' input of the sum processor. One then gets a reference to \texttt{sum}'s (only) \texttt{Pullable}, and start pulling events from that chain. The piping is done through the \texttt{connect()} method; when a processor has two inputs or outputs, the symbolic names \texttt{LEFT}/\texttt{RIGHT} and \texttt{TOP}/\texttt{BOTTOM} can be used instead of 0 and 1. The symbolic names \texttt{INPUT} and \texttt{OUTPUT} refer to the (only) input or output of a processor, and stand for the value 0.

\begin{figure}
\begin{lstlisting}
Fork f = new Fork(2);
FunctionProcessor sum = 
  new CumulativeFunctionProcessor(Addition.instance);
CountDecimate decimate = new CountDecimate(n);
Connector.connect(fork, LEFT, sum, LEFT)
         .connect(fork, RIGHT, decimate, INPUT)
         .connect(decimate, OUTPUT, sum, RIGHT);
\end{lstlisting}
  % ...
\caption{Java code creating the chain of processors corresponding to Figure \ref{fig:simple-pipe}.}
\label{fig:java-glue}
\end{figure}

Another way of creating queries is by using BeepBeep's query language, called the Event Stream Query Language (\esql). \esql{} is the result of a careful process that went along with the development of BeepBeep's processors. Contrarily to some other systems, where the query language is built-in and inseparable from the underlying processing model, in BeepBeep the query language is just another means of instantiating a chain of processors. Rather than programmatically creating and piping processor instances, an \verb+Interpreter+ object can be used to convert a structured text string into that same piping. This means that processors themselves are unaware of the way they have been created. Moreover, we shall see in Section \ref{subsec:custom-lang} that even the basic grammar of the language is completely user-modifiable.

\subsection{Basic Constructs}

\esql{}'s grammar follows a modular design: each element of BeepBeep's architecture (processors, functions) comes with its own syntactical rules. The composition of two processors is expressed by enclosing an expression within another one.
%

\vskip 4pt
\textbf{Top production rule
}
\vskip 4pt

\begin{grammar}

<S> ::= <processor> | <processor-def> ;
\end{grammar}

\vskip 4pt
\textbf{Definition of a processor
}
\vskip 4pt

\begin{grammar}

<processor>          ::= <p-placeholder> | <userdef-proc>  ; 

<p-placeholder>      ::= * ;
\end{grammar}

%
%
%
%
%
%
%


\vskip 4pt
\textbf{User-defined processors. Rules get dynamically added here
}
\vskip 4pt

\begin{grammar}

<userdef-proc>       ::= gnarfnfar ;
\end{grammar}

\vskip 4pt
\textbf{Functions. Rules are added by grammar extensions.
}
\vskip 4pt

\begin{grammar}

<c-function>         ::= arfarfarf 
\end{grammar}


Table \ref{tab:grammar-functions} shows the basic language rules for manipulating processors and functions; two important non-terminal symbols defined there are \nt{processor} and \nt{function}.\footnote{The grammar shown in this section's tables is a direct formatting of BeepBeep's grammar files, taken from its source code. No modification or adaptation to the files was made for this paper.} Creating a constant function out of a constant symbol $c$ is done by writing \textsf{CONSTANT} $c$. Applying a function named $f$ on an event trace is done by writing \textsf{APPLY} $f$ \textsf{WITH} $L$, where $L$ is a comma-separated list of expressions that should each parse as a \nt{processor}. Applying a cumulative processor out of a function $f$ and an input trace $P$ is written \textsf{COMBINE} $P$ \textsf{WITH} $f$.

\begin{table}

\vskip 4pt
\textbf{Top production rule
}
\vskip 4pt

\begin{grammar}

<S> ::= <processor> | <processor-def> ;
\end{grammar}

\vskip 4pt
\textbf{Definition of a processor
}
\vskip 4pt

\begin{grammar}

<processor>          ::= <p-placeholder> | <userdef-proc>  ; 

<p-placeholder>      ::= * ;
\end{grammar}

%
%
%
%
%
%
%


\vskip 4pt
\textbf{User-defined processors. Rules get dynamically added here
}
\vskip 4pt

\begin{grammar}

<userdef-proc>       ::= gnarfnfar ;
\end{grammar}

\vskip 4pt
\textbf{Functions. Rules are added by grammar extensions.
}
\vskip 4pt

\begin{grammar}

<c-function>         ::= arfarfarf 
\end{grammar}


\vskip 4pt
\textbf{Processors
}
\vskip 4pt

\begin{grammar}

<processor>             ::= <fct-coll> | <fct-cp> | <cumulative> ;

<fct-coll>              ::= APPLY <function> <p-collator> ;

<fct-cp>                ::= CONSTANT <constant> ;

<cumulative>            ::= COMBINE <processor> WITH <fct-name> ;
\end{grammar}

\vskip 4pt
\textbf{Functions
}
\vskip 4pt

\begin{grammar}

<function>              ::= <fct-and> | <fct-or> | <fct-not> | <fct-eq> || <constant> ; 

<fct-and>               ::= <constant> $\wedge$ <constant> | <constant> $\wedge$ ( <function> ) | ( <function> ) $\wedge$ <constant> | ( <function> ) $\wedge$ ( <function> ) ;

<fct-or>                ::= <constant> $\vee$ <constant> | <constant> $\vee$ ( <function> ) | ( <function> ) $\vee$ <constant> | ( <function> ) $\vee$ ( <function> ) ;

<fct-not>               ::= $\neg$ <constant> | $\neg$ ( <function> ) ;

<fct-eq>                ::= <constant> = <constant> | <constant> = ( <function> ) | ( <function> ) = <constant> | ( <function> ) = ( <function> ) ;

<constant>              ::= <number> | <boolean> ;
\end{grammar}

\vskip 4pt
\textbf{Function names; only functions from T x T -> T need to have a name
}
\vskip 4pt

\begin{grammar}

<fct-name>              ::= <fctn-and> | <fctn-or> ;

<fctn-and>              ::= CONJUNCTION ;

<fctn-or>               ::= DISJUNCTION ;
\end{grammar}

\vskip 4pt
\textbf{Constants
}
\vskip 4pt

\begin{grammar}

<boolean>           ::= true | false 
\end{grammar}

\caption{Grammar for functions and function processors}
\label{tab:grammar-functions}
\end{table}

BeepBeep comes with only a handful of built-in functions: classical Boolean connectives and equality. Logical conjunction and disjunction can also be referred to by their names, so that they can be used inside a \textsf{COMBINE} expression. These constructs can be freely mixed, so that one can compute the cumulative sum of events from an input trace $P$ as:

\begin{snippet}
\textsf{COMBINE $P$ WITH ADDITION}
\end{snippet}

\begin{table}

\vskip 4pt
\textbf{Types
}
\vskip 4pt

\begin{grammar}

<number>             ::= \^{}\textbackslash{}d+;

<boolean>            ::= TRUE | FALSE ;

<var-name>           ::= \^{}\textbackslash{}\$[\textbackslash{}w\textbackslash{}d]+;
\end{grammar}

\vskip 4pt
\textbf{Processors
}
\vskip 4pt

\begin{grammar}

<processor>          ::= <p-freeze> | <p-window> | <p-decimate> | <p-prefix> 
\alt <p-trim> | <p-slicer> | <p-collator> 
\alt <var-name> ;

<p-freeze>           ::= FREEZE <processor> ;

<p-trim>             ::= TRIM <number> OF <processor> ;

<p-window>           ::= GET <processor> FROM <processor> ON A WINDOW OF <number> ;

<p-decimate>         ::= EVERY <number> <number-suffix> OF <processor> ;

<number-suffix>      ::= ST | ND | RD | TH ;

<p-prefix>           ::= THE FIRST <number> OF <processor> ;

<p-slicer>           ::= SLICE <processor> WITH <processor> ON <fct-name> ;
\end{grammar}

\vskip 4pt
\textbf{Processor list
}
\vskip 4pt

\begin{grammar}

<p-collator>         ::= WITH <proc-list> ;

<proc-list>          ::= <proc-def> , <proc-list> | <proc-def> ;

<proc-def>           ::= <proc-def-named> | <proc-def-anonymous> ;

<proc-def-named>     ::= <processor> AS <var-name> | ( <processor> ) AS <var-name> | ( <processor> AS <var-name> ) ;

<proc-def-anonymous> ::= ( <processor> ) | <processor> ;

<constant>           ::= <map-placeholder> ;

<map-placeholder>    ::= <var-name> 
\end{grammar}

\caption{Grammar for built-in processors}
\label{tab:grammar-tmf}
\end{table}

Table \ref{tab:grammar-tmf} shows the syntax for the basic built-in processors included in BeepBeep's core. The syntax for the Freeze, Decimate, Prefix and Trim processors is straightforward; from example, picking one event in every four from some trace $P$ is written as \textsf{EVERY 4TH OF} $P$.

The Window processor is slightly more involved. As defined in Section \ref{subsubsec:window}, the window requires an input trace $P$, a window width $n$, and another processor $P'$ to run on each window. This is done by writing \textsf{GET $P'$ FROM $P$ ON A WINDOW OF $n$}. Since $P'$ is itself a processor, its expression contains somewhere a reference to its input trace; this reference is replaced by the special placeholder \textsf{*}. For example, the following expression computes the sum of three successive events:

\begin{snippet}
\textsf{GET (COMBINE * WITH ADDITION) FROM $P$ ON A WINDOW OF 3}
\end{snippet}

The slicer works in a similar way. It requires an input trace $P$, a slicing function $f$, and a processor $P'$ to be applied on each slice of the input trace.

The last processor shown in Table \ref{tab:grammar-tmf} is the Collator. This processor is declared using the \textsf{WITH} keyword, followed by a list of expressions that should parse as \nt{processor}; each can be given a name using the optional keyword \textsf{AS}. The collator can be used to apply a computation from more than one input trace. 

For example, if $P$ and $P'$ are two expressions that produce streams of numbers, the pairwise sum of windows of length 3 from these input streams would be written as:

\begin{snippet}
\textsf{APPLY \$A + \$B WITH}\\
\sz\textsf{GET (COMBINE * WITH ADDITION) FROM $P$ ON A WINDOW OF 3 AS \$A,}\\
\sz\textsf{GET (COMBINE * WITH ADDITION) FROM $P'$ ON A WINDOW OF 3 AS \$B}
\end{snippet}
\noindent
The expression defines two placeholders for events from each input trace, named \$A and \$B, which are then used in an expression involving a function. The use of the dollar sign (\$) is only a convention; placeholders do not necessarily have to start with this symbol.

\subsection{Creating Definitions}\label{subec:esql-defs}

One notable feature of \esql{} is the capability for a user to extend the grammar dynamically through expressions, using the \textsf{WHEN} keyword. The corresponding syntactical rules are described in Table \ref{tab:grammar-definitions}. 
For example, the following code snippet shows how a new processor counting the elements of a trace can be defined by the user:

\begin{snippet}
\WHEN{} @P \ISA{} \textsf{PROCESSOR:}\\
\textsf{THE COUNT OF} @P \textsf{IS THE PROCESSOR}\\
\sz \COMBINE{}\\
\sz\sz \textsf{APPLY CONSTANT} 1 \WITH{} @P \\
\sz \WITH{} \textsf{ADDITION}.
\end{snippet}

\begin{table}
\begin{grammar}

<definition> ::= \WHEN{} <type-list> <pattern> \textsf{IS THE} <grammar-symbol> <S>

<pattern> ::= \^{}.*?(?=\textsf{IS})

<type-list> ::= <type-definition> | <type-definition> , <type-list>

<type-definition> ::= <var-name> \textsf{IS A} <grammar-symbol>

<var-name> ::= @<string>

\end{grammar}


\caption{Grammar for adding new syntactical constructs}
\label{tab:grammar-definitions}
\end{table}

The second line of the expression declares a new rule for the non-terminal symbol $\langle\mbox{\textit{processor}}\rangle$ in the grammar of Table~\ref{tab:grammar-functions}. It gives the syntax for that new rule; in that case, it is the expression \textsf{THE COUNT OF}, followed by the symbol ``@P''. The first line declares that ``@P'' must be a grammatical construct whose parsing matches the non-terminal $\langle\mbox{\textit{processor}}\rangle$. Finally, the remainder of the expression describes what \textsf{THE COUNT OF} @P should be replaced with when evaluating an expression; in this case, it is an \textsf{APPLY} statement. From that point on, \textsf{THE COUNT OF} @P can be used anywhere in an expression where a grammatical construct of type $\langle\mbox{\textit{processor}}\rangle$ is required, and this expression itself can accept for @P any processor expression.

This mechanism proves to be much more flexible than user-defined functions provided by other languages, as \emph{any} element of the original grammar can be extended with new definitions, themselves involving any other grammatical element. For example, one can easily define a numerical constant with an expression like
\textsf{PI IS THE NUMBER} 3.1416. This is a special case of the \nt{processor-def} grammatical construct in Table \ref{tab:grammar-definitions}, where the \textsf{WHEN} clause is empty.




\section{Extending Basic Functionalities}\label{sec:palettes} 

BeepBeep was designed from the start to be easily extensible. Any functionality beyond the few built-in processors presented in Section \ref{sec:theory} is implemented through custom processors and grammar extensions, grouped in packages called \emph{palettes}. Concretely, a palette is implemented as a JAR file that is loaded with BeepBeep's main program to extend its functionalities in a particular way, through the mechanisms described in this section.
This modular organization is a flexible and generic means to extend the engine to various application domains, in ways unforeseen by its original designers. Palettes make the engine's core (and each palette individually) relatively small and self-contained, easing the development and debugging process. Moreover, for any given application, only the engine and a small number of palettes need to be loaded; this results in fewer lines of dead code than what a monolithic piece of software would achieve. 
Finally, it is hoped that BeepBeep's palette architecture, combined with its simple extension mechanisms, will help third-party users contribute to the BeepBeep ecosystem by developing and distributing extensions suited to their own needs.

\subsection{Creating Custom Processors}\label{subsec:custom-proc}

Sometimes, creating a new processor cannot easily be done by combining existing ones  using the \textsf{WHEN} construct. BeepBeep also allows users to define their own processors directly as Java objects, using no more than a few lines of boilerplate code. The simplest way to do so is to extend the \texttt{SingleProcessor} class, which takes care of most of the ``plumbing'' related to event management: connecting inputs and outputs, looking after event queues, etc. All that is left to do is to define its input and output arity, and to write the actual computation that should occur, i.e.\ what output event(s) to produce (if any), given an input event.
We illustrate this process on a small example.

The minimal working example for a custom processor is made of six lines of code, and results in a processor that accepts no inputs, and produces no output:

\begin{lstlisting}
import ca.uqac.lif.cep.*;
public class MyProcessor extends SingleProcessor {
  public MyProcessor() {
    super(0, 0);
  }
  public Queue<Object[]> compute(Object[] inputs) {
    return null;
  }
}
\end{lstlisting}

\subsubsection{Example 1: Euclidean Distance}

Consider a processor that takes as input two traces. The events of each trace are instances of a user-defined class \texttt{Point}, which contains member fields \texttt{x} and \texttt{y}. We will write a processor that takes one event (i.e.\ one \texttt{Point}) from each input trace, and return the Euclidean distance between these two points. The input arity of this processor is therefore 2 (it receives two points at a time), and its output arity is 1 (it outputs a number). Specifying the input and output arity is done through the call to \texttt{super()} in the processor's constructor: the first argument is the input arity, and the second argument is the output arity.

The actual functionality of the processor is written in the body of method \texttt{compute()}. This method is called whenever an input event is available, and a new output event is required. Its argument is an array of Java \texttt{Object}s; the size of that array is that of the input arity that was declared for this processor (in our case: 2).

\begin{lstlisting}
public class EuclideanDistance extends SingleProcessor {
  public EuclideanDistance() { super(2, 1); }

  public Queue<Object[]> compute(Object[] inputs) {
    Point p1 = (Point) inputs[0];
    Point p2 = (Point) inputs[1];
    float distance = Math.sqrt(Math.pow(p2.x - p1.x, 2)
      + Math.pow(p2.y - p1.y, 2));
    return Processor.wrapObject(distance);
}}
\end{lstlisting}

The \texttt{compute()} method must return a queue of arrays of objects. If the processor is of output arity $n$, it must put an event into each of its $n$ output traces. It may also decide to output more than one such $n$-uplet for a single input event, and these events are accumulated into a queue ---hence the slightly odd return type. However, if the processor outputs a single element, the tedious process of creating an array of size 1, putting the element in the array, creating a queue, putting the array into the queue and returning the queue is encapsulated in the static method \texttt{Processor.wrapObject()}, which does exactly that.

\subsubsection{Example 2: Maximum}
As a second example, we create a processor that outputs the maximum between the current event and the previous one. That is, given the following input trace $5, 1, 2, 3, 6, 4, \dots$, the processor should output: (nothing), $5, 2, 3, 6, 6, \dots$. Notice how, after receiving the first event, the processor should not return anything yet, as it needs two events before saying something. Here is a possible implementation:

\begin{lstlisting}
public class MyMax extends SingleProcessor {
  Number last = null;
  public MyMax() { super(1, 1); }

  public Queue<Object[]> compute(Object[] inputs) {
    Number current = (Number) inputs[0], output;
    if (last != null) {
      output = Math.max(last, current);
      last = current;
      return Processor.wrapObject(output);
    } else {
      last = current;
      return Processor.getEmptyQueue();
    }
}}
\end{lstlisting}

This example, as well as the previous one, are meant to illustrate how to create custom processors. However, in both cases, it is possible to achieve the same functionality by composing basic processors already provided by BeepBeep. In the first case, one could define a binary function \texttt{Distance}, and encase it into a \texttt{FunctionProcessor}; in the second case, one could apply a \texttt{WindowFunction}, using \texttt{MaxFunction} as the function to evaluate.

\subsection{Grammar Extensions}\label{subsec:custom-lang}

By creating a custom processor, it is possible to pipe it to any other existing processor, provided that its input and output events are of compatible types. We have seen in Section \ref{subec:esql-defs} how a combination of existing processors can be defined directly within \esql{}; it is also possible to extend the grammar of the \esql{} language for a custom \texttt{Processor} object, so that it can be used directly in \esql{} queries.

As an example, let us consider the following processor, which repeats every input event $n$ times, where $n$ is a parameter decided when the processor is instantiated. Its implementation is as follows:

\begin{lstlisting}
public class Repeater extends SingleProcessor {
  private final int numReps;
  
  public Repeater(int n) {
    super(1, 1);
    this.numReps = n;
  }

  public Queue<Object[]> compute(Object[] inputs) {
    Queue<Object[]> queue = new LinkedList<Object[]>();
    for (int i = 0; i < this.numReps; i++) {
      queue.add(inputs);
    }
    return queue;
}}
\end{lstlisting}

The first step is to decide what syntax one shall use to invoke the processor. 
A possibility could be: ``\texttt{REPEAT (}$p$\texttt{) n TIMES}''. In this syntax, $p$ refers to any other \esql{} expression that builds a processor, and $n$ is a number. The result of this expression is itself another object of type \texttt{Processor}.

The second step is to tell the BeepBeep interpreter to add to its grammar a new case for the parsing of the existing $\langle\mbox{\textit{processor}}\rangle$ rule. This rule should correspond to the parsing of the newly-defined \verb+Repeater+ processor. This is done as follows:

\begin{lstlisting}
Interpreter my_int = new Interpreter();
my_int.addCaseToRule("<processor>", "<repeater>");
my_int.addRule("<repeater>", 
  "REPEAT ( <processor> ) <number> TIMES");
my_int.addAssociation("<repeater>", "my.package.Repeater");
\end{lstlisting}

The second instruction tells the interpreter that $\langle\mbox{\textit{processor}}\rangle$ can be parsed as a $\langle\mbox{\textit{repeater}}\rangle$. The parsing pattern for this non-terminal is then added with the call to \verb+addRule()+. This allows the interpreter to know that \texttt{REPEAT xxx n TIMES} corresponds to a processor. The last call tells the interpreter that encountering the \texttt{<repeater>} rule will result in the instantiation of a Java object of the class \texttt{Repeater}. This second argument should be the fully qualified name of the class. That is, if Repeater is located in package my.package, then one should write my.package.Repeater in the call to \verb+addAssociation()+. 

Upon parsing the $\langle\mbox{\textit{processor}}\rangle$ rule, the interpreter will look for a method called \verb+build(Stack<Object>)+ in the corresponding class. The task of the \verb+build()+ method is to consume elements of the parse stack to build a new instance of the object to create, and to put that new object back on the stack so that other objects can consume it during their own construction. 
Creating a new instance of Repeater is therefore straightforward. One simply has to \verb+pop()+ the stack to fetch the value of $n$ and the \verb+Processor+ object to use as input, and discard all ``useless'' keywords. 
One can then instantiate a new \verb+Repeater+, pipe the input into it (using \verb+Connector.connect()+), and put the resulting object on the stack.

\begin{lstlisting}
public static void build(Stack<Object> stack) {
  stack.pop(); // TIMES
  Number n = (Number) stack.pop();
  stack.pop(); // )
  Processor p = (Processor) stack.pop();
  stack.pop(); // (
  stack.pop(); // REPEAT
  Repeater r = new Repeater(n.intValue());
  Connector.connect(p, r);
  stack.push(r);
}
\end{lstlisting}

The possibility of extending \esql{}'s grammar in such a way is a feature unique to the BeepBeep event stream query engine. Adding new grammatical \emph{constructs} is actually more powerful than simply allowing user-defined \emph{functions}, as is done in some other ESP engines. It allows \esql{} to be extended to become a Domain-Specific Language (DSL). As a matter of fact, even the grammar for the built-in processors it soft-coded: it can be completely rewritten at runtime. Therefore, \esql{}, as described in this paper, is only a ``suggestion'' of syntax. Altering the basic grammar of any one of the other systems described in this paper is simply not offered to the user.

This feature required the development of a special parser called Bullwinkle\footnote{\url{https://github.com/sylvainhalle/Bullwinkle}}. Commonly used libraries, such as Yacc or Bison, are parser \emph{generators}: given a grammar, they generate the code corresponding to a parser for that grammar, which can then be included within another application. However, changing this grammar requires re-generating the parser, and hence recompiling the application that uses it. It is clear that such libraries are ill-suited for use cases where new rules can be dynamically added during execution. In contrast, Bullwinkle reads a grammar and parses expressions at run time, making it possible for the grammar to be modified at will by a user.

\subsection{Existing Palettes}

We describe a few of the palettes that have already been developed for BeepBeep in the recent past. These palettes are available alongside BeepBeep from a companion software repository.\footnote{\url{https://github.com/liflab/beepbeep-3-palettes}}

\subsubsection{Tuples and JDBC}

Of particular interest to this paper is the palette manipulating events that are associative maps of scalar values ---in other words, \emph{tuples} in the relational sense of the term. In addition, the palette includes a few utility functions for manipulating tuples.

The \texttt{Select} processor allows a tuple to be created by naming and combining the contents of multiple input events. The \texttt{From} processor transforms input events from multiple traces into an array (which can be used by \texttt{Select}), and the \texttt{Where} processor internally duplicates an input trace and sends it into a \texttt{Filter} evaluating some function. Combined together, these processors provide the same kind of functionality as the SQL-like \textsf{SELECT} statement of other CEP engines.

To this end, the palette defines a new grammatical construct, called \textsf{SELECT}, that allows an output tuple to be created by picking and combining attributes of one or more input tuples. The grammar extension for the \textsf{SELECT} statement is given in Table~\ref{tab:grammar-eml-tuples}. For the sake of simplicity, we only show a few arithmetical functions that manipulate numerical values; the actual syntax of \textsf{SELECT} can easily be made to accommodate functions manipulating other types of scalar values.

\begin{table}
\begin{grammar}

<processor>                      ::= <tuples-where> | <tuples-select> | <tuples-from> 
\alt <tuple-reader> ;

<tuple-name>                     ::= \^{}[a-zA-Z][\textbackslash{}w-]*; 

<constant>                       ::= <get-attribute> ;
\end{grammar}

\vskip 4pt
\textbf{Other processors
}
\vskip 4pt

\begin{grammar}

<tuple-reader>                   ::= THE TUPLES OF <processor> ;
\end{grammar}

\vskip 4pt
\textbf{"SELECT" statement
}
\vskip 4pt

\begin{grammar}

<tuples-select>                  ::= SELECT <attribute-expression-list> <processor> ;

<attribute-expression-list>      ::= <attribute-expression> , <attribute-expression-list>
\alt <attribute-expression> ;

<attribute-expression>           ::= <named-attribute-expression> 
\alt <anonymous-attribute-expression> ;

<named-attribute-expression>     ::= <function> AS <tuple-name> ;

<anonymous-attribute-expression> ::= <function> ;
\end{grammar}

\vskip 4pt
\textbf{"FROM" statement
}
\vskip 4pt

\begin{grammar}

<tuples-from>                    ::= FROM <tuple-expression-list> ;

<tuple-expression-list>          ::= <tuple-expression> , <tuple-expression-list>
\alt <tuple-expression> ;

<tuple-expression>               ::= <named-tuple-expression> 
\alt <anonymous-tuple-expression> ;

<named-tuple-expression>         ::= <processor> AS <tuple-name> ;

<anonymous-tuple-expression>     ::= <processor> ;
\end{grammar}

\vskip 4pt
\textbf{"WHERE" statement
}
\vskip 4pt

\begin{grammar}

<tuples-where>                   ::= ( <processor> ) WHERE <function> ;
\end{grammar}

\vskip 4pt
\textbf{Tuple functions
}
\vskip 4pt

\begin{grammar}

<get-attribute>                  ::= <get-attribute-qual> | <get-attribute-unqual> ;

<get-attribute-qual>             ::= <tuple-name> . <tuple-name> ;

<get-attribute-unqual>           ::= <tuple-name> ;
\end{grammar}

\caption{Grammar for the tuple palette}
\label{tab:grammar-eml-tuples}
\end{table}

One can see how this syntax precisely mirrors the basic form of SQL's command of same name. In contrast to the \textsf{SELECT} statement found in other ESP tools, \esql{}'s only manipulates tuples, and not traces. Operations such as filtering or windowing are obtained by \emph{composing} this statement with other constructs from BeepBeep's grammar. For example, selecting tuples that match some condition is done by piping the output of \textsf{SELECT} into BeepBeep's \texttt{Filter} processor, which is invoked syntactically through the \textsf{WHERE} keyword, as the grammar of Table \ref{tab:grammar-definitions} has already shown. This, as it turns out, results in an expression that reads exactly like SQL's \textsf{SELECT \dots WHERE}, ensuring the backward compatibility that was one of the design goals stated in Section~\ref{sec:principles}.

This palette also allows BeepBeep to be used through Java's JDBC API, as shown in Figure \ref{fig:jdbc}. This makes it possible to access the BeepBeep interpreter like any other relational database engine. This is also in line with one BeepBeep' design goal of \emph{relational transparency}. Surprisingly, despite their obvious roots in database theory, few of the other CEP engines considered in this study (and none of the runtime monitors) provide the same functionality.

\begin{figure}
\begin{lstlisting}
Class.forName("ca.uqac.lif.cep.jdbc.BeepBeepDriver");
Connection con = DriverManager.getConnection("jdbc:beepbeep:",
  "user","password");
Statement stmt = con.createStatement();
ResultSet rs = stmt.executeQuery(
  "SELECT a + b AS c FROM THE TUPLES OF FILE 'myfile.csv'");
while (rs.next()) {
  System.out.println(rs.getInt("c"));
}
rs.close();
con.close();
\end{lstlisting}
\caption{Running a BeepBeep query using its JDBC connector. The API is identical to that of a classical RDBMS.}
\label{fig:jdbc}
\end{figure}

\subsubsection{First-Order Linear Temporal Logic}

This palette provides processors for evaluating all operators of Linear Temporal Logic (LTL), in addition to the first-order quantification defined in LTL-FO$^+$ (and present in previous versions of BeepBeep) \cite{DBLP:journals/tsc/HalleV12}. Each of these operators comes in two flavours: Boolean and ``Troolean''.

Boolean processors are called \texttt{Globally}, \texttt{Eventually}, \texttt{Until}, \texttt{Next}, \texttt{ForAll} and \texttt{Exists}. If $a_0 a_1 a_2 \dots$ is an input trace, the processor \texttt{Globally} produces an output trace $b_0 b_1 b_2 \dots$ such that $b_i = \bot$ if and only there exists $j \geq i$ such that $b_j = \bot$. In other words, the $i$-th output event is the two-valued verdict of evaluating $\mbox{\bf G}\,\varphi$ on the input trace, starting at the $i$-th event. A similar reasoning is applied to the other operators.

Troolean processors are called \texttt{Always}, \texttt{Sometime}, \texttt{UpTo}, \texttt{After}, \texttt{Every} and \texttt{Some}. Each is associated to the Boolean processor with a similar name. If $a_0 a_1 a_2 \dots$ is an input trace, the processor \texttt{Always} produces an output trace $b_0 b_1 b_2 \dots$ such that $b_i = \bot$ if there exists $j \leq i$ such that $b_j = \bot$, and ``?'' (the ``inconclusive'' value of LTL$_3$) otherwise. In other words, the $i$-th output event is the three-valued verdict of evaluating $\mbox{\bf G}\,\varphi$ on the input trace, after reading $i$ events.

Note that these two semantics are distinct, and that both are necessary in the context of event stream processing. Consider the simple LTL property $a \rightarrow \mbox{\bf F}\,b$. In a monitoring context, one is interested in Troolean operators: the verdict of the monitor should be the partial result of evaluating an expression for the current prefix of the trace. Hence, in the case of the trace $accb$, the output trace should be $???\top$: the monitor comes with a definite verdict after reading the fourth event.

However, one may also be interested in using an LTL expression $\varphi$ as a filter: from the input trace, output only events such that $\varphi$ holds. In such a case, Boolean operators are appropriate. Using the same property and the same trace as above, the expected behaviour is to retain the input events $a$, $c$, and $c$; when $b$ arrives, all four events can be released at once, as the fate of $a$ becomes defined (it has been followed by a $b$), and the expression is true right away on the remaining three events. This behaviour is similar to that of an \emph{enforcement automaton} \cite{DBLP:journals/fmsd/FalconeMFR11}.

First-order quantifiers are of the form $\forall x \in f(e) : \varphi$ and $\exists x \in f(e) : \varphi$. Here, $f$ is an arbitrary function that is evaluated over the current event; the only requirement is that it must return a collection (set, list or array) of values. An instance of the processor $\varphi$ is created for each value $c$ of that collection; for each instance, the processor's context is augmented with a new association $x \mapsto c$. Moreover, $\varphi$ can be any processor; this entails it is possible to perform quantification over virtually anything.

The LTL palette provides its own extensions to \esql{}, shown in Table \ref{tab:grammar-ltl}.

\begin{table}
\begin{grammar}

<processor>              ::= <ltl-operator> ;

<ltl-operator>           ::= <globally> | <eventually> | <next> | <until> 
\alt <ltl-not> | <ltl-and> | <ltl-or> ;

<ltl-not>                ::= NOT ( <processor> ) ;

<ltl-and>                ::= ( <processor> ) AND ( <processor> ) ;

<ltl-or>                 ::= ( <processor> ) OR ( <processor> ) ;

<globally>               ::= G ( <processor> ) ;

<eventually>             ::= F ( <processor> ) ;

<next>                   ::= X ( <processor> ) ;

<until>                  ::= ( <processor> ) U ( <processor> ) ;
\end{grammar}

\vskip 4pt
\textbf{Truth values
}
\vskip 4pt

\begin{grammar}

<constant>               ::= <troolean> ;

<troolean>               ::= $\top$ | $\bot$ | ? 
\end{grammar}

\caption{Grammar for the LTL palette}
\label{tab:grammar-ltl}
\end{table}

\subsubsection{Finite-State Machines}

This palette allows one to define a Moore machine, a special case of finite-state machine where each state is associated to an output symbol. This Moore machine allows its transitions to be guarded by arbitrary functions; hence it can operate on traces of events of any type.

Moreover, transitions can be associated to a list of \texttt{ContextAssignment} objects, meaning that the machine can also query and modify its \texttt{Context} object. Depending on the context object being manipulated, the machine can work as a pushdown automaton, an extended finite-state machine \cite{DBLP:conf/dac/ChengK93}, and multiple variations thereof. Combined with the first-order quantifiers of the LTL-FO$^+$ package, a processing similar to Quantified Event Automata (QEA) \cite{DBLP:conf/fm/BarringerFHRR12} is also possible.

\subsubsection{Other Palettes}

Among other palettes, we mention:

\begin{description}

\item[Gnuplot] This palette allows the conversion of events into input files for the Gnuplot application. For example, an event that is a set of $(x,y)$ coordinates can be transformed into a text file producing a 2D scatterplot of these points. An additional processor can receive these strings of text, call Gnuplot in the background and retrieve its output. The events of the output trace, in this case, are binary strings containing image files.\footnote{An example of BeepBeep's plotting feature can be seen at: \url{https://www.youtube.com/watch?v=XyPweHGVI9Q}}


\item[XML, JSON and CSV] The XML palette provides a processor that converts text events into parsed XML documents. It also contains a \texttt{Function} object that can evaluate an XPath expression on an XML document. Another palette provides the same functionalities for events in the JSON and the CSV formats.

\item[Network packets] This palette allows events to be created from traces of network packets captured from a network interface, by making use of the JNetPcap library. It defines a number of functions to extract data from these captured packets, such as their header fields or payload content. Combined with the FSM and LTL palettes, it can be used to express complex sequential patterns over network packets, and form the basis of an Intrusion Detection System (IDS).

\item[Web Sockets] This palette provides a simple way of serializing event data and transmit it through a web socket.  By splitting a query graph across multiple machines and interposing a web socket at their interfaces, a basic form of distribution of computation can be achieved with virtually no configuration required.
\end{description}


\section{Use Cases Revisited}\label{sec:examples} 

The previous sections have shown that BeepBeep's architecture is very generic: it allows arbitrary event types, free mixing of processors from various palettes, windowing over any processor, and an extensible query language.

However, our experience with members of the industry has revealed that the advantages of such genericity may not be immediately obvious. It seems that some of them are somehow conditioned to think only of problems that can be fitted into the system they already use; the non-standard features available in BeepBeep have been frequently dismissed by consequence of this thinking ``inside the box''.
This is why we feel necessary to demonstrate using numerous and explicit examples the range of different problems that can be tackled thanks to BeepBeep's generic architecture. In this section, we revisit every use case shown in Section \ref{sec:scenarios}, and show how each can be handled using the variety of processors and functions described earlier.

\subsection{Stock Ticker}

\begin{figure}
\centering
\subfloat[]{\includegraphics[scale=0.35]{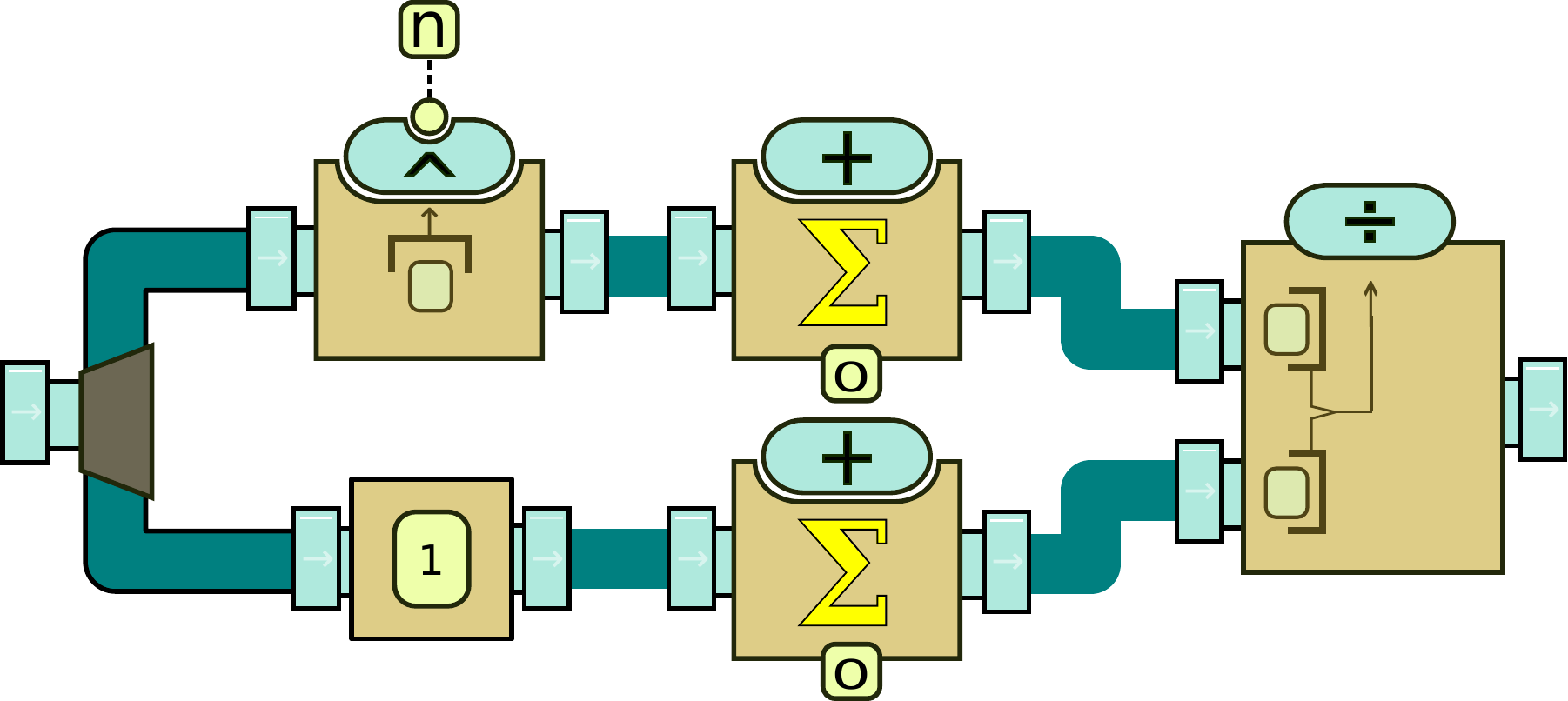}\label{fig:stat-moment}}\\
\subfloat[]{\includegraphics[scale=0.35]{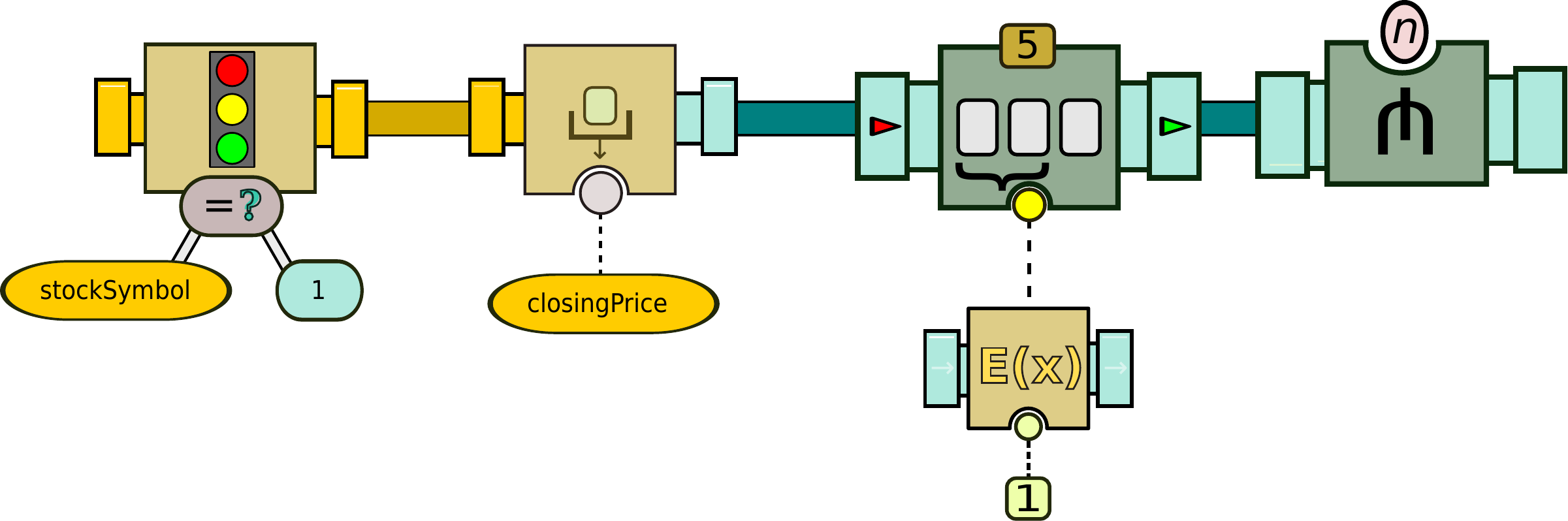}\label{fig:ticker}}
\caption{(a) A chain of function processors for computing the statistical moment of order $n$ on a trace of numerical events; (b) The chain of processors for Query \ref{q:sliding}}
\label{fig:example-ticker}
\end{figure}

Our first example involves processing events from the Stock Ticker scenario. We show how the tumble window of Query \ref{q:sliding} can be written by combining BeepBeep processors. The result is shown in Figure \ref{fig:example-ticker}. In this figure, events flow from the left to the right. First, we calculate the \emph{statistical moment of order $n$} of a set of values, noted $E^n(x)$. As Figure \ref{fig:stat-moment} shows, the input trace is duplicated into two paths. Along the first (top) path, the sequence of numerical values is sent to the \texttt{FunctionProcessor} computing the \nospellcheck{$n$-th} power of each value; these values are then sent to a \texttt{CumulativeProcessor} that calculates the sum of these values. Along the second (bottom) path, values are sent to a \texttt{Mutator} processor that transforms them into the constant 1; these values are then summed into another \texttt{CumulativeProcessor}. The corresponding values are divided by each other, which corresponds to the statistical moment of order $n$ of all numerical values received so far. The average is the case where $n=1$.

Figure \ref{fig:ticker} shows the chain that computes the average of stock symbol 1 over a window of 5 events. Incoming tuples are first filtered according to a function, which fetches the value of the \textsl{stockSymbol} attribute and compares it to the value 1. The processor that is responsible for this filtering is akin to SQL's \textsf{WHERE} processor. The tuples that get through this filtering are then converted into a stream of raw numbers by fetching the value of their \textsl{closingPrice} attribute. The statistical moment of order 1 is then computed over successive windows of width 5, and one out of every five such windows is then allowed to proceed through the last processor, producing the desired hopping window query.

This example introduces colour coding to represent event streams of various types. Orange pipes represent streams of tuples; turquoise pipes contain streams of raw numbers.

\subsection{Healthcare System}

\begin{figure*}
\centering
\includegraphics[scale=0.35]{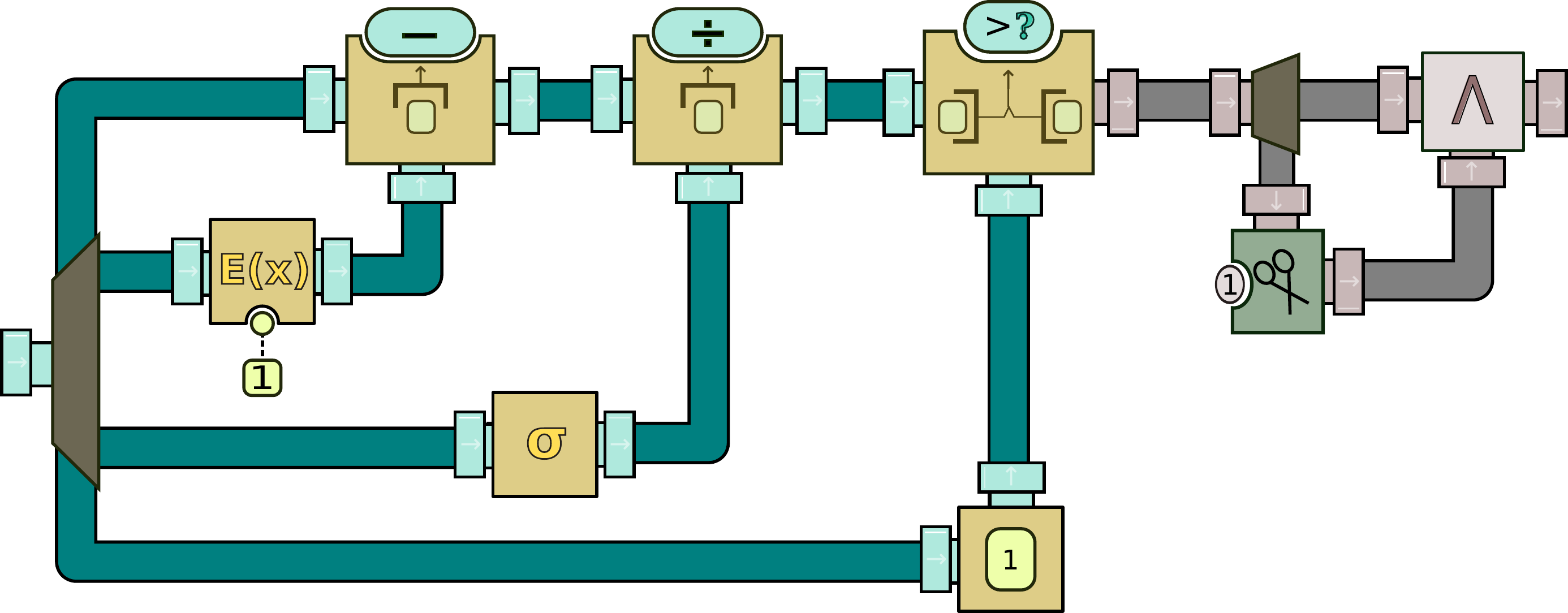}
\caption{The chain of processors for Query \ref{q:trend}}
\label{fig:chain}
\end{figure*}

We show how Query \ref{q:trend} can be computed using chains of function processors. We can reuse the statistical moment processor $E^n(x)$ defined above, and use it for the average ($n=1$) and standard deviation ($n=2$). Equipped with such processors, the desired property can be evaluated by the graph shown in Figure \ref{fig:chain}. The input trace is divided into four copies. The first copy is subtracted by the statistical moment of order 1 of the second copy, corresponding to the distance of a data point to the mean of all data points that have been read so far. This distance is then divided by the standard deviation (computed form the third copy of the trace). A \texttt{FunctionProcessor} then evaluates whether this value is greater than the constant trace with value 1. 

The result is a trace of Boolean values. This trace is itself forked into two copies. One of these copies is sent into a \texttt{Trim} processor, that removes the first event of the input trace; both paths are sent to a processor computing their logical conjunction. Hence, an output event will have the value $\top$ whenever an input value and the next one are both more than two standard deviations from the mean.

Note how this chain of processors involves events of two different types: turquoise pipes carry events consisting of a single numerical value, while grey pipes contain Boolean events.

\subsection{Signal Processing}

\begin{figure*}
\centering
\includegraphics[width=4.5in]{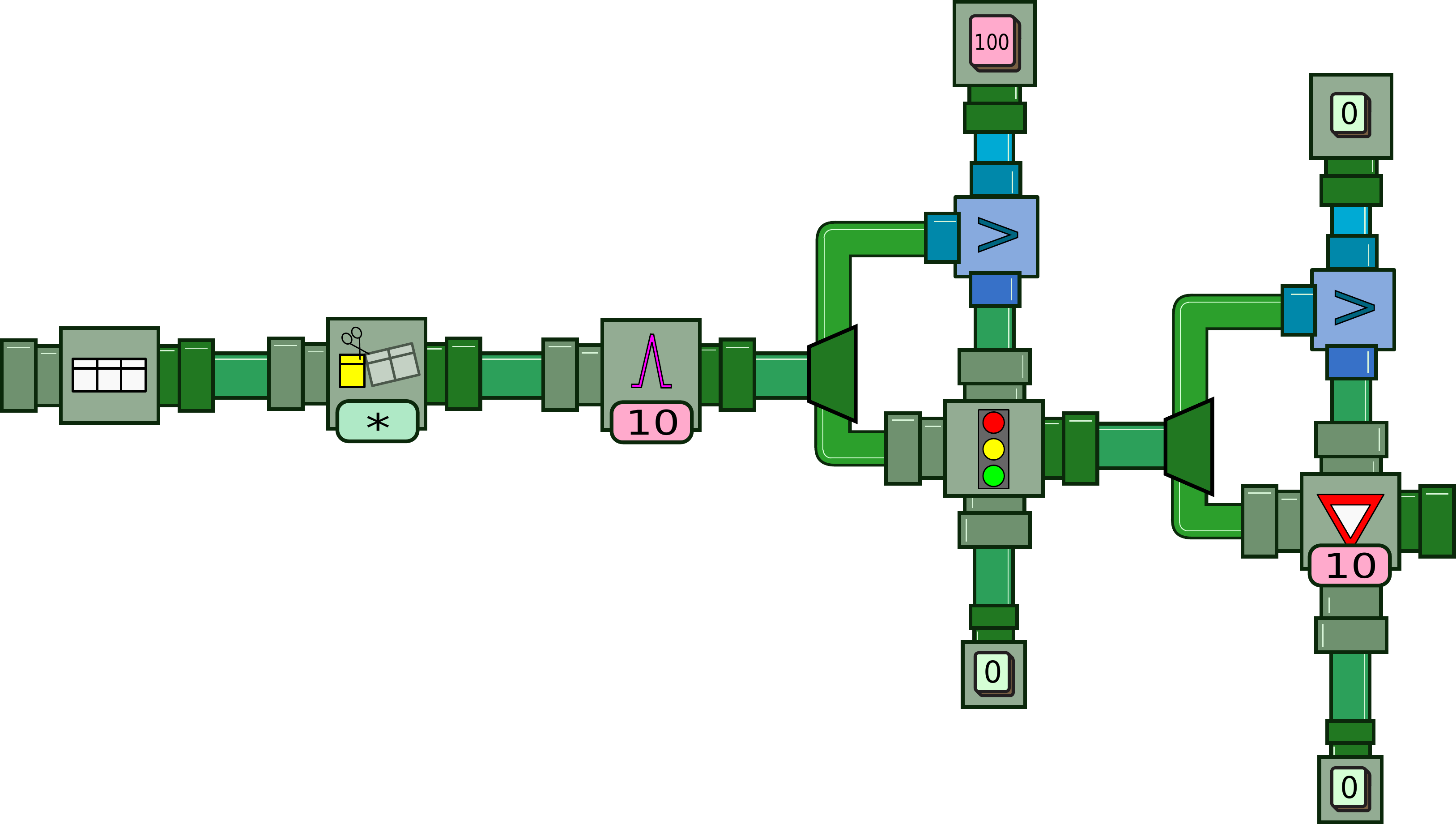}
\caption{The piping of processors for discovering peaks on the original electrical signal. Elements in pink indicate parameters that can be adjusted, changing the behaviour of the pipe.}
\label{fig:piping1}
\end{figure*}

Figure \ref{fig:piping1} describes the chain of basic event processors that are used to discover the peaks on the electrical signal. The signal from the electrical box is sent to a first processor, which transforms raw readings into name-value tuples, one for each time point. Each tuple contains numerical values for various components of the electrical signal; for example, parameter \verb+WL1+ measures the current active power of Phase 1.

The second processor picks one such parameter from the tuple, extracts its value, and discards the rest. The output trace from this processor is therefore a sequence of numbers. This sequence is then fed to the third processor, which detects sudden increases or decreases in a numerical signal. 
%
For each input event, the processor outputs the height of the peak, or the value 0 if this event is not a peak. Since an event needs to be out of the window to determine that it is a peak, the emission of output events is delayed with respect to the consumption of input events.

The next step in the processing takes care of removing some of the noise in the signal. Typical appliances consume at least 100~W and generate a starting peak much higher than that. Therefore, to avoid false positives due to noise, any peak lower than 100~W should be flattened to zero.

In order to do so, the output from the peak detector is replicated in two traces. The first one (top) is sent to a simple comparator, which compares the input value with the constant trace 100, and returns either true or false. This result is the first input of the \emph{dispatcher} processor, represented in Figure \ref{fig:piping1} by traffic lights. The second input of the dispatcher is the output of the peak detector itself, while its third input, in this case, is the constant trace 0. The dispatcher's task is simple: given a triplet of events $(e_1, e_2, e_3)$, (one from each of its inputs), output $e_2$ if $e_1$ is true, and output $e_3$ otherwise. In the present case, this has indeed for effect of replacing all events of the peak detector lower than 100~W to 0.

The resulting trace requires one further cleanup task. Again due to the nature of the electrical signal, two successive peak events may sometimes be reported for the same sudden increase. The last processor takes care of keeping only the first one. 
This \emph{yield} processor behaves like the dispatcher, but with the additional guarantee that the second input will be selected at most once in every $n$ successive events. In the present context, this has for effect of eliminating ``ghost'' peaks in the signal.

Given a feed from an electrical signal, this complete chain of processors produces an output trace of numerical events; most of them should be null, and a few others should indicate the occurrence of an abrupt increase or decrease in the values of the input signal, along with the magnitude of that change. Moreover, the position of these events, relative to the original signal, also indicates the exact moment this change was detected. As an example, Figure \ref{fig:blender} shows the realtime value of three components of the electrical signal, to which the output of the peak detector was superimposed. One can see that the detector behaves as we want, reporting exactly two changes of the appropriate magnitude at the right time.

\begin{figure}
\centering
\includegraphics[width=1.5in]{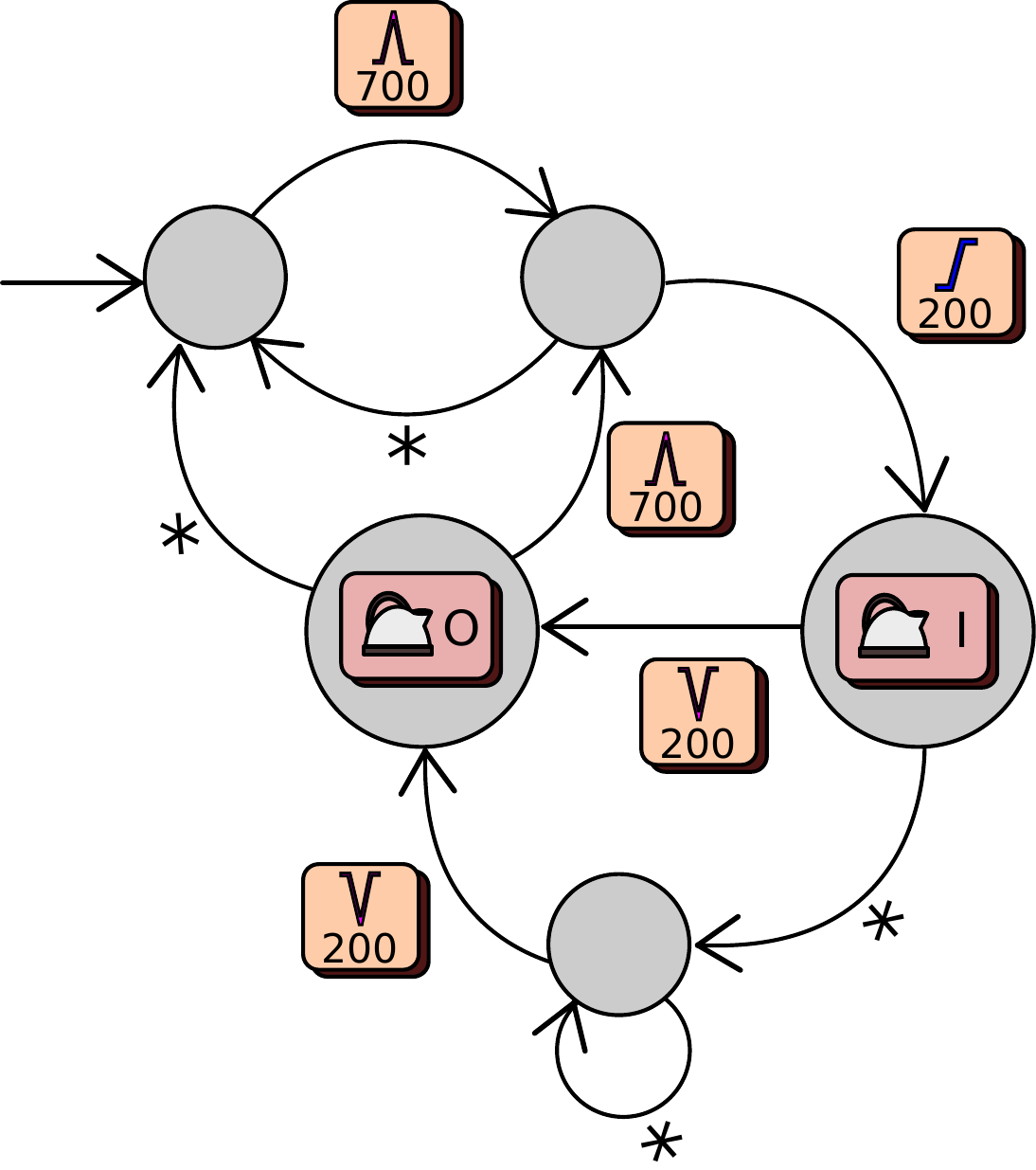}
\caption{The Moore machine for detecting on/off events for a single appliance. 
}
\label{fig:moore}
\end{figure}

The second step is to lift peak and drop events to a yet higher level of abstraction, and to report actual appliances being turned on and off. This is best formalized through the use of a Moore machine, shown in Figure \ref{fig:moore}. From the initial state, the event ``appliance on'' (\textsf{I}) is output only if a peak and a plateau event of the appropriate magnitude are received in immediate succession. At this point, the event ``appliance off'' (\textsf{O}) is emitted only if a drop of the appropriate magnitude is received. All other input events processed by the machine result in no output event being produced. Apart from the actual numerical values, this Moore machine is identical for all appliances.

Notice how the abstraction performed in Step 1 simplifies the problem in Step 2 to the definition of a simple, five-state automaton.

\subsection{Online Auction System}

Our next example is a modified version of the auction system. Rather than simply checking that the sequencing of events for each item is followed, we will take advantage of BeepBeep's flexibility to compute a non-Boolean query: the average number of days since the start of the auction, for all items whose auction is still open and in a valid state.

\begin{figure*}
\centering
\includegraphics[width=4.75in]{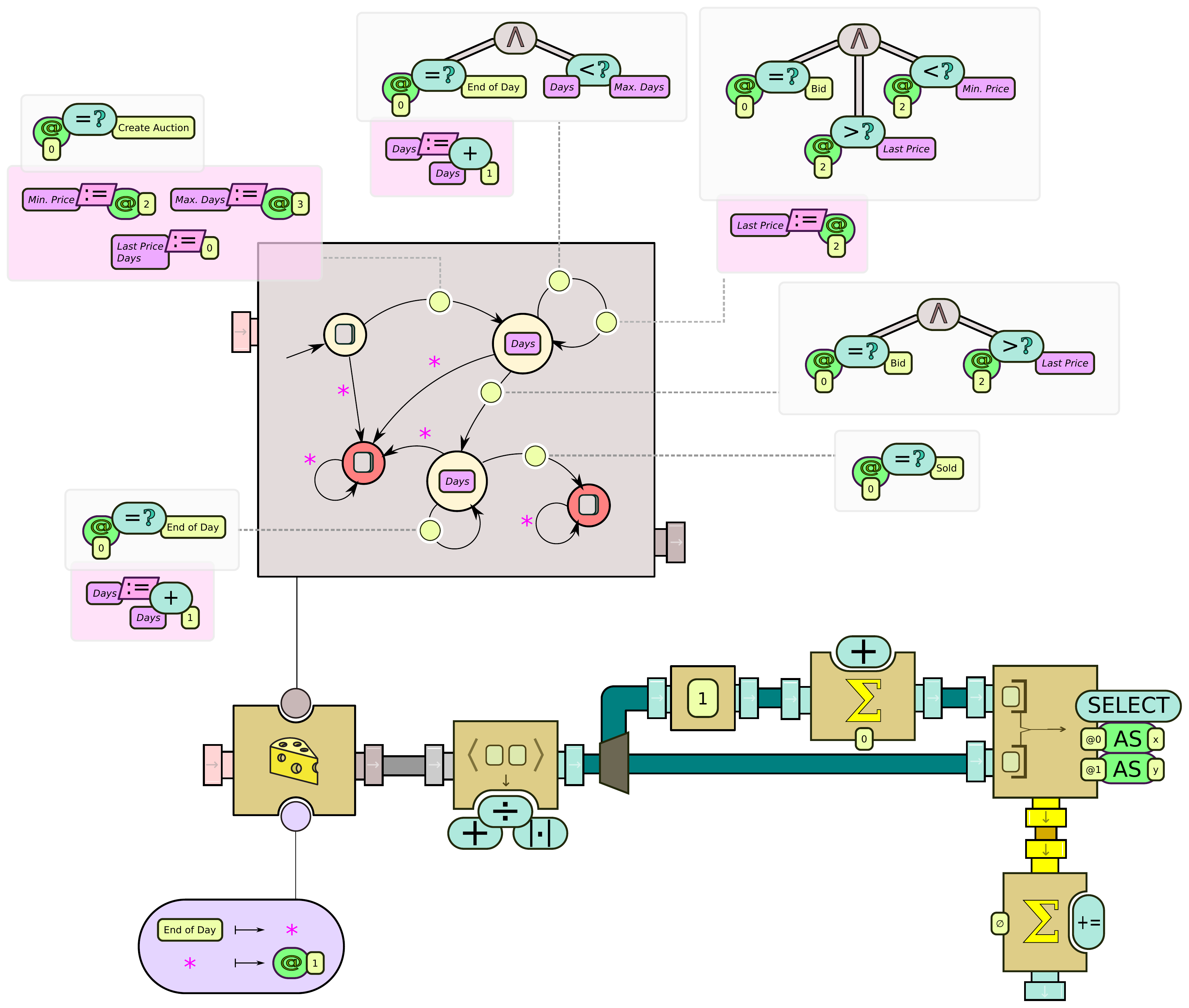}
\caption{Processor graph for the ``Auction Bidding'' query}
\label{fig:auction}
\end{figure*}

The processor graph is shown in Figure \ref{fig:auction}. It starts at the bottom left, with a \texttt{Slicer} processor that takes as input tuples of values. The slicing function is defined in the oval: if the event is \nospellcheck{endOfDay}, it must be sent to all slices; otherwise, the slice is identified by the element at position 1 in the tuple (this corresponds to the name of the item in all other events). For each slice, an instance of a Moore machine will be created, as shown in the top part of the graph.

Each transition in this Moore machine contains two parts: the top part is a function to evaluate on the input event, to decide whether the transition should fire. The bottom part contains instructions on how to modify the \texttt{Context} object of the processor. For example, the top left transition fires if the first element of the event is the string ``Create Auction''. If so, the transition is taken, and the processor's context is updated with the associations \textsl{Last Price} $\mapsto$ 0, \textsl{Days} $\mapsto$ 0. The values of \textsl{Min.\ Price} and \textsl{Max.\ Days} are set with the content of the third and fourth element of the tuple, respectively. The remaining transitions take care of updating the minimum price and the number of days elapsed according to the events received.

Each state of the Moore machine is associated with an output value. For three of these states, the value to output is the empty event, meaning that no output should be produced. For the remaining two states, the value to output is the current content of \textsl{Days}, as defined in the processor's context.

According to the semantics of the \texttt{Slicer}, each output event will consist of a set, formed by the last output of every instance of the Moore machine. Thus, this set will contain the number of elapsed days of all items whose auction is currently open (the Moore machine for the other items outputs no number). This set is then passed to a function processor, which computes the average of its values (sum divided by cardinality).

As a bonus, we show how to plot a graph of the evolution of this average over time. We fork the previous output; one branch of this fork goes into a \texttt{Mutator}, which turns the set into the value 1; this stream of 1s is then sent to a cumulative function processor $\Sigma_+^0$ that computes their sum. Both this and the second branch of the fork are fed into a function processor, that creates a named tuple where $x$ is set to the value of the first input, and $y$ is set to the value of the second input. The result is a tuple where $x$ is the number of input events, and $y$ is the average computed earlier. These tuples are then accumulated into a set with the means of another cumulative function processor, this time performing the set addition operation. The end result is a stream of sets of $(x,y)$ pairs, which could then be sent to a \texttt{Scatterplot} processor to be plotted  with the help of Gnuplot.

One can see again that processors of multiple palettes are involved, and events of various types are mixed: predicates (pink), sets of numbers (grey), numbers (turquoise), and named tuples (yellow).

\subsection{Runtime Verification}

The next example is taken from our previous work on the monitoring of video games \cite{nous-acm-cie}. 
The property we wish to check is that every time a Walker encounters a Blocker, it must turn around and start walking in the opposite direction. An encounter occurs whenever the $(x,y)$ coordinates of the Walker come within 6 pixels horizontally, and 10 pixels vertically, of some Blocker. When this happens, the Walker may continue walking towards the Blocker for a few more events, but eventually turns around and starts walking away. 

\begin{figure}
\centering
\includegraphics[scale=0.35]{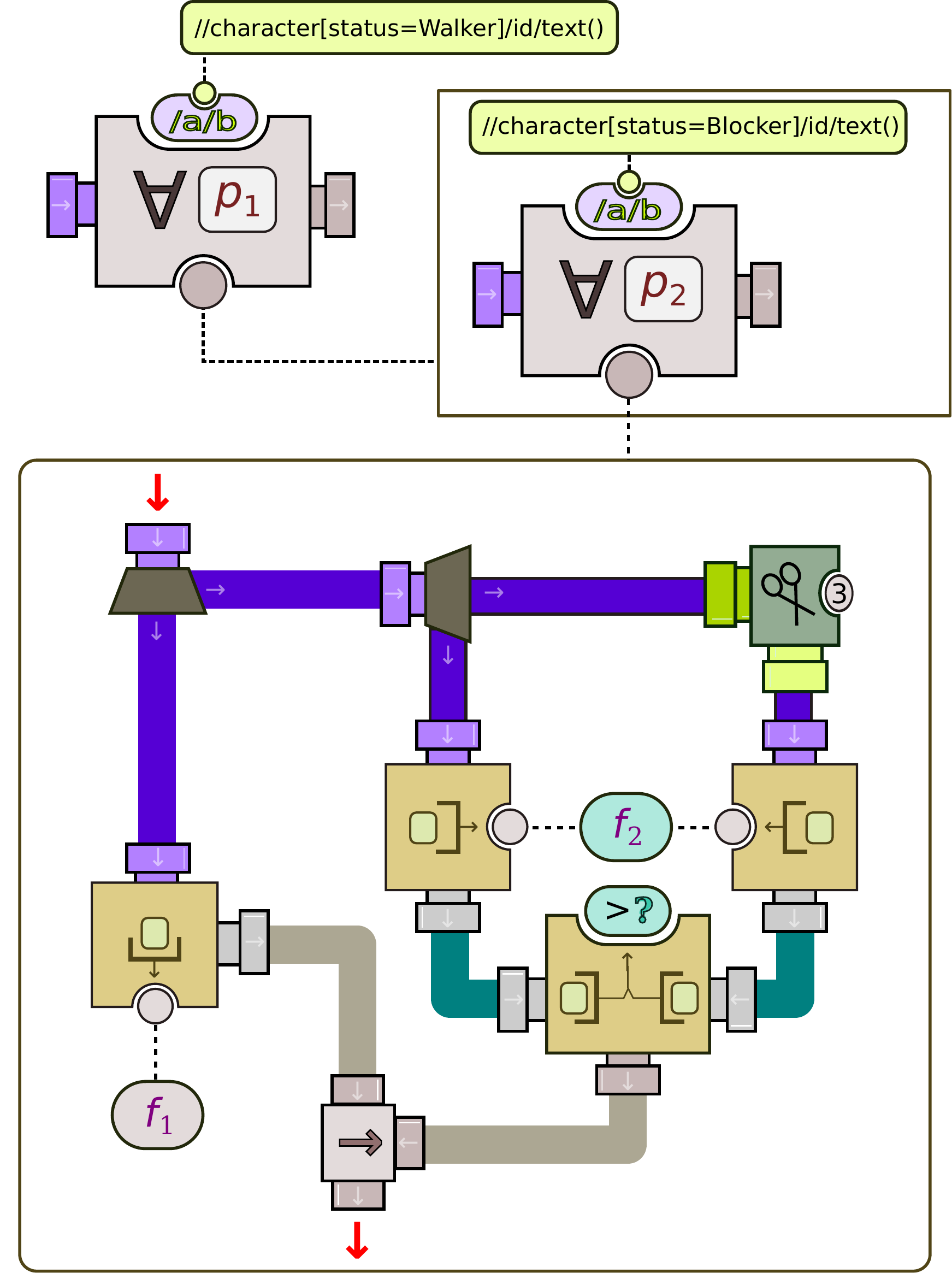}
\caption{Processor graph for property ``Turn Around''}
\label{fig:turnaround}
\end{figure}

Figure \ref{fig:turnaround} shows the processor graph that verifies this. Here, blue pipes carry XML events, turquoise pipes carry events that are scalar numbers, and grey pipes contain Boolean events. The XML trace is first sent into a universal quantifier. The domain function, represented by the oval at the top, is the evaluation of the XPath expression \verb+//character[status=WALKER]/id/text()+ on the current event; this fetches the value of attribute \verb+id+ of all characters whose status is \verb+WALKER+. For every such value $c$, a new instance of the underlying processor will be created, and the context of this processor will be augmented with the association $p_1 \mapsto c$. The underlying processor, in this case, is yet another quantifier. This one fetches the ID of every \verb+BLOCKER+, and for each such value $c'$, creates one instance of the underlying processor and adds to its context the association $p_2 \mapsto c'$.

\begin{figure}
\centering
\subfloat[$f_1$]{\includegraphics[scale=0.6]{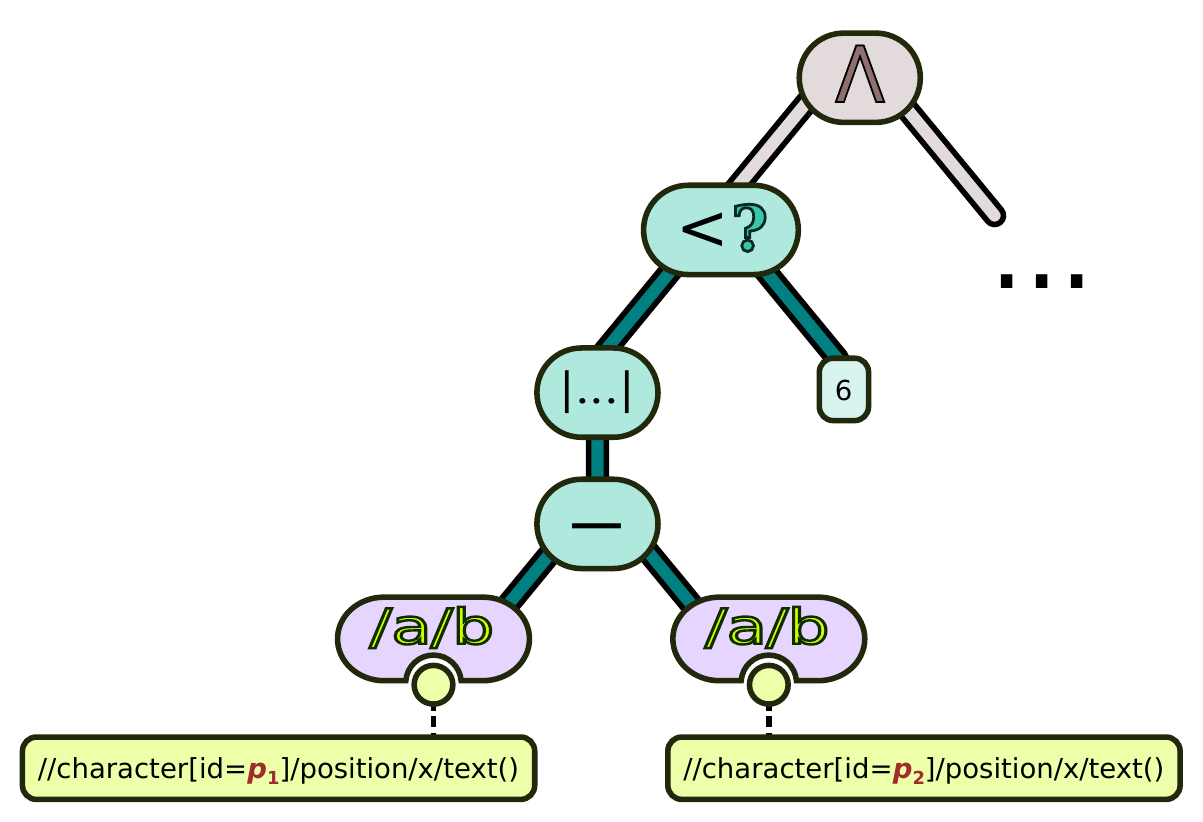}\label{fig:pingus-f1}}\\
\subfloat[$f_2$]{\includegraphics[scale=0.6]{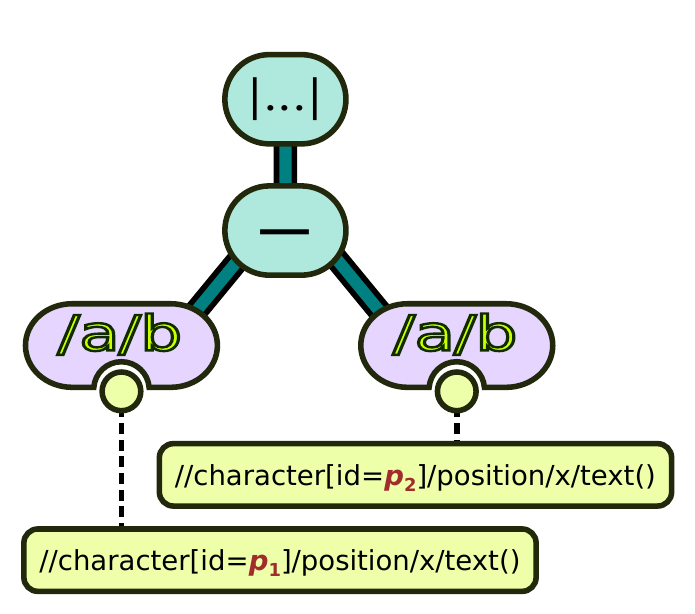}\label{fig:pingus-f2}}
\caption{The two functions used in the processor graph of Figure \ref{fig:turnaround}}
\end{figure}

The underlying processor is the graph enclosed in a large box at the bottom. It creates two copies of the input trace. The first goes to the input of a function processor evaluating function $f_1$ (not shown), on each event. This function evaluates $|x_1 - x_2| < 6 \wedge |y_1 - y_2| < 10$, where $x_i$ and $y_i$ are the coordinates of the Pingu with ID $p_i$. Function $f_1$ is the \texttt{FunctionTree} described in Figure \ref{fig:pingus-f1}. Its left branch fetches the $x$ position of characters with ID $p_1$ and $p_2$, and checks whether their absolute difference is greater than 6. Its right branch (not shown) does a similar comparison with the $y$ position of both characters. Note in this case how the XPath expression to evaluate refers to elements of the processor's context ($p_1$ and $p_2$). The resulting function returns a Boolean value, which is true whenever character $p_1$ collides with $p_2$. 

The second copy of the input trace is duplicated one more time. The first is sent to a function processor evaluating $f_2$, which computes the horizontal distance between $p_1$ and $p_2$. The second is sent to the \texttt{Trim} processor, which is instructed to remove the first three events it receives and lets the others through. The resulting trace is also sent into a function processor evaluating $f_2$. Finally, the two traces are sent as the input of a function processor evaluating the condition $>$. Therefore, this processor checks whether the horizontal distance between $p_1$ and $p_2$ in the current event is smaller than the same distance three events later. If this is true, then $p_1$ moved away from $p_2$ during that interval.

The last step is to evaluate the overall expression. The ``collides'' Boolean trace is combined with the ``moves away'' Boolean trace in the \texttt{Implies} processor. For a given event $e$, the output of this processor will be $\top$ when, if $p_1$ and $p_2$ collide in $e$, then $p_1$ will have moved away from $p_2$ three events later.


\section{Experimental Evaluation}\label{sec:experiments} 

As discussed earlier, BeepBeep was partly designed in reaction to the complexity and heaviness of existing event processing systems; to this end, versatility and simplicity were the primary goals informing all of our design decisions. Therefore, benchmarking BeepBeep against competing CEP tools somehow misses the point: performance, although desirable, was never sought at the price of readable queries or extensibility. Moreover, research papers reporting the use of BeepBeep in various real-world situations (web service testing \cite{DBLP:journals/tsc/HalleV12}, electric load monitoring \cite{DBLP:conf/aaai/HalleGB16}, video game debugging \cite{nous-acm-cie}) have already shown it is ``fast enough'' for these use cases. Nevertheless, we felt fitting to conduct an experimental comparison for two reasons. 

First, few works provide an experimental comparison of CEP tools on the same queries and input data. The most recent and thorough effort of that sort is the RIoTBench platform \cite{riot}, which has measured throughput and resource consumption of Apache Storm on the Microsoft Azure public Cloud. 
However, the benchmark focuses on distributed event stream processing and includes a single system in its analysis. The older BiCEP system seemed to share a similar goal \cite{DBLP:conf/dagstuhl/Bizarro07}; unfortunately, the link provided in BiCEP's paper points to an empty web site, so its implementation does not appear to be extant at the time of this writing. One of the papers describing Siddhi does compare it to Esper on three queries \cite{DBLP:conf/sc/SuhothayanGNCPN11} (filter, sliding window, pattern), and another paper compares Esper's throughput with T-REX \cite{DBLP:journals/jss/CugolaM12} on four. This section is by no means a comprehensive study, but it does provide some empirical substance for the relative merits of each evaluated tool. To the best of our knowledge, the modest empirical review presented in this section is the first published account of a comparison of more than two CEP engines on the same queries.

Second, based on actual discussions and presentations we had with members of both industry and academia, BeepBeep's features have frequently been dismissed on the grounds that ``surely, this can also be done with software X''. We shall go to some lengths to provide detailed evidence to the contrary, in some of the use cases we exposed earlier.

\subsection{Experimental Setup}

Our benchmark focuses on \emph{single-machine} event stream processing systems similar to BeepBeep. The query engines included in our benchmark are:

\begin{itemize}
\item SASE (cf.\ Section \ref{subsubsec:sase}). Our benchmark includes version 1.0 of the software. Its documentation states that some advanced features such as processing streams with imprecise timestamps are not included in this release. However, none of our use cases require these features.
\item Siddhi (cf.\ Section \ref{subsubsec:siddhi}). Our benchmark includes version 3.0.3 of the software.
\item Esper (cf.\ Section \ref{subsubsec:esper}). Our benchmark includes version 5.3.0 of the software.
\item MySQL \footnote{\url{http://mysql.com}}. Our benchmark includes version 5.5 of the software. Although MySQL is not an ESP system, the ``this could be done with a database'' argument was raised often enough to warrant its inclusion in our study.
\end{itemize}

All these tools were used with their default settings. Although the latest version of Cayuga (dating from 2009) is publicly available, some libraries required to build it are unavailable as of 2017. Our attempts to obtain help from the authors have unfortunately remained unanswered, which forced us to exclude it from the benchmark. We also purposefully excluded cloud platforms such as Microsoft Azure, Apache Spark and VoltDB. Their use of multiple machines, and the heavy setup they require before being functional,\footnote{SQLstream alone requires a whopping 1 \emph{gigabyte} of disk space for its basic installation. This should be contrasted with Esper, Siddhi and BeepBeep, which are stand-alone bundles of at most a few megabytes.} does not place them on an equal footing with the other systems we consider. 
We also remind that our goal is not to claim that BeepBeep is the fastest CEP software around, but that reasonable performance can be expected for the ease of use it offers. Table \ref{tab:size} shows the relative footprint of each tool, expressed as the cumulative size of the program and all its library dependencies.

\begin{table}
\centering
\usebox{\codebaseSize}
\caption{Codebase size for each tool included in our experiments}
\label{tab:size}
\end{table}


The experiments were implemented using the LabPal testing framework\footnote{\url{https://liflab.github.io/labpal}}. The principle behind LabPal is that all the necessary code, libraries and input data should be bundled within a single self-contained executable file, such that anyone can download and independently reproduce the experiments. The detailed list of all the queries and input streams included in our benchmark cannot be shown in this paper due to lack of space; however all input files are available from our downloadable lab instance\footnote{\url{https://datahub.io/dataset/beepbeep-3-benchmark}}. All the experiments were run on a \machinestring{}, inside a Java 8 VM with \jvmram{} MB of memory. All experiments were given a timeout of \timeout{} seconds.

\subsection{Relative Expressiveness}

Our original intent was to take each of the twelve queries described in Section \ref{sec:scenarios}, and to compare the behaviour of each tool on these queries. Except for very simple queries, attempting to write the same computation in languages with different and sometimes incompatible syntax and semantics is a non-trivial and generally imperfect exercise \cite{DBLP:conf/rv/MradAHB12}. Our plan was cut short by the limitations imposed by other tools, either on the allowed event types or the query language they offer. Our experiments could humorously be summarized as attempts at fitting a square peg in a round hole.


%

\begin{table}
\centering
\scalebox{0.7}{
\begin{tabular}{|c||c|c|c|c|c|c|}
\hline
\textbf{Query} & \textbf{Esper}& \textbf{SASE}& \textbf{Siddhi}& \textbf{MySQL}& \textbf{BeepBeep}\\
\hline\hline
1 & \textcolor{green}{\cmark} & \textcolor{green}{\cmark} & \textcolor{green}{\cmark} & \textcolor{green}{\cmark} & \textcolor{green}{\cmark}\\
\hline
2 & \textcolor{green}{\cmark} & \textcolor{green}{\cmark} & \textcolor{green}{\cmark} & \textcolor{green}{\cmark} & \textcolor{green}{\cmark}\\
\hline
3 & \textcolor{green}{\cmark} & \textcolor{red}{\xmark} & \textcolor{green}{\cmark} & \textcolor{green}{\cmark} & \textcolor{green}{\cmark}\\
\hline
4 & \textcolor{green}{\cmark} & \rpm & \textcolor{green}{\cmark} & \textcolor{green}{\cmark} & \textcolor{green}{\cmark}\\
\hline
5 & \rpm & \textcolor{red}{\xmark} & \rpm & \textcolor{green}{\cmark} & \textcolor{green}{\cmark}\\
\hline
6 & \rpm & \textcolor{red}{\xmark} & \rpm & \textcolor{green}{\cmark} & \textcolor{green}{\cmark}\\
\hline
7 & \textcolor{green}{\cmark} & \textcolor{green}{\cmark} & \textcolor{green}{\cmark} & \textcolor{green}{\cmark} & \textcolor{green}{\cmark}\\
\hline
8 & \textcolor{red}{\xmark} & \textcolor{red}{\xmark} & \textcolor{red}{\xmark} & \textcolor{green}{\cmark} & \textcolor{green}{\cmark}\\
\hline
9 & \textcolor{red}{\xmark} & \textcolor{red}{\xmark} & \textcolor{red}{\xmark} & \rpm & \textcolor{green}{\cmark}\\
\hline
10 & \rpm & \textcolor{red}{\xmark} & \rpm & \textcolor{green}{\cmark} & \textcolor{green}{\cmark}\\
\hline
11 & \textcolor{red}{\xmark} & \textcolor{red}{\xmark} & \textcolor{red}{\xmark} & \rpm & \textcolor{green}{\cmark}\\
\hline
12 & \textcolor{red}{\xmark} & \textcolor{red}{\xmark} & \textcolor{red}{\xmark} & \rpm & \textcolor{green}{\cmark}\\
\hline
\end{tabular}

}
\caption{Tool support for each of the queries introduced in Section \ref{sec:scenarios}.}
\label{tab:support}
\end{table}

Table \ref{tab:support} gives a summary of the support for each query by the tools included in our study. The checkmark (\cmark) symbol indicates that the system can compute the exact result of the query. The ``\rpm'' symbol indicates that limitations in the tool would force us to evaluate a \emph{simplified} version of the query. This is the case, for example, when XML events have to be flattened into fixed-size tuples, or when a query language imposes that the distance between two events in a pattern be bounded by a finite value. Finally, the cross symbol (\xmark) indicates that there is no reasonable way to handle this query with the tool. We had to come to this verdict in cases where the problem would only be solvable in extremely convoluted ways: for example, computing the two-dimensional heat map of Query \ref{q:heatmap} (whose size is unknown in advance) using only tuples with a fixed schema.

In the following, we give further details on the way each query was handled (or not) by each tool.

\subsubsection{Stock Ticker} 

This use case is closest to ``traditional'' CEP problems and presents the least issues in terms of tool support. However, SASE cannot handle some of the ticker queries, due to the fact that its implementation lacks support for aggregate functions. 

\subsubsection{Healthcare Records} This scenario presents more problems. Since some of the tools impose that events be tuples, HL7 events must be replaced by tuples with dummy attribute names $a_1$, \dots, $a_n$. In each event, attribute $a_i$ has for value the $i$-th field of the corresponding HL7 message. However, this brings an additional problem, as the $i$-th field of each message may not be of the same type. 

Moreover, even with such a manual doctoring of the inputs, the expression of these properties is still problematic. Aggregation functions in Siddhi and Esper are over fixed windows. In Siddhi, one can easily compute the standard deviation of a field over multiple events, as well as its mean. However, what is expected in Query \ref{q:trend} is the ratio of these two quantities; alas, an expression like \verb+(x - AVG(x)) / STDEV(x)+ (and variations thereof) are all rejected as a syntax error. A workaround would be to generate a stream computing \verb+(x - AVG(x))+, and another computing \verb+STDEV(x)+; however, matching events from these two streams cannot be merged, apart from computing their join; this, in turn, requires a fake counter to be added to each event to be used as the join attribute.

\subsubsection{Online Auction System}

This use case presents the same problem as the previous one, since events in the trace do not have the same attributes and values. Events should therefore be simplified so that each has a \textsl{name} attribute, and three other ``dummy'' attributes whose meaning differs according to \textsl{name}. Item names have also been turned into numbers so that all these fields can be of the Integer type.

After these simplifications, Query \ref{q:price-increase} can be accommodated in Siddhi, Esper and SASE using their pattern syntax. However, Query \ref{q:forbidden-bid} correlates a value inside an event (the \textsl{Duration} of an auction) to the number of \nospellcheck{endOfDay} events that may be seen before bids of an item become forbidden. Unfortunately, we found no sensible way of expressing this fact using the query language of any other tool. We did manage, however, to write an SQL query achieving this result.

\subsubsection{Electric Load Monitoring} 

All systems but BeepBeep are discarded, as they lack the peak detection algorithms necessary to perform the first level of abstraction of the original input trace. It goes without saying that tuple query languages are very ill-suited for this task; the best one can do using SiddhiQL or EQL queries would be to define a pattern of $n$ successive events, and detect large differences between the first and the $n$-th ---which is a very imprecise characterization of a peak. Note also that it does not suffice to watch the min/max difference over a sliding window, as the same peak may be detected more than once (or not at all in the case of a hopping window query, if the peak occurs across the boundary of two successive windows).

However, even assuming a correct peak input stream, CEP query languages still have trouble expressing the Moore machine required to produce the final output trace from the trace of peaks. The best one can do is write two queries, one producing a ``Toaster On'' event when \verb+x+ is greater than $800$, and a ``Toaster Off'' when \verb+x+ is smaller than $-800$ ---and again multiplex these two streams to produce the desired output. At this point, the original problem has gone through a handful of simplifications and approximations, and some features (such as multiplexing) are still missing to actually run it using a CEP engine. We hope the reader agrees with our conclusion that the load monitoring problem cannot be reasonably solved using other tools.

\subsubsection{Runtime Verification}

At the risk of being tedious, we also show the issues faced when attempting to write the video game queries in SiddhiQL or EQL. First, nested XML events must be converted into a sequence of tuples, one for each Pingu inside the event. The same artificial timestamp is appended to these events so that they can remain grouped. Even though Esper supports nested events, its query language lacks a ``for all'' construct, so the tuple conversion is also necessary. Then, matching a Blocker and a Walker within the same event becomes problematic, as the unrolling of an XML event may sometimes put the Blocker before the Walker, or after; the pattern query has to consider the two possible orderings. 

\subsubsection{Synthetic Traces}

Given that many of the use cases on which BeepBeep was showcased are handled with difficulty (if at all) by other tools, we reversed our experimental evaluation, and instead took BeepBeep to ``their'' field. We focused our experimental evaluation on simple synthetic traces of tuples made of random strings and numerical values; the traces considered contain \longestTrace{} events. We devised a number of ``generic'' queries on these traces, intended to probe the basic query types described in Section \ref{sec:scenarios}. The queries we included are shown in Table \ref{tab:syn-queries}.

\begin{table}
\begin{enumerate}
\item[S1] Passthrough: select all attributes from all events from the input stream
\item[S2] Filter: select all events where \textsl{stockSymbol} = 1
\item[S3] Sliding window 1: compute the average of \textsl{closingPrice} for a sliding window of 5 successive events where \textsl{stockSymbol} = 1
\item[S4] Sliding window 2: select all events whose price is greater than the price average of the past 50 events
\item[S5] Pattern 1: select all events such as the price for stock symbol "2" is less than 2, and such that the next event with stock symbol "1" has the same price
\item[S6] Pattern 2: select all events such as the price for stock symbol "2" is less than 2, that are followed by any number of events where closing price for stock symbol "1" is greater than for stock symbol "2", until the price for stock symbol "2" becomes less than 2 again
\item[S7] Window pattern: find all events for stock 1 such that the price average of stock 1 for the last 50 events is above that of stock 2 for at least 4 out of 5 successive events
\end{enumerate}
\caption{The synthetic queries considered in our benchmark.}
\label{tab:syn-queries}
\end{table}

Even then, the last query is problematic, as no query language (except BeepBeep's) provides an easy way to count events in a sliding window that satisfy a condition. One can select events that satisfy a condition and then create a window on the resulting stream, but this does not yield the desired result (no event is output if the condition is not satisfied). In the present case, since our condition is simply $x > y$, a workaround is to evaluate
\[
\max\left(0, \frac{x-y}{|x-y|}\right)
\]
which returns 1 if $x > y$ and 0 otherwise, and then to sum these values over a window. However, this trick hardly generalizes to more complex conditions.

\subsection{Measured Throughput}

Each tool was run on its own version of each query on randomly generated traces as described above. Since MySQL is not an event processing engine, it cannot operate in a streaming fashion. We converted the input trace into one large \textsf{CREATE TABLE} statement, and then ran an SQL query on that table; we interpret as an output ``stream'' the table that results from that query (with each row of the table being assimilated to an event). Note that in the results given below, running time includes the execution of \textsf{CREATE TABLE}. This is to establish a fair comparison, as all other systems start their processing from a stream that is completely unknown in advance.

\begin{figure}
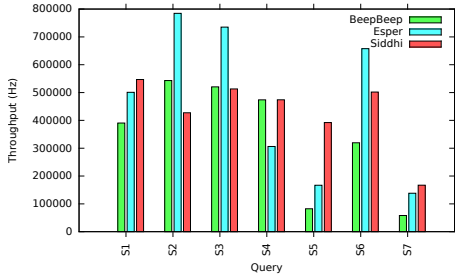

\centering
\scalebox{0.7}{
\usebox{\plotthroughputByTool}
}
\caption{Running time of each tool on each of the queries.}
\label{fig:runtime}
\end{figure}

We measured the elapsed time taken by each system to process the queries, and deduced from this time the throughput, measured in Hz (number of input events consumed per second). The results are plotted in the histogram of Figure \ref{fig:runtime}.

Two of the contenders had to be disqualified from the onset. The first is SASE, which for traces of that size, crashes by running out of heap space. The second is MySQL, which exceeded the timeout on all queries. Figure \ref{fig:sql-long} gives part of the explanation. It shows one of the event stream queries (S6) written as an SQL expression. To simplify the query, a view on the original trace (i.e.\ table) is first created; otherwise, the corresponding expression would have to be repeated three times in the following \textsf{SELECT} statement. Note that this view already involves a self-join. The query itself is far from a one-liner: it must create an intricate condition on timestamps for three copies of the original trace in order to correctly express the sequential pattern to be observed. While the query does compute the correct result, the absence of support for even simple sequential patterns makes it so complex that its evaluation is not practically feasible. A similar argument had already been made in \cite{DBLP:journals/fmsd/BasinKMZ15}. For this reason, our plots do not include MySQL in the experimental results.

\newsavebox{\sqllongbox}
\begin{lrbox}{\sqllongbox}
\begin{lstlisting}[language=sql]
DROP TABLE IF EXISTS ThePrices;

CREATE VIEW ThePrices AS
 SELECT T1.closingPrice AS p1, T2.closingPrice AS p2, T1.timestamp AS timestamp
 FROM stocks AS t1, stocks AS t2
 WHERE t1.timestamp = t2.timestamp;

SELECT COUNT(*) FROM
  (SELECT timestamp FROM ThePrices WHERE p2 < 2) AS T0, (
  SELECT T1.timestamp - 1 AS timestamp 
    FROM ThePrices AS T1,
    (SELECT MAX(TA.timestamp) AS n1, TB.timestamp AS n2
          FROM (SELECT timestamp FROM ThePrices WHERE p1 <= p2) AS TA
            JOIN (SELECT timestamp FROM ThePrices WHERE p2 < 2) AS TB
          WHERE TA.timestamp < TB.timestamp GROUP BY TB.timestamp
      ) AS T2
      WHERE T1.timestamp > T2.n1 AND T1.timestamp <= T2.n2) AS T3
    WHERE T0.timestamp = T3.timestamp
\end{lstlisting}
\end{lrbox}
\newsavebox{\esqllongbox}
\begin{lrbox}{\esqllongbox}
\begin{minipage}{6in}
\begin{verbatim}
WHEN @P IS A PROCESSOR:
  MY PATTERN IN @P IS THE PROCESSOR
  (SELECT closingPrice > 1 FROM @P)
  AND (
    EVENTUALLY (SELECT closingPrice > 2 AND stockSymbol = 1 FROM @P)
  ).
\end{verbatim}
\end{minipage}
\end{lrbox}
\begin{figure*}
\centering
\subfloat[MySQL]{
\usebox{\sqllongbox}
}
\\
\subfloat[\esql]{
\usebox{\esqllongbox}
}
\caption{Query S6: (a) the query in SQL; (b) the same query in eSQL.}
\label{fig:sql-long}
\end{figure*}

A second observation is that, for most of the queries, Siddhi and Esper provide comparable throughput, with BeepBeep having on average half their throughput on our sample of queries. We are actually happily surprised with these results, and expected a much larger difference between commercial-grade CEP systems and our proposed implementation. For example, in BeepBeep, computing the average of a sequence of values is not done by a built-in primitive function, as is the case with Siddhi and Esper; rather, Figure \ref{fig:stat-moment} shows it is a user-defined combination of basic processors, involving a fork, two cumulative functions and a division processor. This clearly impacts performance, but as discussed above, improves genericity: computing the statistical moment of order 3 can be computed by simply changing the value of $n$ (which has no impact on running time), while other tools no longer provide an efficient built-in primitive.

Similarly, the computation of a window in BeepBeep is done in a very naive way: one instance of the processor given as an argument is created for each window, and the contents of the window are ``replayed'' to that processor. This is clearly sub-optimal when the function to compute over the window is simple and known in advance. For example, an average can easily be updated in constant time by subtracting the leftmost value leaving the window and adding the rightmost value entering it. However, as we have already discussed, BeepBeep's windows are completely independent from the processor to evaluate, which can be much more complex than the built-in, stateless arithmetical functions provided by other systems.

As we said earlier, these empirical results are not intended to be a thorough benchmark of multiple CEP systems. The observations made in this section, however, are sufficient to support two claims:

\begin{enumerate}
\item Some use cases exposed in Section \ref{sec:scenarios} are difficult (if not impossible) to model using the query language of some commercial-grade CEP tools or RDBMS.
\item For typical window and pattern queries supported by CEP tools, BeepBeep has a lower, but still reasonable throughput.
\end{enumerate}


\section{Conclusion}\label{sec:conclusion} 

In this paper, we have presented a short introduction to the field of Complex Event Processing, and highlighted the differences between classical CEP problems and properties typically considered in Runtime Verification. In particular, we have seen how CEP problems involve intricate computations and transformations over data fields inside events, while runtime monitors are generally more powerful for evaluating properties that relate to the sequencing of events. Moreover, we have presented various use cases taken from existing literature, in which the traditional conception of CEP is extended by new types of events and queries.

A review of existing solutions has highlighted many of their useful features, but also numerous shortcomings: complex usage, rigid event structure, limited expressiveness, lack of support for user-defined extensions. These observations motivated the development of BeepBeep, an event stream processing engine that attempts to reconcile CEP and RV by providing a general environment that can accommodate queries and specifications from both. In BeepBeep's generic architecture, basic units of computation called \emph{processors} can be freely composed to evaluate a wide range of expressions. Given an appropriate toolbox of processors, properties involving extended finite-state machines, temporal logic, aggregation and various other concepts can be evaluated. Moreover, through the modular mechanism of \emph{palettes}, end users can easily create their own processors, thereby extending the expressiveness of the tool. BeepBeep also proposes its own declarative input language, \esql{}, which provides an alternative to creating processor chains through ``glue'' code.

Despite our efforts for designing a simple and extensible query language, our experiments revealed that very often, a simple manipulation of processors through a GUI would be a much easier way to write processing chains than large blocks of SQL-like text, irrespective of the actual language used. Consequently, work is planned on developing a simple, Aurora-like box interface for creating and modifying queries.

BeepBeep's goal is to occupy a currently vacant niche among event stream processing engines: it lies somewhere in between low-level command line scripts for small trace crunching tasks, on one end, and heavy distributed event processing platforms on the other. The variety of proposed palettes, combined with a simple computational model, makes it suitable for the definition of clean and readable processing chains at an appropriate level of abstraction. While top-notch performance was not the first design goal, an experimental evaluation has shown that reasonable throughput can be achieved for a variety of queries. Rather than try to compete with commercial-grade platforms like Storm or Kinesis, BeepBeep could best be viewed as a toolbox for creating expressive computations \emph{within} these environments. As a matter of fact, the development of (straightforward) adapters from BeepBeep to these environments is currently under way.

Several research problems around BeepBeep's concepts of processors and event streams are also left unexplored. For example, BeepBeep currently does not support \emph{lazy evaluation}; if the output of an $n$-ary processor can be determined by looking at fewer than $n$ inputs, all inputs must still be computed and consumed. Implementing lazy evaluation in a stream processing environment could provide some performance benefits, but is considered at the moment as a non-trivial task. In addition, since each processor represents an independent unit of computation communicating through message passing, chains of processors should be easily amenable to parallelization; whether this would bring tangible improvements in terms of throughput is currently unknown. Other straightforward technical improvements, such as the use of the Disruptor data structure in place of queues to improve performance \cite{disruptor}, will also be considered.

In time, it is hoped that BeepBeep will be adopted as a modular framework under which multiple event processing techniques can be developed and coexist, and that their potential for composition will make the sum greater than its parts.


\bibliographystyle{elsarticle-num}



\end{document}